%% file: main.tex
\documentclass[aps,pra,floatfix,superscriptaddress,notitlepage]{revtex4-1}
\usepackage[T2A]{fontenc}
\usepackage[utf8]{inputenc}
\usepackage[english]{babel}

\usepackage{graphicx}
\usepackage{nicefrac}
\usepackage{siunitx}
\usepackage{braket}
\usepackage{amsmath}
\usepackage{amsfonts}
\usepackage{longtable,booktabs}
\usepackage{dcolumn}
\usepackage{tikz}
\usepackage{siunitx}
\usepackage{multirow}
\usepackage{bigdelim}
\usepackage{amssymb}
\usepackage{amsthm}
\usepackage{float}
\usepackage{verbatim}
\usepackage{slashed}
\usepackage{epsf}
\usepackage{bm}
\usepackage{bbm}
\usepackage{longtable}
\usepackage{graphicx}
\usepackage{hyperref}
\usepackage{feynmp}
\usepackage{indentfirst}
\usepackage{siunitx}
\usepackage{subfigure}
\usepackage[table,xcdraw]{xcolor}
\usepackage{booktabs}
\newcommand{\matr}[3]{\langle #1 | #2 | #3 \rangle}
\newcommand{\braam}[1]{\langle#1|}
\newcommand{\ketam}[1]{|#1\rangle}

\usepackage{soul}
\soulregister\cite7
\soulregister\ref7

\begin{document}

\title{Bound-electron self-energy calculations in Feynman and Coulomb gauges: detailed analysis}

\author{M.~A.~Reiter} 
\affiliation{Department of Physics, St. Petersburg State University, Universitetskaya nab. 7/9, 199034 St. Petersburg, Russia}

\author{E.~O.~Lazarev}
\affiliation{School of Physics and Engineering, ITMO University, Kronverkskiy pr. 49, 197101 St. Petersburg, Russia}

\author{D.~A.~Glazov}
\affiliation{School of Physics and Engineering, ITMO University, Kronverkskiy pr. 49, 197101 St. Petersburg, Russia}
\affiliation{Petersburg Nuclear Physics Institute named by B.P. Konstantinov of NRC ``Kurchatov Institute'', Orlova roscha 1, 188300 Gatchina, Leningrad region, Russia}

\author{A.~V.~Malyshev}
\affiliation{Department of Physics, St. Petersburg State University, Universitetskaya nab. 7/9, 199034 St. Petersburg, Russia}
\affiliation{Petersburg Nuclear Physics Institute named by B.P. Konstantinov of NRC ``Kurchatov Institute'', Orlova roscha 1, 188300 Gatchina, Leningrad region, Russia}

\author{A.~V.~Volotka}
\affiliation{School of Physics and Engineering, ITMO University, Kronverkskiy pr. 49, 197101 St. Petersburg, Russia}

\begin{abstract}
The energy correction associated with the self-energy diagram is the leading (in magnitude) and fundamental (in significance) contribution to the Lamb shift in highly charged ions. Conventional approaches to this correction rely on partial-wave expansions, which is a stumbling block limiting accuracy. To elucidate the issue, we perform a comprehensive comparative analysis of partial-wave-expansion convergence between two gauges: Feynman and Coulomb. Some tricks for improving the convergence are discussed as well.

\end{abstract}

\maketitle

\section{Introduction}

Quantum electrodynamics (QED) effects are indispensable for the accurate theoretical description of various atomic systems. To account for relativistic effects, the corresponding calculations are to be performed non-perturbatively in the nuclear-strength parameter $\alpha Z$ ($\alpha$ is the fine-structure constant and $Z$ is the nuclear charge). In the present work, we employ the related formalism known as the Furry picture of QED~\cite{PhysRev.81.115}. We note that the corresponding approach is relevant not only for highly charged ions~\cite{mohr_main, Yerokhin:2003:47, Sapirstein:2008:25, Glazov:2011:71, volotka_progress_2013, Shabaev:2018:60, indelicato_qed_2019, SHABAEV202494}, but also finds application in the case of light systems including the hydrogen atom, see, e.g., Refs.~\cite{Jentschura:1999:53, Yerokhin:2015:233002, PhysRevLett.133.251803, Mohr:2025:025002}. An alternative, and in some sense complementary, nonrelativistic QED approach~\cite{caswell}, primarily applicable to low-$Z$ systems, is beyond the scope of this paper. One of the leading QED corrections to 
energy levels originates from the one-electron self-energy (SE) diagram, shown in Fig.~\ref{fig:se_dm}.

However, high-precision evaluation of the corresponding contribution holds significance
not only for studying the Lamb shift and transition energies in hydrogen-like ions and other few-electron systems~\cite{Stohlker, PhysRevLett.94.223001, PhysRevLett.66.1434, PhysRevLett.91.073202, PhysRevLett.95.233003, Trassinelli_2009, Trassinelli_2011, Loetzsch2024} (for
related theory, see, e.g., Refs.~\cite{PhysRevLett.97.253004, PhysRevA.77.032501, PhysRevA.83.012504, se_yerokh, PhysRevA.71.062104, PhysRevA.100.062506}), but also for investigating a wide range of atomic properties. Accurate calculations of the SE contribution is a cornerstone in the \textit{ab initio} treatment of 
the one-electron QED corrections to the $g$ factor~\cite{PhysRevA.62.032510, yerokhin_self-energy_2002},
hyperfine structure~\cite{PhysRevA.64.012506, PhysRevA.63.032506, PhysRevLett.100.163001, PhysRevA.81.012502}, 
quadratic Zeeman splitting~\cite{agababaev2025qedeffectsquadraticzeeman}, 
nuclear-magnetic shielding \cite{Yerokhin:2011:043004, Yerokhin:2012:022512}, 
E1~\cite{E1_Kozlov} and M1~\cite{M1_Volotka} transition amplitudes. 
The SE contribution also plays an important role in the treatment of various many-electron QED effects, see, e.g.,~\cite{Yerokhin:1997:361, Persson:1996:2805, PhysRevA.60.3522, PhysRevLett.103.033005, PhysRevLett.112.253004} and references therein. 
In all these applications it is often necessary to calculate off-diagonal matrix elements of the SE operator. However, the one-electron (diagonal) SE contribution is the simplest example that one can and should use to probe different methods and numerical approaches.


Within the Furry-picture formalism, an issue of representing the bound-electron propagator in calculations naturally arises. Unlike the free-electron propagator, the propagator in the presence of external field generally does not have a closed-form expression. However, when the spherical symmetry is present, as is the case of atomic electrons, the partial-wave (PW) expansion helps to construct the propagator. With this in mind, we use two state-of-the-art approaches: the Green's function (GF) method~\cite{MOHR197426, MOHR197452, PhysRevA.38.5066, sym12050800}, and the finite-basis-set (FBS) method~\cite{PhysRevA.37.307, S1, S2, DKB}. 

Within the first approach, the bound-electron propagator is represented in terms of the Dirac-equation solutions bounded at infinity and at origin. For the point-nucleus Coulomb potential, the corresponding solutions are expressed analytically via the Whittaker functions~\cite{PhysRev.101.843}. For more realistic models of nuclear-charge distribution, e.g., the Fermi model, or other spherically symmetric potentials, e.g., some local screening potentials, these solutions can be found numerically. This approach enables one to perform calculations with a high numerical accuracy when large numerical cancellations occur or when the PW expansion does not converge rapidly. Note that the Green's function, considered as a function of any of its radial arguments, is discontinuous, when the arguments coincide. Therefore, obtaining accurate results for the matrix elements of the Green's function requires special care.

Within the second approach, the electron Green's function is represented using the FBS's. The basis sets for the Dirac equation can be constructed, e.g., from the B splines~\cite{J_Sapirstein_1996, johnson, DKB}, Sturm orbitals~\cite{S1, S2, VL} or Gaussians~\cite{boys_gauss, se_gauss}.  The corresponding numerical procedures can easily incorporate any spherically symmetric potential. Moreover, for systems that do not possess spherical symmetry, the FBSs can also be readily prepared~\cite{PhysRevA.89.012514, Artemyev:2015:243004, DIRAC25}.
The advantage of this method is that it allows one to easily separate out and exclude, if necessary, the contribution of a specific bound state to the Green's function. Note also that this method provides an approximation to the Green's function, which is a continuous function of the radial arguments. Nevertheless, the basis-set method has some important drawbacks when compared to the GF approach. Because of an additional parameter, the number of basis functions $N$, the obtained results should be extrapolated to $N \rightarrow \infty$. Typically, this sets a limit on the computational accuracy. Furthermore, the number of partial waves, usually accessible in practical calculations by means of the FBS method, is limited and smaller than that in the Green's function method~\cite{sym12050800}. In this paper, we use the FBS method with the basis obtained from B splines within the dual-kinetic-balance (DKB) approach~\cite{DKB}. The latter method eliminates the so-called spurious states and establishes the correct asymptotics of wave functions in the non-relativistic limit. The effectiveness of this basis 
for calculating the SE correction has been demonstrated, e.g., in Ref.~\cite{DKB} for the Coulomb potential and in Ref.~\cite{Glazov:06:pla} for local screening potentials. 

The SE diagram contains ultraviolet (UV) divergences. In the present work, these divergences are renormalized using the conventional approach
based on the potential expansion of the bound-electron propagator~\cite{snyderman}. Within this expansion, three contributions naturally arise: zero-, one-, and many-potential terms. Only the zero- and one-potential terms, which correspond to the first and second terms of the potential expansion, are UV divergent. They are treated separately in momentum space together with the mass counterterm. The remaining many-potential term
is considered in coordinate space using the PW expansion. The truncation of this expansion and subsequent estimation of the residual is generally the main source of numerical uncertainty of the results.

The total SE contribution is gauge invariant, unlike the individual terms of its potential expansion. It is therefore intriguing to compare the magnitude of these terms in different gauges of the photon propagator. Starting from the pioneering works~\cite{MOHR197426, MOHR197452}, the SE correction is calculated predominantly in the Feynman gauge, see, e.g., Refs.~\cite{MOHR197452, PhysRevA.26.2338, base_1999, PhysRevA.63.042512, PhysRevA.69.064103, PhysRevA.95.060501} and references therein. In the Coulomb gauge, similar calculations were carried out in Refs.~\cite{hen, yer_2025}. These works suggest that the many-potential contribution is significantly smaller in the Coulomb gauge. 
However, a smaller absolute value does not necessarily imply better convergence of the PW expansion. Therefore, a thorough comparative analysis of the many-potential contributions in both gauges is in demand. Note that other covariant gauges, aside from the Feynman one, can also be applied to the SE calculations. For example, the Fried-Yennie gauge~\cite{PhysRevD.47.3647} was also considered in Ref.~\cite{yer_2025}. In the present work, however, we restrict our consideration only to the Feynman and Coulomb gauges.

Another important issue in the SE calculations is to somehow accelerate the convergence of PW expansions mentioned above. It is reasonable to assume that the poor convergence of the many-potential contribution is due to the leading term of its potential expansion. Therefore, the natural first step is to additionally subtract the two-potential term, which should improve the convergence of the remainder. The subtracted term has to be evaluated separately with high precision within some alternative method. In the following, we will refer to this approach as the two-potential scheme. It should be noted, however, that the direct calculation of the two-potential term in momentum space in a closed form, i.e., not resorting to the PW expansion, turns out to be a challenging problem~\cite{PhysRevLett.100.163001, Yer2010}.
 
Another convergence-acceleration approach, hereafter refered to as the Sapirstein-Cheng (SC) scheme, was proposed recently in Ref.~\cite{sap_orig}. Within the framework of this promising scheme, the quasi-two-potential contribution, which is an approximation to the two-potential term, is subtracted from the many-potential term instead. In contrast to the exact two-potential term, its quasi-two-potential counterpart can be easily calculated in momentum space. For the point-nucleus case, calculations of the SE correction using this acceleration scheme have been performed in Ref.~\cite{yer_2025} for different gauges. However, the case of an extended nucleus is also worth examining.

Yet another convergence-acceleration approach was developed earlier for the SE contribution to the Lamb shift in Ref.~\cite{PhysRevA.72.042502}. We refer to it as the Yerokhin, Pachucki, and Shabaev (YPS) scheme. This scheme also improves the convergence by subtracting an approximation to the many-potential contribution, that can be calculated in a closed form. The YPS scheme was initially developed for the calculations in the Feynman gauge, and was later generalized to the Coulomb gauge in Ref.~\cite{yer_2025}. Judging by the results of the latter paper, it is an extremely powerful scheme, however, it is difficult to generalize it to more complex diagrams, unlike other convergence-acceleration approaches. For instance, the two-potential scheme has recently been applied in the QED calculations of the quadratic Zeeman effect~\cite{agababaev2025qedeffectsquadraticzeeman}, and some modifications of the SC scheme have been applied to the calculations of the two-electron SE~\cite{sap_alexey} and two-loop SE~\cite{PhysRevLett.133.251803, PhysRevA.111.042820} contributions. For this reason, we do not consider the YPS scheme in the current study.

Thus we thoroughly examine the application of several different schemes to the SE calculations in both the Coulomb and Feynman gauges. We conduct an independent of Ref.~\cite{yer_2025} analysis, consider in details the issue of the PW-expansion convergence, evaluate the SE correction for finite-size nuclei, and collect the necessary formulas in a convenient form that allow one to make the corresponding calculations.

Throughout this article the relativistic units ($\hbar = c = 1$) and the Heaviside charge unit ($\alpha=e^2/(4\pi), e<0$) are used. We use roman style ($\rm{p}$) for four-vectors, bold style ($\bf{p}$) for space vectors, and italic style ($p$) for scalars. 

\newpage
\section{Self-energy contribution}
\subsection{Basic formalism}

Within the framework of the Furry picture, the description of a bound electron starts with the Dirac equation:  
\begin{equation}
    \left[-i(\pmb{\alpha} \cdot \pmb{\nabla}) + \beta m_e + V(\mathbf{r})\right] |a\rangle = \varepsilon_a |a\rangle,
\label{eq:dirac}
\end{equation}
where $|a\rangle$ and $\varepsilon_a$ are the Dirac wave function and relativistic energy, $\pmb{\alpha}$ and $\beta$ are the Dirac matrices, $V(\mathbf{r})$ is the binding potential, which is a function of the position vector $\mathbf{r}$. In the following, we consider only hydrogen-like ions where the binding potential is spherically symmetric, $V(\mathbf{r}) = V(r)$, $r = |\mathbf{r}|$. Then, the solution of Eq.~(\ref{eq:dirac}) can be represented in the form:
\begin{equation}
    \ketam{a} \leftrightarrow \psi_a(\mathbf{r}) = {\begin{pmatrix}g_a(r) \chi_{\kappa_a m_a} (\mathbf{\hat{r}}) \\ i f_a(r) \chi_{-\kappa_a m_a} (\mathbf{\hat{r}}) \end{pmatrix}},
\label{wf_coord}
\end{equation}
where $g_a(r)$ and $f_a(r)$ are the large and small radial components, $\chi_{\kappa_a m_a} (\mathbf{\hat{r}})$ is the spin-angular spinor, $\mathbf{\hat{r}}=\mathbf{r}/r$, $\kappa_a=(j_a+1/2)(-1)^{j_a+l_a+1/2}$ is the Dirac angular quantum number, $l_a$ is the spatial parity, and $m_a$ is the projection of the total angular momentum $j_a=|\kappa_a|-1/2$. The radial wave functions of the bound-electron state are characterized by $\kappa_a$ and the radial quantum number $n_{r_a}$: $g_a(r) = g_{{n_r}_a \kappa_a}(r)$, $f_a(r) = f_{{n_r}_a \kappa_a}(r)$.

\begin{figure}
\centering
\includegraphics[width=0.2\textwidth]{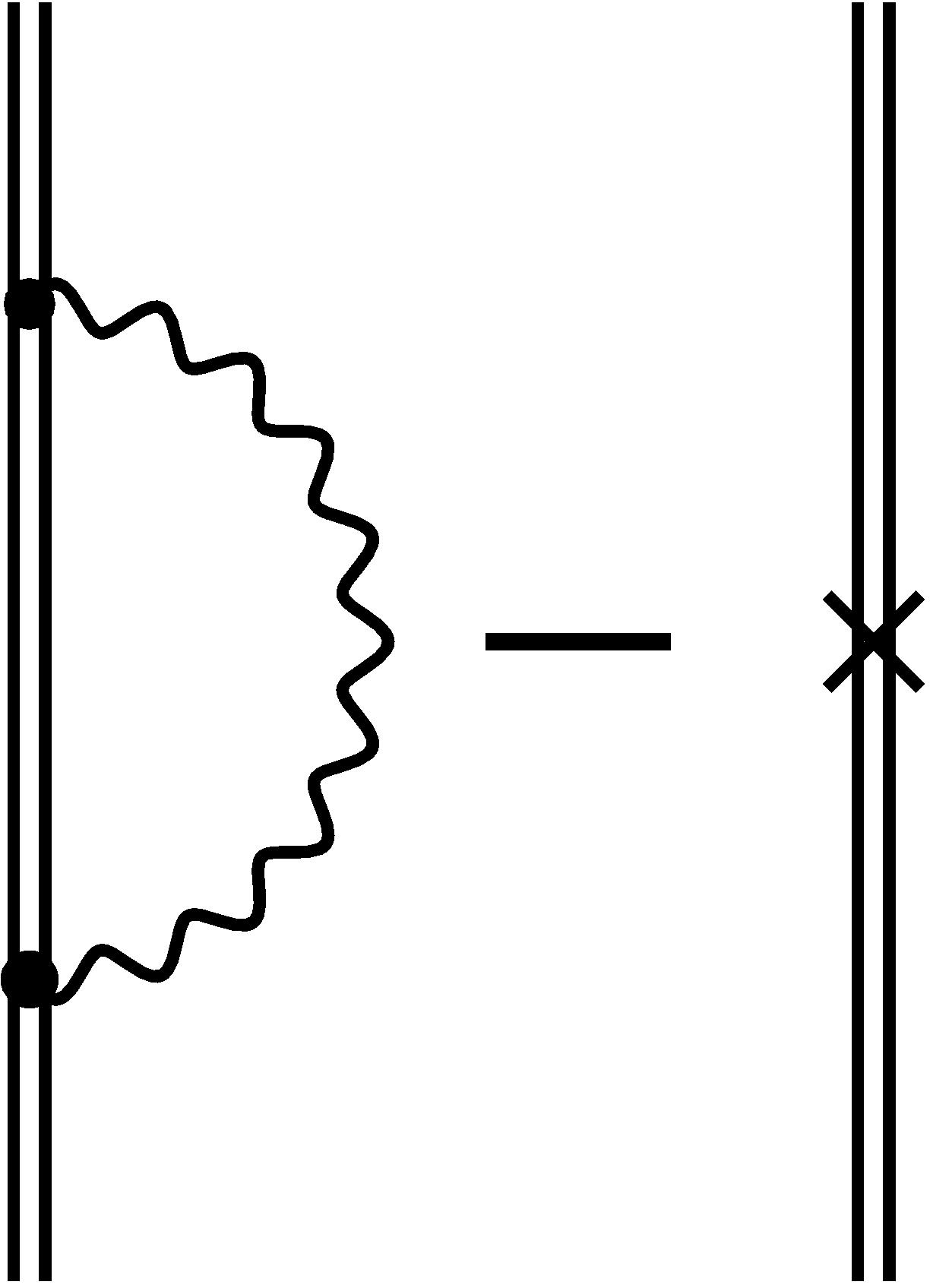}
\caption{Self-energy diagram with the related mass counterterm. The double line indicates the electron propagator in the external field of the nucleus, the wavy line denotes the photon propagator, and the cross stands for the mass counterterm.}
\label{fig:se_dm}
\end{figure}

The energy shift $\Delta \varepsilon_a$ of the bound state $|a\rangle$ due to the first-order self-energy correction, which is graphically represented in Fig.~\ref{fig:se_dm}, is given by the real part of the following expression~\cite{mohr_main}:

\begin{equation}
\begin{gathered}
    \Delta \varepsilon_{a} =2 i \alpha \int_{-\infty}^{\infty} d \omega \int d^{3} \mathbf{r}_1  d^{3} \mathbf{r}_2 \psi^{\dag}_{a} (\mathbf{r}_{1}) \alpha^{\mu} \\
    \times G(\varepsilon_{a}-\omega, \mathbf{r}_1, \mathbf{r}_2) \alpha^{\nu} \psi_{a}(\mathbf{r}_{2}) D_{\mu \nu}(\omega, \mathbf{r}_{12}) \\
    -\delta m \int d^{3} \mathbf{r} {\psi}^\dagger_a(\mathbf{r}) \beta \psi_{a}(\mathbf{r}),
\end{gathered}
\label{se_general_1999}
\end{equation}
where $\mathbf{r}_{12}=\mathbf{r}_{1}-\mathbf{r}_{2}$, $D_{\mu \nu}(\omega, \mathbf{r}_{12})$ is the photon
propagator, $\alpha^\mu = (1, {\pmb{\alpha}})$, $G(\varepsilon, \mathbf{r}_1, \mathbf{r}_2)$ is the bound-electron Green's function, and $\delta m$ is the mass counterterm. Let us define the gauge-dependent operator $I(\omega, \mathbf{r}_1, \mathbf{r}_2)$ according to
\begin{equation}
    I(\omega, \mathbf{r}_1, \mathbf{r}_2) = e^2 \alpha^\nu \alpha^\mu D_{\mu \nu}(\omega, \mathbf{r}_{12}).
\end{equation}
In the Feynman and Coulomb gauges, this operator takes the forms $I_\text{F}$ and $I_\text{C}$, respectively,
\begin{equation}
    I_\text{F}(\omega, \mathbf{r}_1, \mathbf{r}_2)= \alpha \left[ 1 - ({\pmb{\alpha}}_1 \cdot {\pmb{\alpha}}_2) \right] \frac{e^{i \hat{\omega} r_{12}}}{r_{12}},
\label{inter_F}
\end{equation}
\begin{equation}
    I_\text{C}(\omega, \mathbf{r}_1, \mathbf{r}_2)= \alpha \left[ \frac{1}{r_{12}} - ({\pmb{\alpha}}_1 \cdot {\pmb{\alpha}}_2)
    \frac{e^{i \hat{\omega} r_{12}}}{r_{12}} + (\pmb{\alpha}_1 \cdot \pmb{\nabla}_1)(\pmb{\alpha}_2 \cdot \pmb{\nabla}_2) \frac{e^{i \hat{\omega} r_{12}}-1}{\omega^2r_{12}} \right],
\end{equation}
where $r_{12}=|\mathbf{r}_{12}|$, $\hat{\omega} = \sqrt{\omega^2+i 0}$, and the branch of the square root is fixed with the condition $\text{Im \,} \sqrt{\omega^2+i 0} > 0$. 

The bound-electron propagator $G$ can be expressed using the spectral representation, 
\begin{equation}
    G(\omega, \mathbf{r}_1, \mathbf{r}_2) = \sum_n \frac{\psi_n(\mathbf{r}_1) \psi_n^\dagger (\mathbf{r}_2)}{\omega - \varepsilon_n(1-i0)},
\label{eq:el_prop_general}
\end{equation}
where the summation runs over the complete Dirac spectrum. 
In shortened form, the formula (\ref{eq:el_prop_general}) can be written as follows:
\begin{equation}
    G(\omega) = \sum_n \frac{|n \rangle \langle n|}{\omega - \varepsilon_n^-},
\label{el_prop}
\end{equation}
where the notation $\varepsilon_n^- = \varepsilon_n (1 - i 0)$ was introduced.
Then, the energy shift (\ref{se_general_1999}) can be expressed as:
\begin{equation}
    \Delta \varepsilon_a = \langle a | \gamma^0  \Big[\Sigma(\varepsilon_a) - \delta m \Big] | a \rangle,
\label{SE_full}
\end{equation}
where $\gamma^0 \equiv \beta$ and the diagonal matrix element of the one-loop self-energy operator $\Sigma(E)$ is given by:
\begin{equation}
    \langle a | \gamma^0 \Sigma(E) | a \rangle = \frac{i}{2 \pi} \int_{-\infty}^{\infty} d \omega \sum_n \frac{ \langle a n | I(\omega) | n a \rangle}{E - \omega - \varepsilon_n^-}.
\label{SE_general}
\end{equation}

The expression (\ref{SE_general}) contains the UV divergences. To properly treat them, we expand the bound-electron propagator $G$ in powers of the binding potential $V$~\cite{blundell_snyderman}. The potential expansion can be conveniently written as:
\begin{equation}
\begin{gathered}
    G(\omega) = \sum_f \frac{\ketam{f} \braam{f}}{\omega - \varepsilon_f^-} + \sum_{f_1, f_2} \frac{\ketam{f_1} \matr{f_1}{V}{f_2} \braam{f_2}}{(\omega - \varepsilon_{f_1}^-)(\omega - \varepsilon_{f_2}^-)} + \\ + \sum_{f_1, f_2, n} \frac{\ketam{f_1} \matr{f_1}{V}{n} \matr{n}{V}{f_2} \braam{f_2}}{(\omega - \varepsilon_{f_1}^-)(\omega - \varepsilon_{n}^-)(\omega - \varepsilon_{f_2}^-)},
\end{gathered}
\label{potential_exp}
\end{equation}
where the states $\ketam{n}$ correspond to the bound-electron spectrum, the states $\ketam{f}$, $\ketam{f_1}$, and $\ketam{f_2}$ correspond to the free-electron spectrum, and the last term is expressed via the bound-electron Green's function $G(\omega)$ itself and encompasses an infinite number of terms corresponding to the two or more interactions with the nucleus. As before, all summations in Eq.~(\ref{potential_exp}), are carried out over the complete spectra for the corresponding Dirac Hamiltonians.
\begin{figure}
\centering
\includegraphics[width=0.5\textwidth]{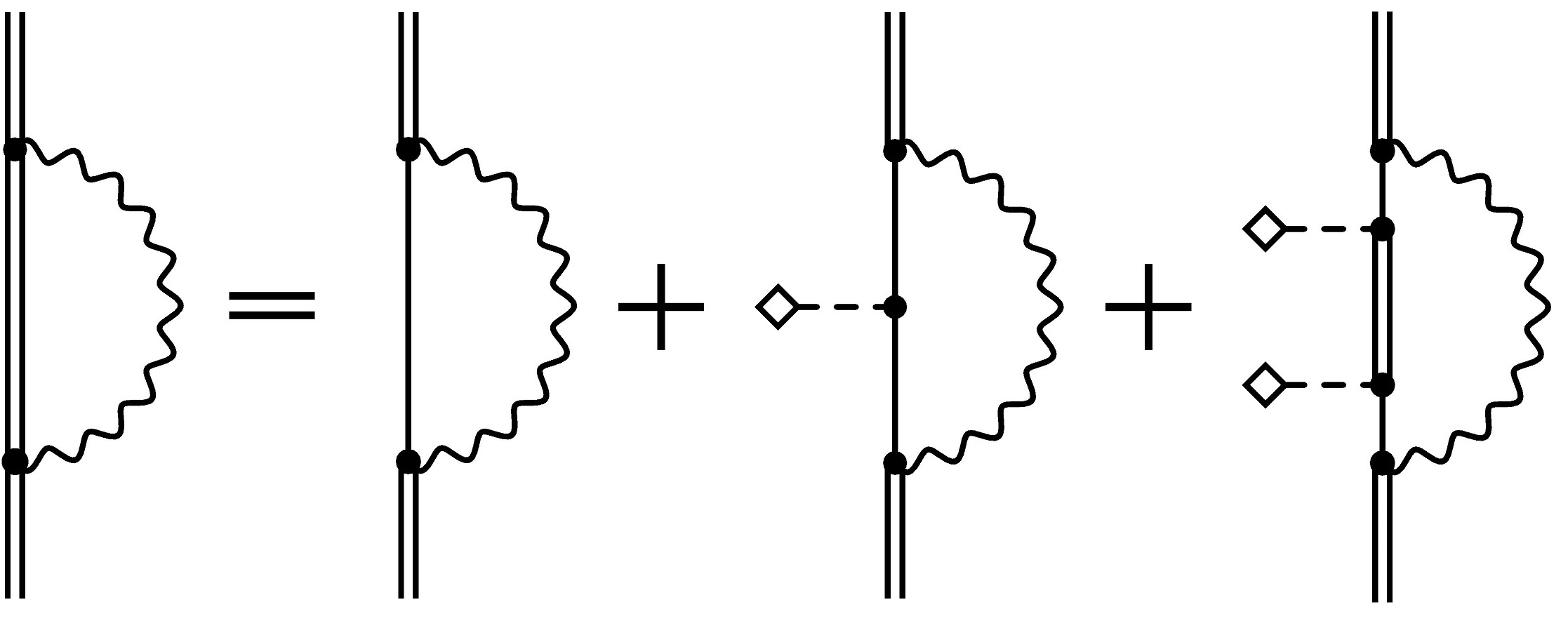}
\caption{
Potential expansion of the self-energy diagram. The dashed line ended by a rhombus denotes the interaction with the binding potential. The single line denotes the free-electron propagator. The mass counterterm is not shown.
}
\label{fig:prop_decomp}
\end{figure}
Substituting the potential expansion (\ref{potential_exp}) into Eq.~(\ref{SE_general}) results in the SE diagram being split into three contributions, as shown graphically in Fig.~\ref{fig:prop_decomp}. These contributions are commonly referred to as the zero-, one-, and many-potential terms. The zero- and one-potential terms diverge, whereas the many-potential term is UV finite. The UV-divergent terms have to be renormalized along with the mass counterterm. For this aim, they are considered in momentum space, and the dimensional regularization is applied to isolate and eliminate the UV divergences. This renormalization procedure yields the UV-finite contributions $\Delta \varepsilon_a^{\text{0p}}$ and $\Delta \varepsilon_a^{\text{1p}}$ to the Lamb shift. The many-potential term $\Delta \varepsilon_a^{\text{Mp}}$ is calculated in the coordinate space. Finally, the SE correction to  energy levels is expressed as the gauge-invariant sum of three contributions:
\begin{equation}
\begin{gathered}
    \Delta \varepsilon_a
    = \Delta \varepsilon_a^{\text{0p}} + \Delta \varepsilon_a^{\text{1p}} + \Delta \varepsilon_a^{\text{Mp}}.
\end{gathered}
\end{equation}
In the following subsections we consider in detail these terms and some specific methods to improve the convergence of the partial-wave expansion in the many-potential part.


\subsection{Zero- and one-potential terms}

The Fourier transform of the wave function (\ref{wf_coord}), which solves the Dirac equation (\ref{eq:dirac}) in a spherically-symmetric potential, reads as:
\begin{equation}
    \psi_a(\mathbf{p}) = \int d^3 {\mathbf{r}} e^{-i ({\mathbf{p \cdot r}})} \psi_a({\mathbf{r}}) = i^{-l_a} {\begin{pmatrix}\widetilde{g}_a({p}) \chi_{\kappa_a m_a} (\mathbf{\hat{p}}) \\ \widetilde{f}_a(p) \chi_{-\kappa_a m_a} (\mathbf{\hat{p}}) \end{pmatrix}},
\label{wf_momenta}
\end{equation}
where $\mathbf{\hat{p}}=\mathbf{p}/p$, $l_a=|\kappa_a+1/2|-1/2$, and for the bound state $\widetilde{g}_a(p) = \widetilde{g}_{{n_r}_a \kappa_a}(p)$ and $\widetilde{f}_a(p) = \widetilde{f}_{{n_r}_a \kappa_a}(p)$. In the point-nucleus case, there are analytical expressions for $\widetilde{g}_a(p)$ and $\widetilde{f}_a(p)$, which can be found, e.g., Ref.~\cite{mohr_main}. For an arbitrary potential, they can be obtained numerically according to:
\begin{equation}
    \widetilde{g}_a(p) = 4 \pi \int_0^\infty dr r^2 j_{l_a}(pr) g_a(r),
\end{equation}
\begin{equation}
    \widetilde{f}_a(p) = - 4 \pi \frac{\kappa_a}{|\kappa_a|} \int_0^\infty dr r^2 j_{2j_a-l_a}(pr) f_a(r), 
\label{ftr_wf_radial}
\end{equation}
where $j_l$ is the spherical Bessel function of the first kind. 

The zero-potential term is expressed as:
\begin{equation}
\begin{gathered}
    \Delta \varepsilon_a^{\text{0p}} = \int \frac{d^3 \mathbf{p}}{(2 \pi)^3} \psi_a^\dagger(\mathbf{p}) \gamma^0 \Sigma^{(0)}_R (\mathrm{p}) \psi_a(\mathbf{p}),
\end{gathered}
\label{SE0praw}
\end{equation}
where $\Sigma^{(0)}_R$ is the renormalized self-energy operator in free-particle QED and $\mathrm{p} = (\varepsilon_a, \mathbf{p})$. The explicit form of the operator $\Sigma_R^{(0)}$ is presented in Appendix~\ref{ap:zero_pot} in Eq.~(\ref{sigma0ren}) with the gauge-dependent coefficients $a$, $b$, and $c$ defined in Eqs.~(\ref{abc_f}) and (\ref{abc_c}).
After substituting Eqs.~(\ref{sigma0ren}) and (\ref{wf_momenta}) into Eq.~(\ref{SE0praw}), it becomes possible to analytically perform the integration over the angular variables using the relation $(\pmb{\sigma} \cdot \mathbf{\hat{ p}}) \chi_{\kappa\mu}(\mathbf{\hat{p}}) = - \chi_{-\kappa\mu}(\mathbf{\hat{p}})$ and the orthonormality condition for the spin-angular spinors. This results in an expression for the zero-potential contribution, which is convenient for numerical calculations:
\begin{equation}
    \Delta \varepsilon_a^{\text{0p}} = \frac{\alpha}{4 \pi} \int_0^\infty \frac{p^2 dp}{(2 \pi)^3} \{a(\varepsilon_a,p) [\widetilde{g}_a^2-\widetilde{f}_a^2] \\ + b(\varepsilon_a, p) [\varepsilon_a (\widetilde{g}_a^2+\widetilde{f}_a^2) + 2 p \widetilde{g}_a \widetilde{f}_a ] 
    + c(\varepsilon_a, p) [\widetilde{g}_a^2 + \widetilde{f}_a^2]\},
\label{SE0p}
\end{equation}
where the dependence of $\widetilde{g}_a$ and $\widetilde{f}_a$ on $p$ is omitted for brevity.

The one-potential term is expressed as:
\begin{equation}
\begin{gathered}
    \Delta \varepsilon_a^{\text{1p}} = \int \frac{d^3 \mathbf{p}' d^3 \mathbf{p}}{(2 \pi)^6} \psi_a^\dagger(\mathbf{p}') \gamma^0 \Gamma^{(0)}_R (\mathrm{p}', \mathrm{p}) \widetilde{V}(|\mathbf{p}' - \mathbf{p}|) \psi_a(\mathbf{p}),
\end{gathered}
\label{SE1praw}
\end{equation}
where $\Gamma^{(0)}_R$ is the renormalized free-particle vertex function, the explicit form of which with the gauge-dependent coefficients $A$-$G_2$ is given in the Appendix~\ref{ap:one_pot} in Eq.~(\ref{vertex_function}), $\widetilde{V}$ is the Fourier transform of the binding potential, and $\mathrm{p} = (\varepsilon_a, \mathbf{p})$ and $\mathrm{p}' = (\varepsilon_a, \mathbf{p}')$. In this case, the angular integration can be performed using the identity: 
\begin{equation}
    \frac{1}{2 j + 1} \sum_m \chi_{\kappa m}^\dagger (\mathbf{\hat{p}}') \chi_{\kappa m} (\mathbf{\hat{p}}) = \frac{1}{4 \pi} P_l(z),
\end{equation}
where $z$ is the cosine of the angle between the vectors $\mathbf{p}$ and $\mathbf{p}'$, namely, $z = (\mathbf{\hat{p}} \cdot \mathbf{\hat{p}'})$, and $P_l$ is the Legendre polynomial. The final expression for the one-potential term reads as
\begin{equation}
\begin{gathered}
    \Delta \varepsilon_a^{\text{1p}} = \frac{\alpha}{2 (2 \pi)^6} \int_0^{\infty} p'^2 dp' \int_0^{\infty} p^2 dp \int_{-1}^1 dz \widetilde{V}(q) \{X_1 P_{l_a}(z) + X_2 P_{2j_a - l_a}(z)\},
\end{gathered}
\label{SE1p}
\end{equation}
where $q^2 = p^2 + p'^2 - 2 p p' z$. The coefficients $X_1$ and $X_2$ are defined as: 
\begin{align}
    X_1 &= A \widetilde{g}'_a \widetilde{g}_a + \varepsilon_a (B_1+B_2) K_1' \widetilde{g}_a
    +  \varepsilon_a   (C_1+C_2)   \widetilde{g}'_a K_1 \nonumber
\\ 
    &+  D K_1 K_1' + \varepsilon_a   (H_1+H_2) \widetilde{g}'_a \widetilde{g}_a +G_1 K_1' \widetilde{g}_a + G_2 \widetilde{g}'_a K_1,\\
    X_2 &= A \widetilde{f}'_a \widetilde{f}_a
    +  \varepsilon_a (B_1+B_2) K_2' \widetilde{f}_a
    +  \varepsilon_a (C_1+C_2) \widetilde{f}'_a K_2 \nonumber
\\ 
    &+ D K_2 K_2' - \varepsilon_a (H_1+H_2) \widetilde{f}'_a \widetilde{f}_a - G_1 K_2' \widetilde{f}_a - G_2 \widetilde{f}'_a K_2, 
\end{align} 
where
\begin{equation}
    K_1 = \varepsilon_a \widetilde{g}_a + p \widetilde{f}_a, \; K_1' = \varepsilon_a \widetilde{g}'_a + p' \widetilde{f}'_a, \; K_2 = \varepsilon_a \widetilde{f}_a + p \widetilde{g}_a, \; K_2' = \varepsilon_a \widetilde{f}'_a + p' \widetilde{g}'_a.
\end{equation} 
For brevity, the dependence of wave functions on $p$ and $p'$ is omitted. For the functions of $p'$, an additional prime is added. Thus, the shorthand notations are $\widetilde{g}_a = \widetilde{g}_a(p)$, $\widetilde{f}_a = \widetilde{f}_a(p)$, $\widetilde{g}'_a = \widetilde{g}_a(p')$, and $\widetilde{f}'_a = \widetilde{f}_a(p')$.

\subsection{Many-potential term}
\label{formalism:mpot}

As noted above, the remaining many-potential contribution is calculated in coordinate space using the PW expansion.
Therefore, let us start with the PW expansion of the bound-electron propagator~(\ref{eq:el_prop_general}), inspired by the explicit form of the Dirac-equation solution~(\ref{wf_coord}):
\begin{equation}
    G(\omega, \mathbf{r}_1, \mathbf{r}_2) = \sum_{\kappa_n}{\begin{pmatrix}G_{{\kappa}_n}^{11}(\omega, {r_1, r_2}) \pi_{\kappa_n}^{++} (\mathbf{\hat{r}}_1, \mathbf{\hat{r}}_2)&-iG_{\kappa_n}^{12}(\omega, {r_1, r_2}) \pi_{\kappa_n}^{+-} (\mathbf{\hat{r}}_1, \mathbf{\hat{r}}_2)\\i G_{\kappa_n}^{21}(\omega, {r_1, r_2}) \pi_{\kappa_n}^{-+} (\mathbf{\hat{r}}_1, \mathbf{\hat{r}}_2)&G_{\kappa_n}^{22}(\omega, {r_1, r_2}) \pi_{\kappa_n}^{--} (\mathbf{\hat{r}}_1, \mathbf{\hat{r}}_2) \end{pmatrix}},
\label{eq:G_kp_exp}
\end{equation}
where $\pi^{\pm \pm} (\mathbf{\hat{r}}_1, \mathbf{\hat{r}}_2) = \sum_{m_n} \chi_{\pm {\kappa_n} {m_n}} (\mathbf{\hat{r}}_1) \chi_{\pm {\kappa_n} {m_n}}^\dagger (\mathbf{\hat{r}}_2)$ and the components of the radial Green's function $G_{\kappa_n}^{ij}$ are obtained in different ways depending on the method used: FBS or GF. 

For a fixed $\kappa_n$, the  spectrum of the radial Dirac equation is non-degenerate, so it is convenient to introduce the index $i_n$ that enumerates solutions for the given angular symmetry:
\begin{equation}
\label{eq:rad:Dirac}
\begin{cases}
    \left( -\dfrac{d}{dr} + \dfrac{\kappa_n-1}{r} \right)f_{{i_n} {\kappa_n}} + \left(  V+m_e\right)g_{{i_n} {\kappa_n}} = \varepsilon_{{i_n} {\kappa_n}}g_{{i_n} {\kappa_n}} \, , \\
    \left( \dfrac{d}{dr} + \dfrac{\kappa_n+1}{r} \right)g_{{i_n} {\kappa_n}} + \left(  V-m_e\right)f_{{i_n}, {\kappa_n}} = \varepsilon_{{i_n} {\kappa_n}}f_{{i_n} {\kappa_n}} \,.
\end{cases}             
\end{equation}
Then, Eq.~(\ref{eq:el_prop_general}) results for, e.g., $G^{11}$ in
\begin{equation}
    G_{\kappa_n}^{11}(\omega, {r_1, r_2}) = \sum_{i_n} \frac{g_{{i_n} {\kappa_n}}(r_1) g_{{i_n} {\kappa_n}}(r_2)}{\omega - \varepsilon_{{i_n} {\kappa_n}}^-}.
\label{eq:g11}
\end{equation}
We stress that the index $i_n$ in Eq.~(\ref{eq:g11}) and similar expressions runs over both the positive- and negative-energy Dirac continua as well as all bound states for the chosen $\kappa_n$. 
When using the DKB approach, for each $\kappa_n$ the  spectrum of the radial Dirac equation (\ref{eq:rad:Dirac}) is replaced with the finite set of solutions. In this case, the index $i_n$ runs over these finite sets. When using the GF approach, the radial Green's function $G_{\kappa_n}^{ij}$ can be constructed from the  solutions of the homogeneous Dirac equation, which are bounded at zero, $\phi_{\kappa_n}^0=(g_{\kappa_n}^0\;\;f_{\kappa_n}^0)^\text{{T}}$, and at infinity, $\phi_{\kappa_n}^\infty=(g_{\kappa_n}^\infty \;\; f_{\kappa_n}^\infty)^\text{{T}}$, (see, e.g., Ref.~\cite{sym12050800}):
\begin{equation}
    G_{{\kappa_n}}(\omega,r_1,r_2)= \phi_{{\kappa_n}}^{\infty} (r_1) {\phi_{{\kappa_n}}^{0}}^\text{{T}} (r_2) \theta (r_1-r_2) + \phi_{{\kappa_n}}^{0} (r_1) {\phi_{{\kappa_n}}^{\infty}}^\text{{T}} (r_2) \theta (r_2-r_1).
\label{eq:G_phi_th}
\end{equation}
Note that the solutions $\phi^0$ and $\phi^\infty$ are supposed here to be normalized so that their Wronskian equals one. Expressions similar to Eqs.~(\ref{eq:G_kp_exp})-(\ref{eq:G_phi_th}) can also be obtained for the free-electron Green's function, which we will denote as $G^{(0)}$.

With this in mind, we construct the many-potential part of the bound-electron Green's function which involves two or more interactions with the binding potential and corresponds to the last term in the expression (\ref{potential_exp}): 
\begin{equation}
    G^{(2+)}(\omega,\mathbf{r}_1, \mathbf{r}_2) = \int d^3 \mathbf{x} \int d^3 \mathbf{y} G^{(0)} (\omega,\mathbf{r}_1, \mathbf{x}) V(x) G (\omega,\mathbf{x}, \mathbf{y}) V(y) G^{(0)} (\omega,\mathbf{y}, \mathbf{r}_2). 
\label{pot_decomp_gf}
\end{equation}
Since the potential $V$ is assumed to be spherically symmetric, 
angular integrations in Eq.~(\ref{pot_decomp_gf}) can be easily carried out. As a result, one obtains the same PW expansion for $G^{(2+)}$ as that for $G$ in Eq.~(\ref{eq:G_kp_exp}). The radial components of $G^{(2+)}$ read as
\begin{equation}
    G^{(2+)ij}_{\kappa_n}(\omega,r_1,r_2) = \sum_{k,m} \int dx x^2 \int dy y^2 G^{(0)ik}_{\kappa_n} (\omega,r_1,x) V(x) G^{km}_{\kappa_n} (\omega,x,y) V(y) G^{(0)mj}_{\kappa_n} (\omega,y,r_2).
\label{eq:g_2p_ij}
\end{equation}
In the GF case, because of the $\theta$-functions, radial integrations need to be handled carefully in Eq.~(\ref{eq:g_2p_ij}), see, e.g., Ref.~\cite{Art_trick}. In the framework of the DKB approach, it is convenient to rewrite $G^{(2+)}$ in the form:
\begin{equation}
    G^{(2+)}(\omega) = \sum_{n} \frac{|\check{n}\rangle \langle\check{n}|}{\omega - \varepsilon_{n}^-},
\label{g2p}
\end{equation}
where the new states $\ketam{\check{n}}$ are defined by:
\begin{equation}
    |\check{n}\rangle = \sum_{f} \frac{|f\rangle \langle f | V | n \rangle}{\omega-\varepsilon_{f}^-} \equiv {\begin{pmatrix}\check{g}_{i_n \kappa_n}(r) \chi_{\kappa_n m_n} (\mathbf{\hat{r}}) \\ i \check{f}_{i_n \kappa_n}(r) \chi_{-\kappa_n m_n} (\mathbf{\hat{r}}) \end{pmatrix}}.
\label{eq:n_bird}
\end{equation}
 Then, using Eq.~(\ref{eq:n_bird}), the many-potential contribution is expressed as:
\begin{equation}
    \Delta \varepsilon_a^{\text{Mp}} = \frac{i}{2 \pi} \int_{-\infty}^{\infty} d \omega \sum_n \frac{ \langle a \check{n} | I(\omega) | \check{n} a \rangle}{\varepsilon_a - \omega - \varepsilon_n^-}.
\label{se_mp_bwave}
\end{equation}
The expression~(\ref{se_mp_bwave}) can be readily adjusted to be used within the GF approach. For this aim, one needs to reverse an expression similar to  Eq.~(\ref{eq:g11}):
\begin{equation}
    \sum_{i_n} \frac{\check{g}_{{i_n} {\kappa_n}}(r_1) \check{g}_{{i_n} {\kappa_n}}(r_2)}{\omega - \varepsilon_n^-} \rightarrow G^{(2+)11}_{\kappa_n}(\omega, r_1, r_2).
\end{equation}

The matrix element of the operator $I(\omega)$ can be expressed as follows: 
\begin{equation}
\begin{gathered}
    \langle ab|I( \omega )|cd \rangle= \sum_{JM} (-1)^ {j_{a}-m_{a} + J - M + j_{b} - m_{b}}
    {\begin{pmatrix}j_a&J&j_c\\-m_{a}&M&m_{c} \end{pmatrix}}
    {\begin{pmatrix}j_b&J&j_d\\-m_{b}&-M&m_{d} \end{pmatrix}} \langle a b || I(\omega) || c d\rangle_J,
\end{gathered}
\label{M_el_expansion}
\end{equation}
where the reduced matrix elements, $\langle a b || I(\omega) || c d\rangle_J$, on the right-hand side do not depend on the angular-momentum projections. Their explicit forms in the Feynman and Coulomb gauges are given in Appendix~\ref{ap:m_pot} in Eqs.~(\ref{M_el_f}) and (\ref{M_el_c}), respectively. Finally, the summation over the projections $m_n$ in Eq.~(\ref{se_mp_bwave}) can be done analytically, and we obtain:
\begin{equation}
    \Delta \varepsilon_a^{\text{Mp}} = \frac{i}{2 \pi} \sum_{\kappa_n} \int_{-\infty}^{\infty} d \omega \sum_{J, i_n} \frac {(-1)^{j_n-j_a+J}}{2 j_a+1} \frac{\langle a \check{n} || I(\omega) || \check{n}  a\rangle_J}{\varepsilon_{a} - \omega -\varepsilon_n^-}.
\label{se_mp}
\end{equation}
The sums over $J$ and $\kappa_n$ in Eq.~(\ref{se_mp}) are not independent due to the triangular inequality. In our calculations, the sum over $\kappa_n$ 
is considered as the primary one. As a result, the many-potential contribution is represented by the sum of the often poorly converging PW-expansion series in $\kappa_n$. To calculate it, one has to resort to the procedure of extrapolation in $k=|\kappa_n|$. In what follows, we for brevity will use $\kappa$ instead of $\kappa_n$, if this does not lead to misunderstandings.

\subsection{Partial-wave expansion acceleration schemes}

The convergence of the PW-expansion series could be improved using different acceleration schemes. Their general idea is to separate out and subtract the term $\Delta \varepsilon_{a,x}^{\text{subtr}}$, which contains the slowly converging part of the PW expansion of the many-potential contribution. The index ``$x$'' in $\Delta \varepsilon_{a,x}^{\text{subtr}}$ indicates that the subtracted term should be calculated exactly in the same manner as  $\Delta\varepsilon_a^{\text{Mp}}$, that is in coordinate space using the PW expansion. Naturally that the same term, evaluated separately, but not necessarily within the same approach, so we denote it as $\Delta \varepsilon_{a}^{\text{subtr}}$ without the additional index ``$x$'', should then be added back. The discussed idea can be illustrated by the formula:
\begin{equation} \label{eq:subtraction_scheme}
    \Delta \varepsilon_a^{\text{Mp}} \rightarrow (\Delta \varepsilon_a^{\text{Mp}} - \Delta \varepsilon_{a,x}^{\text{subtr}}) + \Delta \varepsilon_{a}^{\text{subtr}}.
\end{equation}
It is implied that the PW-expansion series for the two contributions in the brackets in Eq.~(\ref{eq:subtraction_scheme}) are subtracted term by term and only then extrapolated in $k$. The resulting difference is expected to demonstrate a better convergence (at least, in terms of absolute values) than the initial series for the many-potential contribution. Therefore, if one can calculate $\Delta \varepsilon_{a}^{\text{subtr}}$ with high accuracy, it will determine the overall improvement in the calculations of the many-potential contribution to the SE correction.

Within the two-potential acceleration scheme (see the Introduction section), the subtracted term is the two-potential contribution, $\Delta \varepsilon_{a,x}^{\text{subtr}}=\Delta \varepsilon_{a,x}^{2\text{p}}$. When adding this term back, one must evaluate it with high precision in order to benefit from applying the scheme. A natural idea is to perform the corresponding calculations in momentum space without using the PW expansions. However, this problem is complicated and leads to complex multidimensional integrals, see, e.g., Ref.~\cite{Yer2010}, where a similar approach was employed in the point-nucleus case for the SE correction to the hyperfine splitting. For this reason, an alternative was proposed in Refs.~\cite{PhysRevLett.98.173004, Art_trick}, where the difference between the many- and two-potential terms was addressed using the FBS method, while the two-potential term itself was also considered in coordinate space but employing the GF method and truncating the corresponding PW summations at higher values of angular momenta. We note that in this case, it is not necessary to solve the system of differential equations numerically for an arbitrary potential, since the PW expansion of the free-electron propagator can be expressed in terms of the spherical Bessel functions \cite{mohr_main} and the desired two-potential term can be obtained by combining three such propagators. In the present work, we study this two-potential scheme but for simplicity the difference between the many- and two-potential terms is also treated based on the GF approach. 
The PW expansion for this difference arises from the Green's function
\begin{align}
    G^{(3+)} = G^{(2+)} - \sum_{f_1, f_2, f_3} \frac{\ketam{f_1} \matr{f_1}{V}{f_2} \matr{f_2}{V}{f_3} \braam{f_3}}{(\omega - \varepsilon_{f_1}^-)(\omega - \varepsilon_{f_2}^-)(\omega - \varepsilon_{f_3}^-)} \,.
\end{align}
We refer to the term corresponding to $G^{(3+)}$ as the three-plus-potential contribution, $\Delta \varepsilon_a^{\text{(3+)p}} \equiv ( \Delta \varepsilon_a^{\text{Mp}} - \Delta \varepsilon_{a,x}^{2\text{p}} )$.


\begin{figure}[h]
\centering
\includegraphics[width=0.5\textwidth]{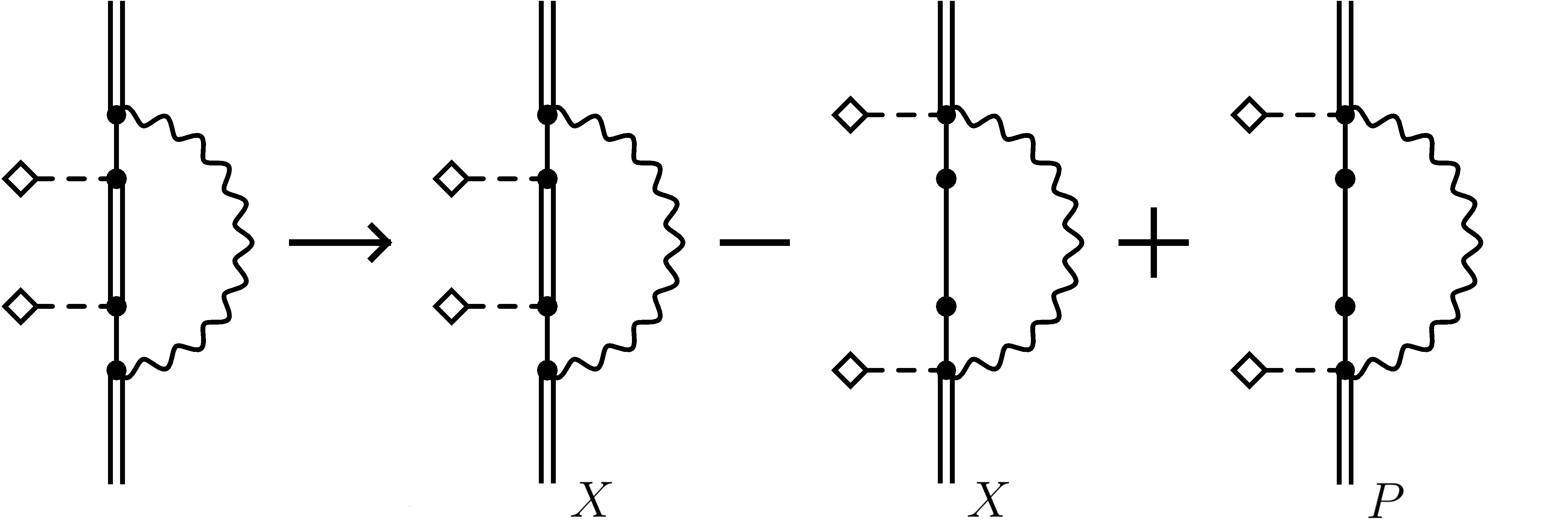}
\caption{Diagrammatic representation of the quasi-two-potential term separation. 
The last term on the right side is calculated in momentum space without using any PW expansion.}
\label{fig:sap_trick}
\end{figure}
The Sapirstein-Cheng (SC) scheme~\cite{sap_orig} is based on the idea that the dominant contribution from the free-electron propagators $G^{(0)}(\omega,\mathbf{r}_1, \mathbf{r}_2)$ comes from the region where both their spatial arguments are close to each other~\cite{MOHR197426}, $\mathbf{r}_1 \approx \mathbf{r}_2$. Therefore, one can obtain an approximation to the two-potential term by ``transposing'' the potentials from the inner electron line to the vertices, where the photon propagator is attached, see Fig.~\ref{fig:sap_trick}. We refer to this approximation as the quasi-two-potential term and denote it as $\Delta \varepsilon_{a,x}^{\text{subtr}}=\Delta \tilde{\varepsilon}_{a,x}^{2\text{p}}$.
The difference between the many- and quasi-two-potential terms, $\Delta \tilde{\varepsilon}_{a, x}^{(3+)\text{p}} \equiv \Delta \varepsilon_a^{\text{Mp}} - \Delta \tilde{\varepsilon}_{a,x}^{2\text{p}}$, referred to as the quasi-three-plus-potential contribution, can be studied in the coordinate space using the PW expansion for the Green's function
\begin{equation}
    \widetilde{G}^{(3+)} = G^{(2+)} - \sum_{f} \frac{\ketam{Vf} \braam{fV}}{(\varepsilon_a - \omega - \varepsilon_{f}^-)^3}
\end{equation}
Here, $\ketam{Vf}=V\ketam{f}$. A solid advantage of the SC scheme is the fact that it is fairly easy to calculate the quasi-two-potential contribution with high accuracy in momentum space. We denote the result as $\Delta \tilde{\varepsilon}_{a,p}^{2\text{p}}$ and use it as $\Delta \varepsilon_{a}^{\text{subtr}}$.
The potentials that have been transposed into vertices can be formally taken into account by multiplying the wave function of the external state $|a\rangle$ by them. An issue arises of calculating the Fourier transform of such products. Since $V$ is assumed to be spherically symmetric, this can be done by a straightforward generalization of Eqs.~(\ref{wf_momenta})-(\ref{ftr_wf_radial}).
We represent the result in the following form:  
\begin{equation}
    \phi_a(\mathbf{p}) = \int d^3 {\mathbf{r}} e^{-i ({\mathbf{p \cdot r}})} V(r) \psi_a({\mathbf{r}}) = i^{-l_a} {\begin{pmatrix}\widetilde{t}_a(p) \chi_{\kappa_a m_a} (\mathbf{\hat{p}}) \\ \widetilde{s}_a(p) \chi_{-\kappa_a m_a} (\mathbf{\hat{p}}) \end{pmatrix}},
\label{wf_momenta_new}
\end{equation}
\begin{equation}
    \widetilde{t}_a(p) = 4 \pi \int_0^\infty dr r^2 j_{l_a}(pr) V(r) g_a(r),
\end{equation}
\begin{equation}
    \widetilde{s}_a(p) = - 4 \pi \frac{\kappa_a}{|\kappa_a|} \int_0^\infty dr r^2 j_{2j_a-l_a}(pr) V(r) f_a(r).
\label{ftr_wf_radial_new}
\end{equation}
Using the identity
\begin{equation}
    \int d \mathbf{y} d \mathbf{z} G^{(0)}(\omega, \mathbf{x_1}, \mathbf{y}) G^{(0)}(\omega, \mathbf{y}, \mathbf{z}) G^{(0)}(\omega, \mathbf{z}, \mathbf{x_2}) = \frac{1}{2} \frac{\partial^2}{\partial \omega^2} G^{(0)}(\omega, \mathbf{x_1}, \mathbf{x_2}),
\end{equation}
one obtains
\begin{equation}
    \Delta \tilde{\varepsilon}_{a,p}^{2\text{p}} = \frac{1}{2} \int \frac{d^3 \mathbf{p}}{(2 \pi)^3} \bar{\phi}_a (\mathbf{p}) \frac{\partial^2 \Sigma_R^{(0)} (\mathrm{p})}{\partial p_0^2} \Bigr|_{p_0 = \varepsilon_a} \phi_a (\mathbf{p}),
\end{equation}
where $\mathrm{p}=(p_0,\mathbf{p})$. The explicit form of the derivatives of the free-electron self-energy operator $\Sigma_R^{(0)}$ with respect to $p_0$ in both gauges is given in Appendix~\ref{ap:zero_pot_deriv}. After performing the angular integration, which is carried out similarly to that in the zero-potential contribution, one can obtain:
\begin{equation}
\begin{gathered}
    \Delta \tilde{\varepsilon}_{a,p}^{2\text{p}} = \frac{\alpha}{8 \pi} \int \frac{dp p^2}{(2 \pi)^3} \left[ N_1 \left(\widetilde{t}_a^2 - \widetilde{s}_a^2 \right) + N_2 \left(\widetilde{t}_a^2  + \widetilde{s}_a^2 \right) + 2 p N_3 \widetilde{t}_a \widetilde{s}_a \right]
\end{gathered}
\end{equation}
where $N_1$, $N_2$, and $N_3$ are defined in Appendix~\ref{ap:zero_pot_deriv}, $\widetilde{t}_a=\widetilde{t}_a(p)$ and $\widetilde{s}_a=\widetilde{s}_a(p)$.

\section{Computational details}


The calculations are carried out by means of two approaches: the FBS method within the DKB basis and the GF method. The methods differ only in the treatment of the terms evaluated in coordinate space. The momentum-space contributions
are the same in both approaches.


\subsection{Momentum-space calculations}

The contributions evaluated in momentum space include the zero- and one-potential terms resulting from the renormalization procedure,
as well as the quasi-two-potential terms corresponding to the SC convergence-acceleration scheme. All the integrals in the zero-, one-, and quasi-two-potential terms are calculated using the Gauss-Legendre quadratures. Let us first briefly discuss an evaluation of the zero-potential terms. The domain of integration over $p$ is divided into two parts, $p \leq p_0$ and $p \geq p_0$, where $p_0 = \alpha Z$. The integration over the second domain can be carried out using a change of variables of the form $t = \alpha Z/p$, as a result of which the ray $[\alpha Z, \infty)$, is mapped onto the interval $[1, 0]$. The first domain is finite, so its treatment causes no problems. Compared to the Feynman gauge, the zero-potential contribution in the Coulomb gauge contains an additional integration over the Feynman parameter $x$, see Eqs.~(\ref{SE0p}), (\ref{abc_c}), and (\ref{fi_zp_cg}). The coefficient $F_2$ in Eq.~(\ref{fi_zp_cg}) must be calculated as a principal-value integral. From a numerical point of view, it is helpful to isolate the singularity. We do that by adding and subtracting the same integrand but with the numerator evaluated at the singularity point $x=a$:
\begin{equation}
    \int dx \frac{F(x)}{x-a+i 0} = \int dx \frac{F(x)-F(a)}{x-a+i 0} + \int dx \frac{F(a)}{x-a+i 0}.
\label{0p_trick}
\end{equation}
The first term in the right-hand side of Eq.~(\ref{0p_trick}) becomes regular. To avoid numerical problems at $x=a$, the exact expression can be replaced with its Taylor series in a small vicinity of this point. The second term in Eq.~(\ref{0p_trick}) is calculated analytically.  
Similar methods can be used to calculate $\Delta \tilde{\varepsilon}_a^{2\text{p}}$ in momentum space. To avoid possible numerical problems due to large cancellations, one should use the Taylor-series expansion for the corresponding integrands.

The evaluation of the one-potential contributions involves the integrations over the variables $p$, $p'$,  $z$, and a number of Feynman parameters. When integrating over the Feynman parameters in both gauges, to avoid possible numerical errors, the Taylor series for the integrands are employed. In the Coulomb-gauge case we use the results of Ref.~\cite{OP_CG}, but expand all the coefficients $F_1$ - $F_{21}$ at ``suspicious''{} points. 
The expression (\ref{SE1p}) has an integrable singularity for $q=0$ when $p=p'$ and $z = 1$. Therefore it is necessary to properly choose integration nodes for the variable $z$ to ensure the integral convergence. For this goal, we use the transformation $z = 1-2t^2$ with $t\in[0,]$. To calculate the integrals over the variables $p$ and $p'$, we change the variables according to $P=(p+p')/\sqrt{2}$ and $P'=(p-p')/\sqrt{2}$. The integral over $P$ is calculated similar to the integral over $p$ for the zero-potential contribution, but the separation point $p_0 = 2 \alpha Z$ is chosen instead. The integral over $P'$ is taken within the finite interval $[-P,P]$.

To estimate an uncertainty of the momentum-space calculations, the number of quadrature nodes and position of the separation points $p_0$ are varied.

\subsection{Coordinate-space calculations}
The calculation of all coordinate-space contributions, many-potential, two-potential, and (quasi-) three-plus-potential terms, has many common points. Let us illustrate them with an example of the many-potential contribution. Its evaluation essentially reduces to the calculation of  three integrals: over the radial variables $r_1$ and $r_2$, see Eqs.~(\ref{M_el_f}) and (\ref{M_el_c}), and over the energy parameter $\omega$. Taking into account the exponential decay of bound-electron wave functions at large distances, all radial integrations are performed over the finite interval $[0,R_\text{max}]$. The value of the parameter $R_\text{max}$ is chosen so that the results are independent of it within the desired accuracy. All the radial integrations are carried out using the Gauss–Legendre quadratures. However, the specific details differ slightly in the DKB and GF approaches. 

In the DKB approach, the domain $[0,R_\text{max}]$ is divided into a finite number of intervals by the nodes used to construct the B splines. The integration method for the double integral over $r_1$ and $r_2$ depends on whether these variables belong to the same interval or not. If not, the integrations are carried out independently using the same quadrature for both variables. However, when $r_1$ and $r_2$ fall into the same interval, a more delicate integration is employed to take into account the fact that the integrand depends on $r_< = \min \{r_1, r_2\}$ and $r_> = \max \{r_1, r_2\}$. Namely, for some fixed value of $r_1$ the corresponding interval is divided into two subitervals lying to the left and to the right from $r_1$. The integration over $r_2$ is performed by using the Gauss-Legendre quadratures for both subintervals. Note that within the DKB approach, the integration over radial variables turns out to be ``tied'' to the size of the basis $N$. As the FBS size grows, the number of  nodes also increases, that leads to the improvement of the integration accuracy. Thus, the single parameter $N$ affects both the accuracy of the Green's function representation and the integration accuracy. 

In the GF approach, the radial integrations become more complicated due to the aforementioned discontinuities of the electron Green's functions. The integrations over $x$ and $y$ in Eq.~(\ref{pot_decomp_gf}) cannot be performed independently of the integrations over $r_1$ and $r_2$. To evaluate the radial integrals, we employ a variation of the method described, e.g., in Ref.~\cite{sym12050800}. Namely, a hierarchy of the radial integration grids is prepared, with each subsequent grid becoming finer and ``lying'' inside the previous one.
Then, all the radial integrations are arranged in the following order: $r_1$, $r_2$, $x$, and $y$. Therefore, the finest grid corresponds to the variable $y$. The accuracy of the GF approach is determined mainly by the accuracy of the radial integrations.

\begin{figure}[h]
\centering
\include{contour}
\caption{Deformed contour (blue solid line) in the complex plane $\omega$. Singularities and branching cuts of the electron Green's function (black lines and black dots) and photon propagator (black dashed lines) are shown.}
\label{fig:contour}
\end{figure}

The photon and electron propagators have a complicated analytical structure  as functions of the energy parameter $\omega$ in the complex plane. They are two-valued functions and are defined on the corresponding Riemann surfaces with two sheets (for each propagator). Fig.~\ref{fig:contour} shows the complex plane of the integration variable $\omega$, the branch points of both propagators with the cuts starting from them, and the poles of the electron propagator corresponding to the bound states. In the expression for the SE correction (\ref{se_general_1999}), the integration contour goes along the real axis, but it can be bended in the $\omega$ complex plane to ``simplify'' the calculations. Because of the zero photon mass, the contour is squeezed between two infinitely close branching points near zero. Introducing a finite mass would allow them to be pushed  apart, which can be easily demonstrated in the Feynman gauge. Consequently, the contour must always pass through $\omega=0$. In principle, there is no other restrictions on the contour. Thus, it can be changed as needed, respecting the analytical structure of the integrand.

In this work, the approach proposed in Ref.~\cite{base_1999} is used. Namely, we employ the transformation that results in the contour shown in Fig.~\ref{fig:contour}. The original contour along the real axis is rotated and deformed. The contributions from the circular arcs vanish as the radius of the circle tends to infinity, so we do not consider them. The rotation is done to avoid fast oscillations of the integrand encountered on the real axis: on the vertical parts of the deformed contour the integrand decays exponentially.  The deformation of the contour near zero and its shift to the right from the imaginary axis is done to avoid going close to the poles of the electron Green’s function, which correspond to the bound states with binding energies no less than that of the state under consideration. For large enough values of $\kappa$, the contour simplifies and can be taken along the imaginary axis. The integration over $\omega$ is evaluated using the Simpson quadratures. The number of integration points is increased until the required accuracy is achieved.

\subsection{Extrapolation of the PW-expansion sums}\label{sec:extr}

We obtain a coordinate-space contribution as a PW series in $\kappa$, which should inevitably be truncated at some $|\kappa_{\text{max}}|$. The remainder of the series can then be estimated using an extrapolation procedure. The extrapolation methods discussed below are applied to all treated coordinate-space contributions.
Here and in what follows, when discussing partial sums, we will denote by the index $k>0$ the sum of all terms of the PW expansion with $|\kappa|\le k$. 

When using the DKB method, the individual contributions strongly depend on the FBS size $N$, therefore, an extrapolation over this parameter is also required. In this case, the most stable results are obtained if one extrapolates with respect to the basis size first. Let us denote as $S_{k}(N)$ the partial sum obtained for the given FBS size $N$. For each $k$, the extrapolation over $N$ is performed using a non-linear least-squares method using an ansatz of the form:
\begin{equation}
\label{eq:na_anzatz}
    S_{k}(N) = A_{k} \left( \frac{1}{N} \right)^{B_{k}} + S_{k},
\end{equation}
where $S_{k}$ are the extrapolated to the infinite basis set partial sums, and $A_k$ and $B_k$ are adjustable numerical coefficients. In the GF approach, the sums $S_k$ are obtained directly from the calculations, and this extrapolation is not required.

To extrapolate over $k$ in both DKB and GF approaches, we implement three following ansatzes for the partial sums~$S_k$:
\begin{equation}
\label{eq:kappa_expansion}
    \tilde{S}(k) = S_\infty + \frac{C_2}{k^2} + \frac{C_3}{k^3} + ... + \frac{C_m}{k^m}~,
\end{equation}
where the parameter $S_\infty$ and coefficients $C_i$ for $i=2, \ldots, m$ (with $m$ being $4$, $5$ and $6$) are obtained by a kind of statistical method. Namely, for each $m$, $N_{\text{iter}}$ times a subset of $k$'s, $\{k_i\}_{i=1}^{m}$, is randomly selected. The parameter $N_{\text{iter}}$ is fixed by the condition that all possible subsets are considered for the largest $m$ ($m=6$). For each element of the above subsets we construct and solve the system of $m$ linear equations of the form 
\begin{equation}
    \tilde{S}(k_i)=S_{k_i},\quad i=1, \dots, m
\end{equation}
to determine $m$ variables $S_\infty$, $C_2$, \ldots, $C_m$. 
Repeating this step $3N_{\text{iter}}$ times yields a set of approximations for $S_\infty$. Evaluating the mean value $\overline{S}_\infty$ and variance $\sigma_{k}$, one obtains the desired extrapolated sum and an estimation for its uncertainty. We note that we approximate the tails only (by excluding the first few $k$'s from consideration). Changing the tail size produces results that are consistent with each other.




Within the DKB approach, the uncertainties of extrapolation in $k$, $\sigma_k$, and extrapolation to an infinite basis set, $\sigma_N$, are summed quadratically
\begin{equation}
    \sigma_{\text{tot}} = \sqrt{\sigma_{\kappa}^2 + \sigma_{N}^2}.
\end{equation}
Example of the use of this two-step extrapolation procedure can be found in the ``results'' section below.

\newpage

\section{Results}

In the present work, we conduct two sets of SE-correction calculations. First, we consider the $1S_{1/2}$ state in three hydrogen-like ions: argon ($Z=18$), xenon ($Z=54$), and uranium ($Z=92$). We adopt the results from Ref.~\cite{hen} as a benchmark for  comparison. For this reason, in the first part of calculations, the values of root-mean-square (rms) nuclear radii as well as the values of fundamental constants, e.g., the fine-structure constant $\alpha$ and the Hartree energy in eV, are taken to be the same as there. The obtained data are used to analyze the convergence of the many-potential contribution.
 Second, we study the SE corrections for the $1S_{1/2}$, $2S_{1/2}$, $2P_{1/2}$, and $2P_{3/2}$ states in a set of hydrogen-like ions with the nuclear charge $Z$ in the range from $Z=10$ to $Z=100$ and compare the obtained results with the ones presented in Ref.~\cite{se_yerokh}. In this case, the rms radii and models of the nuclear-charge distribution are chosen to coincide with  those in the corresponding work, however, for $\alpha$ the numerical value  from the  CODATA (2018)~\cite{codata_2018} is used. In the second part, the SE correction is represented in terms of the dimensionless function $F(\alpha Z)$, defined as:
\begin{equation}
\label{eq:FaZ}
    \Delta \varepsilon_a = \frac{\alpha}{\pi} \frac{(\alpha Z)^4}{n^3} F(\alpha Z) m c^2.
\end{equation} 
The data obtained in the second part of the work are used to analyze the acceleration-convergence schemes.

\subsection{Many-potential term: convergence analysis}

\begin{figure}[H]
\begin{minipage}[h]{0.49\textwidth}
\center{\includegraphics[width=0.99\textwidth]{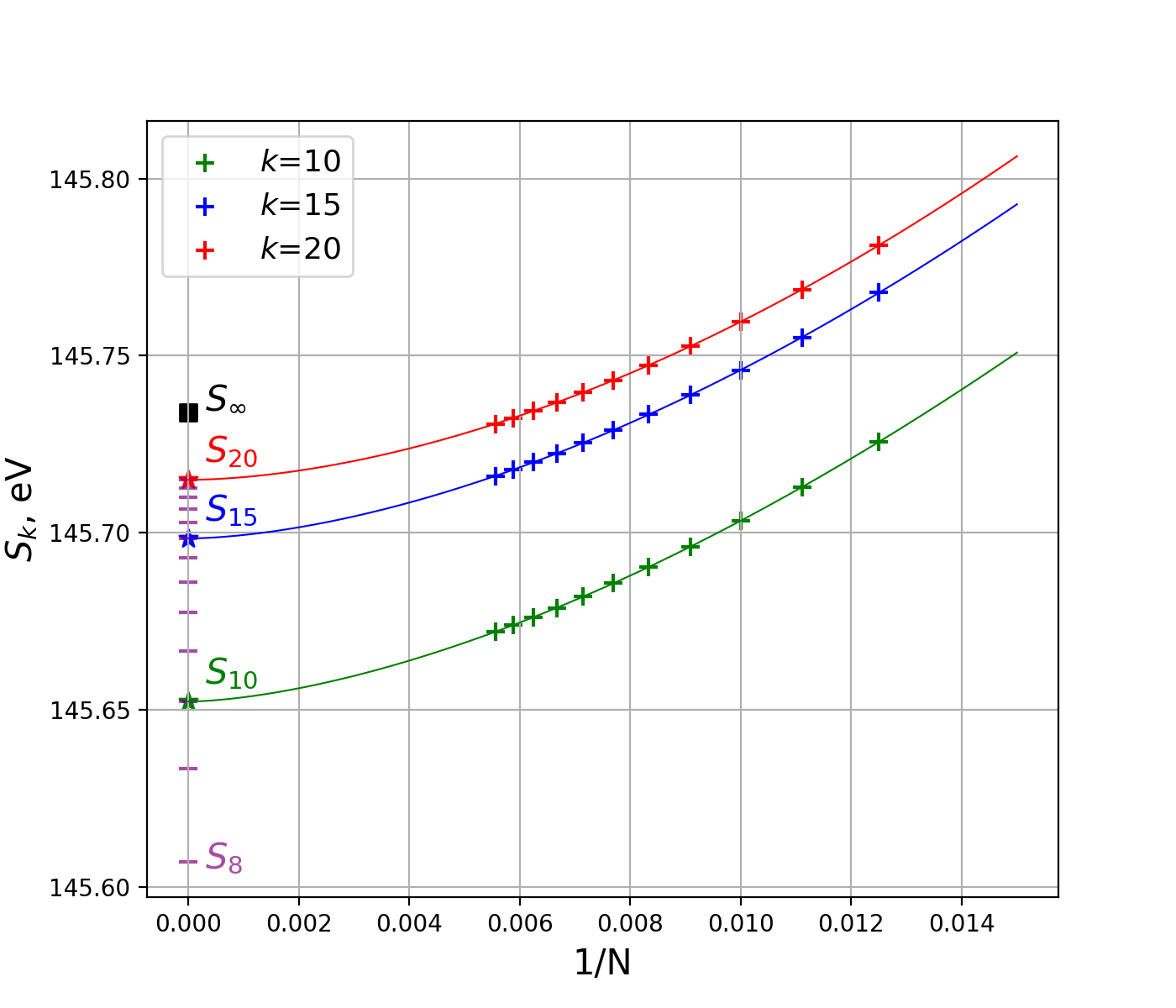} \\ (a)}
\end{minipage}
\hfill
\begin{minipage}[h]{0.49\textwidth}
\center{\includegraphics[width=0.99\textwidth]{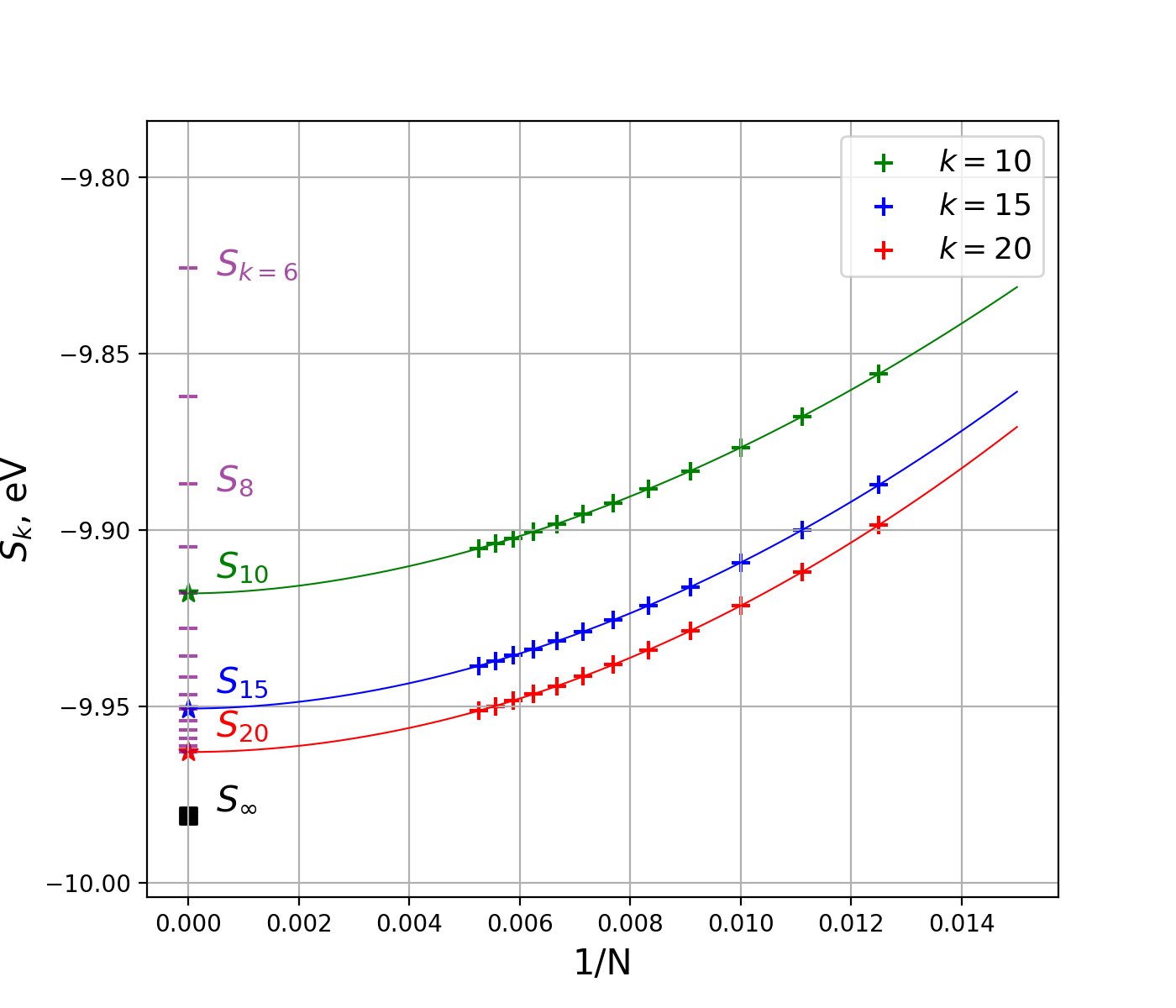} \\ (b)}
\end{minipage}
\caption{The extrapolation of the many-potential contribution to the self-energy correction for the $1S_{1/2}$ state of hydrogen-like xenon ($Z=54$) obtained within the framework of the DKB approach in the Feynman (a) and Coulomb (b) gauges.}
\label{fig:na_extrapol_example}
\end{figure}

Let us start with illustrating  the two-step extrapolation procedure, described in the previous section, using  the ground state of hydrogen-like xenon ($Z=54$) as an example, see Fig.~\ref{fig:na_extrapol_example}. The many-potential term is considered, and the results for both gauges are presented. At the first stage, the extrapolation of the partial sums $S_k(N)$ with respect to the size of the basis $N$ is performed. For convenience, only the extrapolation curves for $k= 10$, $15$, and $20$  are shown as functions of $1/N$. The corresponding extrapolated sums $S_{10}$, $S_{15}$, and $S_{20}$ are depicted as stars at $1/N \rightarrow 0$. The   extrapolated  partial sums $S_k$ obtained for other values of $k$ are depicted with dashes as well. At the second stage, the extrapolation over $k$ is performed. The final values for $\Delta \varepsilon_a^{\text{Mp}} = S_\infty$ are shown with black squares. Note that the GF method yields results that are indistinguishable at this scale from those obtained by the DKB method. Based on our experience, a small difference in the FBS size extrapolation in the two gauges appears in the small-$Z$ region, where the convergence behavior is slightly better in the Coulomb gauge. In the region of medium- and large-$Z$ there is no significant difference between the gauges. This can be seen, e.g.,  from Fig.~\ref{fig:na_extrapol_example}, where the absolute values of the many-potential terms in two gauges differ by approximately one order of magnitude, while the rate of convergence in the FBS size is approximately the same.



\begin{figure}[H]
\begin{minipage}[h]{0.49\textwidth}
\center{\includegraphics[width=0.99\textwidth]{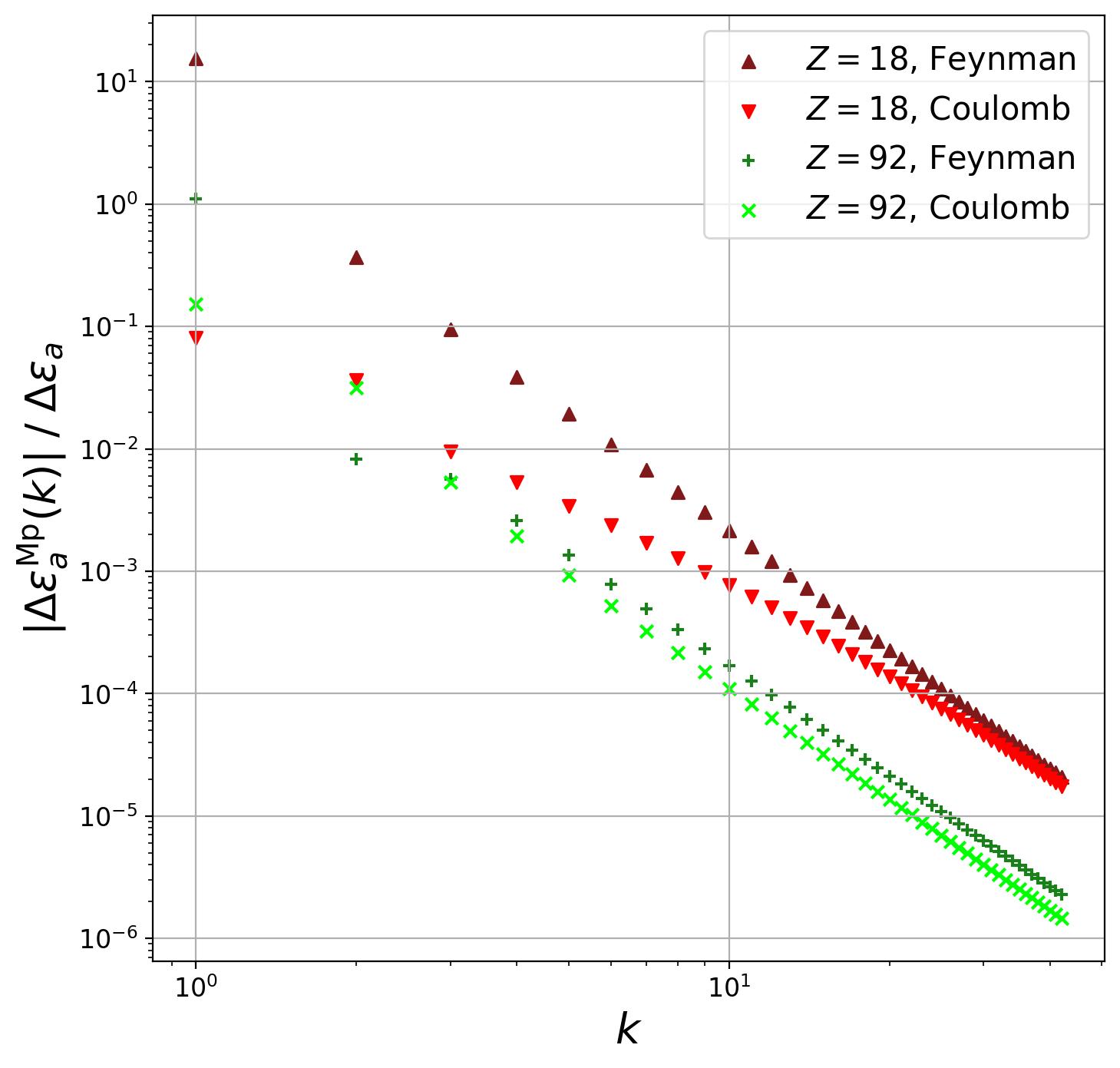} \\ (a)}
\end{minipage}
\hfill
\begin{minipage}[h]{0.49\textwidth}
\center{\includegraphics[width=0.99\textwidth]{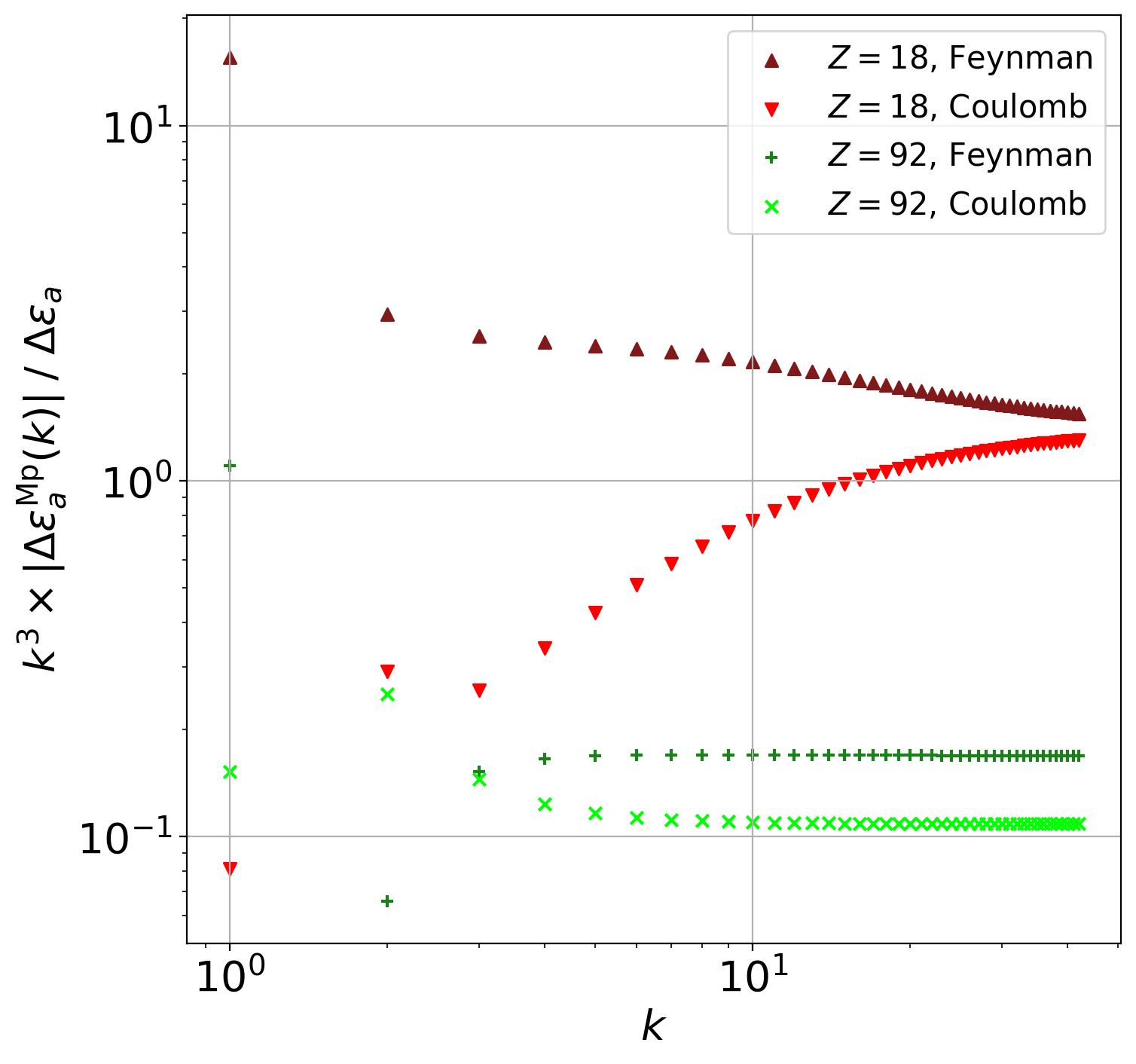} \\ (b)}
\end{minipage}
\caption{The individual many-potential contributions $\Delta \varepsilon_a^{\text{Mp}}(k)$: (a) normalized to the total self-energy correction $\Delta \varepsilon_a$ and (b) additionally multiplied by $k^3$. Note the log-log scale.}
\label{fig:mp_contributions_new}
\end{figure}

 Let us now examine in details the PW convergence of the many-potential contributions and perform a comparative analysis within two gauges. The zero-, one-, and many-potential contributions to the SE correction for the $1S_{1/2}$ state in hydrogen-like argon ($Z=18$) and uranium ($Z=92$) in Feynman and Coulomb gauges are presented in Table~\ref{tab:contribs_mp}. For the many-potential contributions, the individual terms of the PW expansion, $\Delta \varepsilon_a^{\text{Mp}}(k)$, obtained by means of the GF approach are shown. These terms represent sums of contributions with $\kappa=k$ and $\kappa=-k$, i.e., the partial sums are given by $S_k = \sum_{i=1}^k\Delta \varepsilon_a^{\text{Mp}}(i)$.  The row labeled ``$\sum_{k=1}^{24}$'' gives the partial sums $S_{24}$, while the subsequent row ``$\sum_{k=25}^{\infty}$ [extr.]'' provides our estimations for the remainders of PW series obtained by applying the statistical extrapolation procedure discussed above. The total SE corrections are presented in the last line. 

For convenience, the absolute values of the individual terms $\Delta \varepsilon_a^{\text{Mp}}(k)$ from Table~\ref{tab:contribs_mp}, normalized to the total self-energy correction $\Delta \varepsilon_a$, are plotted in Fig.~\ref{fig:mp_contributions_new}(a).
From Table~\ref{tab:contribs_mp} and Fig.~\ref{fig:mp_contributions_new}(a), it can be noticed that for both considered $Z$ the many-potential contributions as well as the first terms of the PW expansions  are significantly smaller in the Coulomb gauge (in the case of  $Z=18$,  by two orders of magnitude) than in the Feynman gauge. This difference, however, becomes much less noticeable for large values of $k$, although the Feynman-gauge contributions  still exceed their Coulomb-gauge counterparts.

Multiplying the individual terms $\Delta \varepsilon_a^{\text{Mp}}(k)$ obtained for two gauges by the  factor of $k^3$ provides a more informative comparison of their decay rate with the growth of $k$, that is demonstrated in Fig.~\ref{fig:mp_contributions_new}(b). Let us discuss the leading asymptotic behavior of the PW-expansion terms for $k \rightarrow \infty$: $\Delta \varepsilon_a^{\text{Mp}}(k) \sim 1/{k^p}$.
Our numerical results indicate  that in both  the  Feynman and Coulomb gauges,  the exponent $p$ exceeds 2 over a wide range of $Z$, with $p$ being close to $3$ in the large-$Z$ region. These conclusions do not contradict those from  Ref.~\cite{PhysRevA.72.042502}, where it was claimed that in the Feynman gauge $p=3$ regardless of the nuclear charge $Z$. The latter observation justifies the fact that the anzatz for the extrapolation over $k$ in~(\ref{eq:kappa_expansion}) starts from ${1}/{k^2}$, since $\sum_{i=1}^{k} {1}/{i^3} \sim {1}/{k^2}$. 

We also note that the numerical coefficients for the higher powers of $1/k$ in the expansion~(\ref{eq:kappa_expansion}) in the Coulomb gauge are larger than their Feynman-gauge counterparts. That fact is also illustrated in the Fig.~\ref{fig:mp_contributions_new}(b) -- convergence to asymptotics in the Feynman gauge is significantly slower. As a result, within the $k$-range studied in the present work ($k\le 20$ or $k\le 24$), the leading asymptotic behavior is less pronounced in the Coulomb gauge. The latter fact apparently balances out the smallness of the absolute value of the Coulomb-gauge many-potential contribution. In practice, within the GF approach, given the same resource costs (number of partial waves, number of integration nodes for radial variables, etc.), the Coulomb and Feynman gauges are equally suitable for estimating the many-potential contribution to the SE correction within a wide range of nuclear charges $Z$.
This conclusion slightly contradicts what was claimed in Ref.~\cite{yer_2025}. Nevertheless, we agree with the fact that the advantages of the Coulomb gauge are the improved convergence of the many-potential contributions over the basis-set size in the small-$Z$ region, relative smallness of the many-potential contribution~\cite{hen, yer_2025}, and the absence of cancellation between different parts of the SE correction.


\subsection{Application of the convergence-acceleration schemes}

\begin{figure}[h!]
\begin{minipage}[h]{0.49\textwidth}
\center{\includegraphics[width=0.99\textwidth]{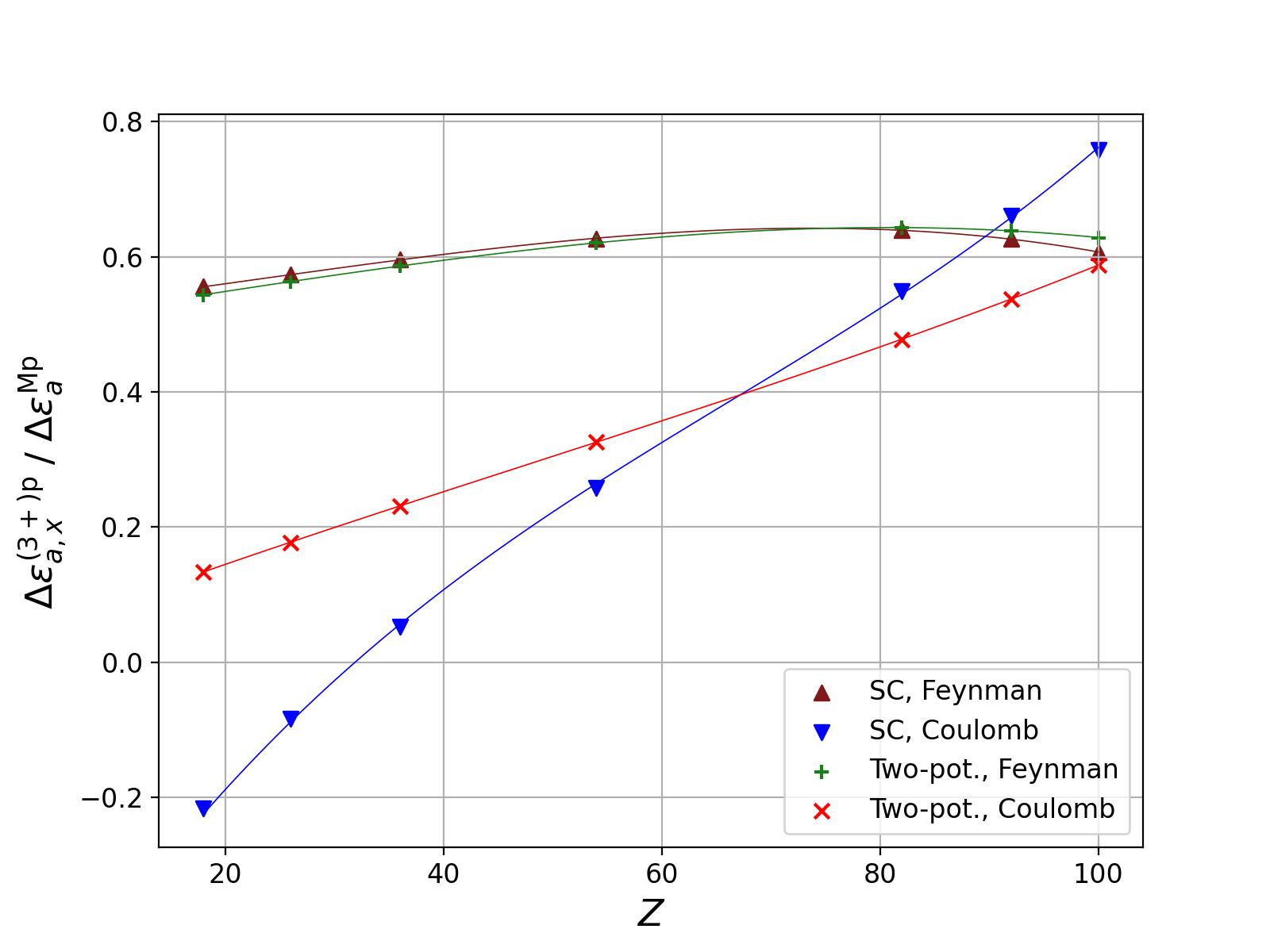} \\ (a)}
\end{minipage}
\hfill
\begin{minipage}[h]{0.49\textwidth}      \center{\includegraphics[width=0.99\textwidth]{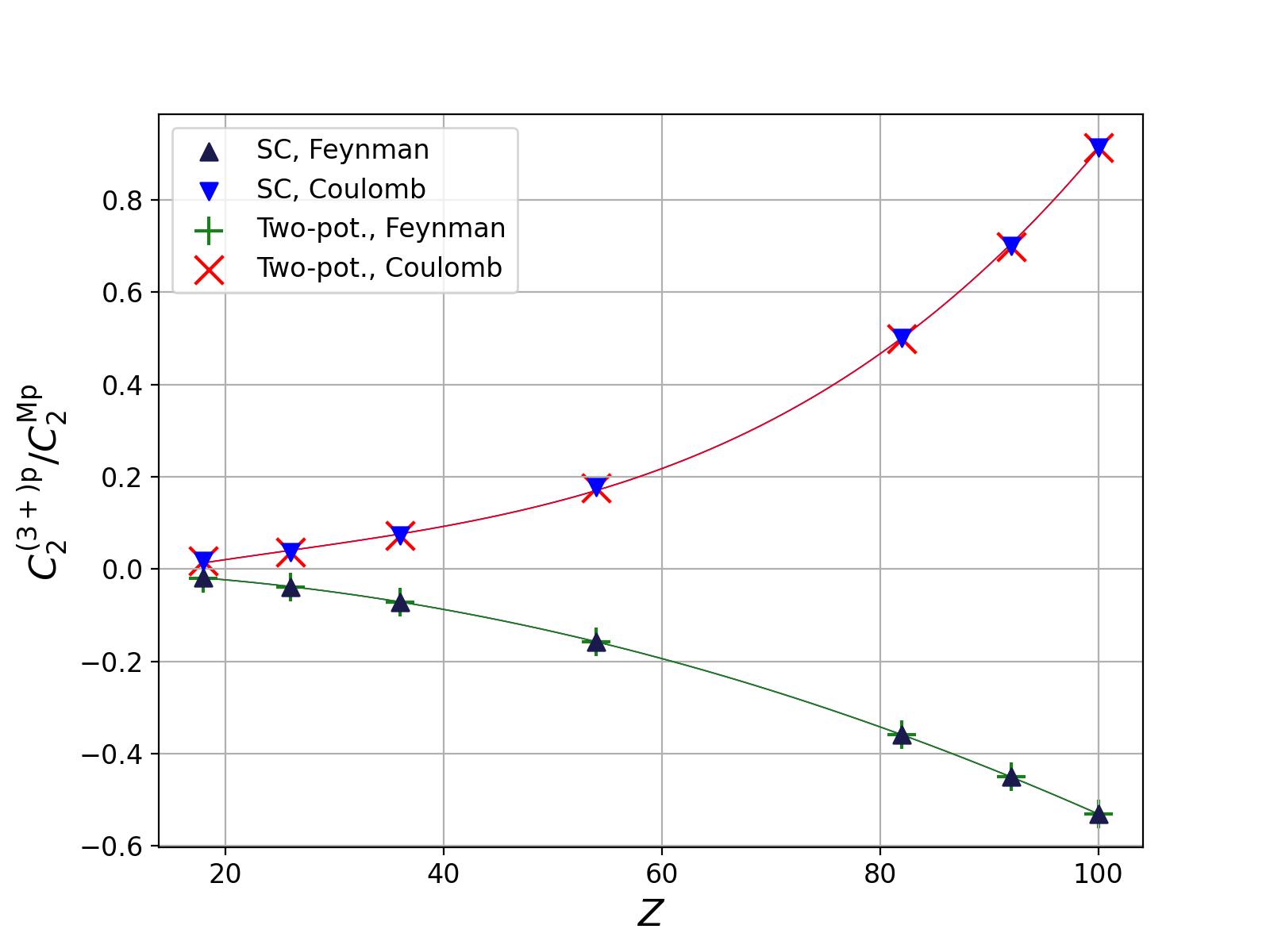} \\ (b)}
\end{minipage}
\caption{Comparison of two convergence-acceleration schemes in two gauges, Feynman and Coulomb ones: (a) three-plus-potential $\Delta \varepsilon_{a,x}^\text{(3+)p}$ and quasi-three-plus-potential $\Delta \tilde{\varepsilon}_{a,x}^\text{(3+)p}$ contributions normalized to the total many-potential contributions $\Delta \varepsilon_{a}^\text{Mp}$; (b) the same for the coefficients $C_2$ defined in Eq.~(\ref{eq:kappa_expansion}).}
\label{fig:extrapol_coeffs}
\end{figure}

Now let us turn to the comparative analysis of the SC and two-potential schemes used to accelerate the convergence of the PW expansions for the many-potential contributions in the Feynman and Coulomb gauges. In Fig.~\ref{fig:extrapol_coeffs}(a),  ratios of the  three-plus-potential $\Delta \varepsilon_{a,x}^\text{(3+)p}$ and quasi-three-plus-potential $\Delta \tilde{\varepsilon}_{a,x}^\text{(3+)p}$ contributions to the total many-potential contributions $\Delta \varepsilon_{a}^\text{Mp}$ are plotted. 
In the low-$Z$ region, the two-potential scheme in Coulomb gauge yields the smallest remainder which is to be extrapolated. However, in this case the overall accuracy  is limited by the two-potential contribution, which also has to be treated within the PW expansion.  

Let us turn to the coefficients $C_2$ from Eq.~(\ref{eq:kappa_expansion}) obtained for $\Delta {\varepsilon}_{a,x}^\text{(3+)p}$ and $\Delta \tilde{\varepsilon}_{a,x}^\text{(3+)p}$ (designated as $C_2^{\text{(3+)p}}$) by means of our statistical extrapolation approach. 
These coefficients normalized to the  coefficients $C_2^{\text{Mp}}$ for the corresponding many-potential contributions are shown in Fig.~\ref{fig:extrapol_coeffs}(b). These data partially explain the convergence-acceleration mechanism. In the area of low $Z$, the coefficients $C_2$ for the (quasi-)three-plus-potential contributions decrease by one or two orders of magnitude compared to their counterparts for the many-potential contributions. In the region of high $Z$, they also decrease, but less effectively. It should be noted that for each gauge the reduction of the coefficient $C_2$ for both studied acceleration schemes is almost identical, which confirms that the SC trick indeed provide a good approximation to the two-potential contribution.

\begin{figure}[h!]
\begin{minipage}[h]{0.49\textwidth}
\center{\includegraphics[width=0.99\textwidth]{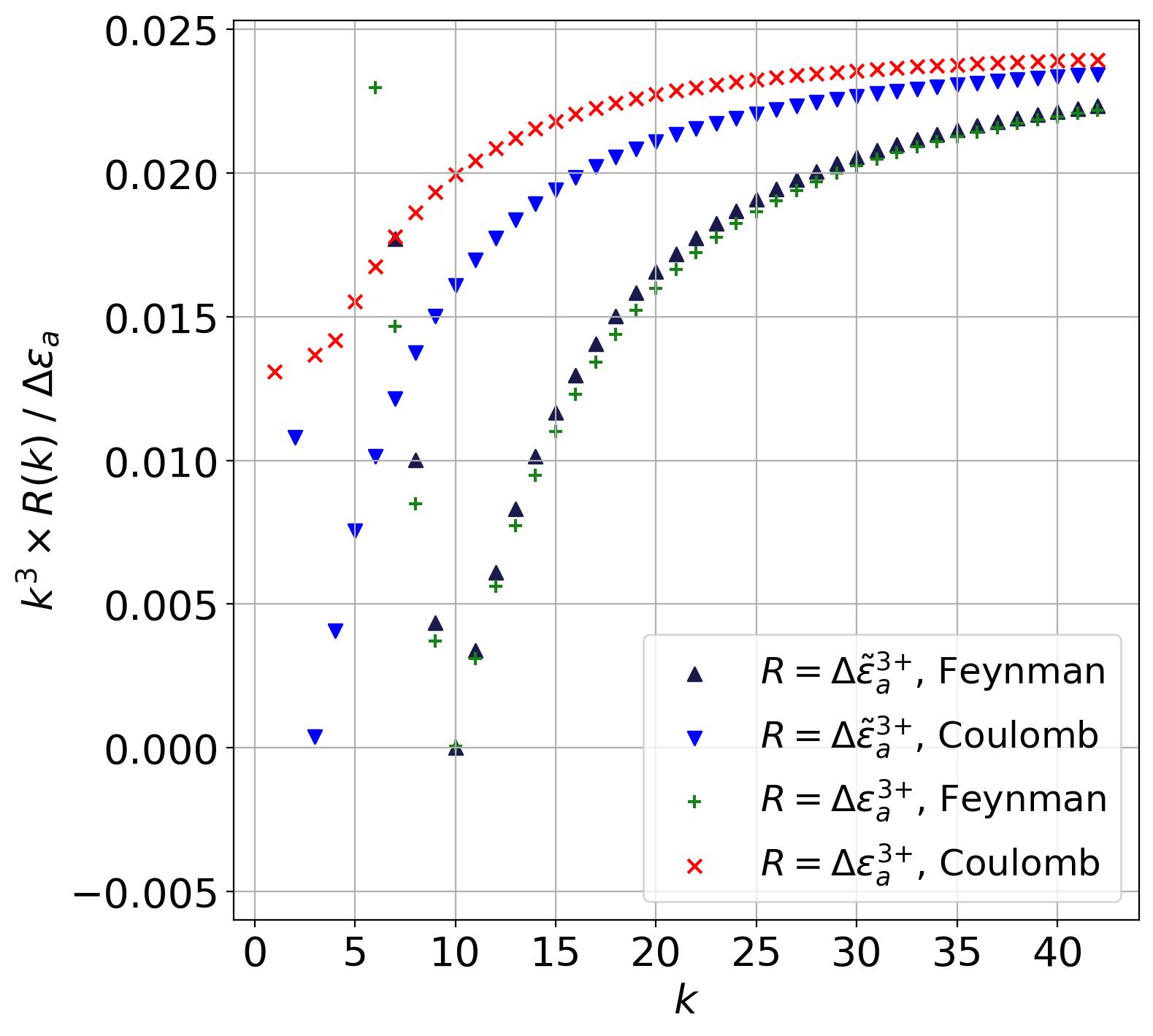} \\ (a)}
\end{minipage}
\hfill
\begin{minipage}[h]{0.49\textwidth}      \center{\includegraphics[width=0.99\textwidth]{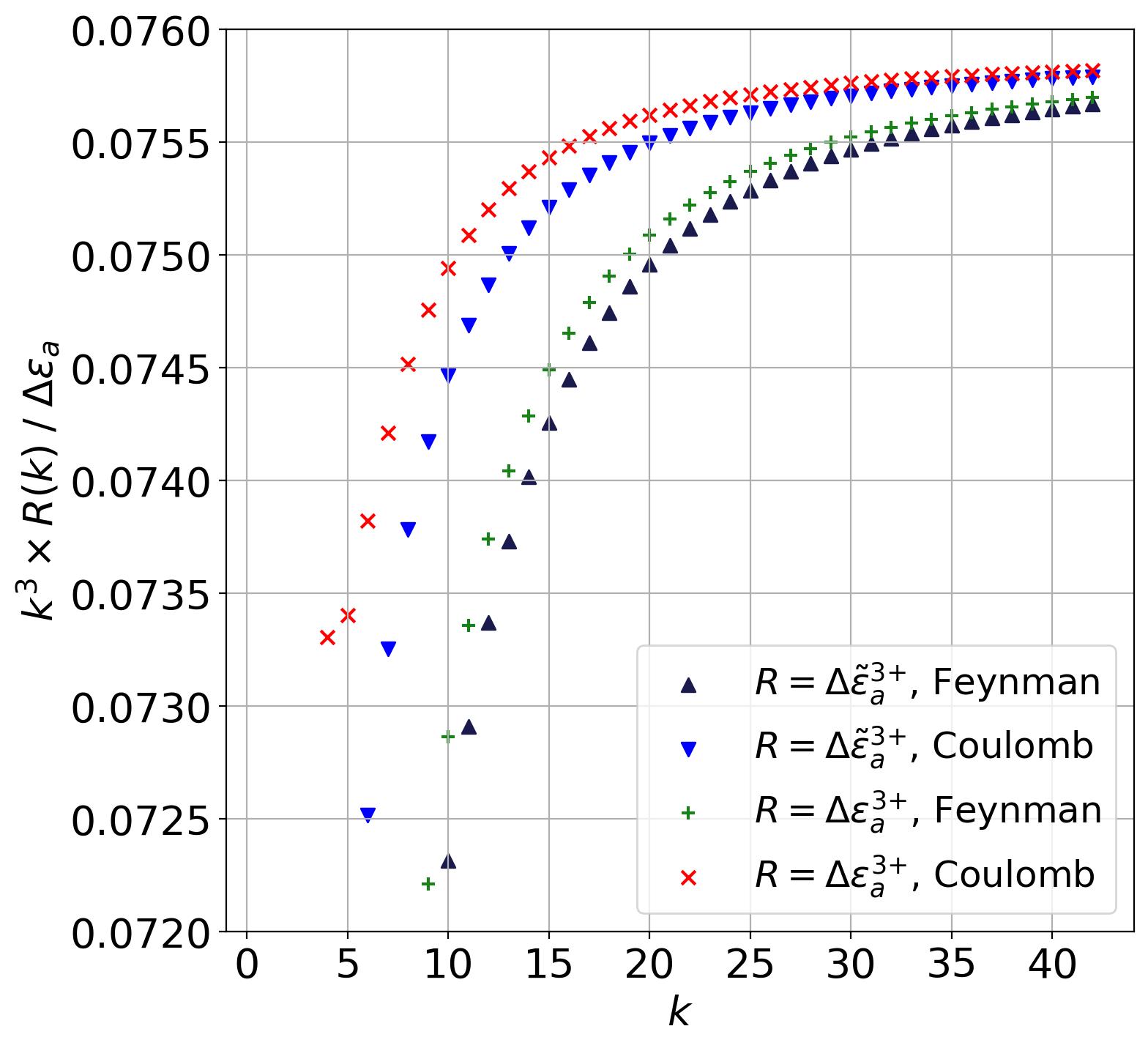} \\ (b)}
\end{minipage}
\caption{
The individual contributions $\Delta \varepsilon_a^{\text{(3+)p}}(k)$ and $\Delta \tilde{\varepsilon}_a^{\text{(3+)p}}(k)$ normalized to the total SE correction and multiplied by the factor of $k^3$: (a) for $Z=18$ argon; (b) for $Z=92$ uranium.}
\label{fig:mp_contributions1}
\end{figure}

The subtraction of the two-potential or quasi-two-potential contributions from the many-potential contribution apparently does not change the leading asymptotic behavior of the PW-expansion terms for the remainders $\Delta \varepsilon^\text{(3+)p}_a$ and $\Delta \tilde{\varepsilon}^\text{(3+)p}_a$ and only decreases the absolute values of the corresponding coefficients. To illustrate this, in Fig.~\ref{fig:mp_contributions1} we plot the individual terms of the PW expansions for these remainders normalized to the total SE correction and additionally multiplied by $k^3$, exactly as it was done in Fig.~\ref{fig:mp_contributions_new} for the many-potential contributions. One can see that the individual terms tend to constants in all the cases, i.e. $p=3$ for (quasi-)three-plus potential terms. That statement has also been verified empirically: the extrapolation procedure gives obviously incorrect results if one omits the term $C_2/k^2$ in the Eq.~(\ref{eq:kappa_expansion}). We stress that our conclusion contradicts the one from work~\cite{sap_orig}.  Fig.~\ref{fig:mp_contributions1} also demonstrates that  for both acceleration schemes the approach to the asymptotics in the Coulomb gauge  is much faster than in the Feynman gauge. The last statement makes it easier to extrapolate the corresponding contributions and certainly makes the Coulomb gauge more preferable for SE calculations within the convergence-acceleration schemes for a wide range of $Z$.

\subsection{Data: summary}

In this section, we present and discuss the results of our calculations of SE corrections to the $n=1$ and $n=2$ energy levels in hydrogen-like ions. 

First, in Table~\ref{tab:we_sw} we study the SE correction for the ground $1S_{1/2}$ state of hydrogen-like argon ($Z=18$), xenon ($Z=54$), and uranium ($Z=92$). As was noted above, Ref.~\cite{hen} is considered to be the benchmark in this case. Therefore, in this part of the calculations, we adopt the values of the fundamental constants and nuclear parameters to be the same as those used in that work. For each ion, we consider two gauges, the Coulomb gauge and the Feynman gauge, which are indicated in the second column as ``C'' and ``F'', respectively. The corresponding zero- and one-potential contributions, calculated in momentum space, are shown in the third and forth columns. For each gauge we compare different numerical approaches to the evaluation of the the many-potential contribution $\Delta \varepsilon_a^\text{Mp}$: (i) the standard method, labeled in the column ``Scheme'' as ``DIR'', in which the many-potential contribution is treated as a whole; (ii) the two-potential acceleration scheme, labeled as ``TP'', where this contribution is represented as the sum $\Delta \varepsilon_a^\text{2p}+\Delta \varepsilon_a^\text{(3+)p}$, and (iii) the Sapirstein-Cheng acceleration scheme, labeled as ``SC'', where it is given by the sum $\Delta \tilde{\varepsilon}_a^\text{2p}+\Delta \tilde{\varepsilon}_a^\text{(3+)p}$. We also compare two methods for representing the electron propagators: the finite-basis-set method within the DKB basis and the Green's function approach, which are indicated as ``DKB'' and ``GF'', respectively. All the coordinate-space contributions, $\Delta \varepsilon_a^\text{Mp}$, $\Delta \varepsilon_a^\text{2p}$, $\Delta \varepsilon_a^\text{(3+)p}$, and $\Delta \tilde{\varepsilon}_a^\text{(3+)p}$, are obtained by applying the statistical extrapolation procedure described in Sec.~\ref{sec:extr}. Within the DKB approach, the PW-expansion series are truncated at $|\kappa_\text{max}|=20$. When using the GF approach, these series are truncated at $|\kappa_\text{max}|\sim45$ in the case of the two-potential contribution $\Delta \varepsilon_a^\text{2p}$ and at $|\kappa_\text{max}|=24$ in the other cases. Finally, the total SE corrections obtained by means of different approaches are given in the last column. For each ion and gauge, the results from Ref.~\cite{hen} are shown for comparison as well.

As can be seen from Table~\ref{tab:we_sw}, the main source of numerical errors, regardless of the gauge, is the many-potential contribution. Even when applying the acceleration schemes, this contributions still limits the accuracy of the total SE corrections. 
We note that the many-potential contribution is significantly smaller in the Coulomb gauge than in the Feynman gauge (see the penultimate column in Table~\ref{tab:we_sw}), which was also noticed in Refs.~\cite{hen,yer_2025}. However, for the reasons described above, this does not provide a significant increase in accuracy unless acceleration schemes are involved. Although obtaining the zero- and one-potential contributions with sufficient accuracy is not a technically difficult task, an attractive feature of the Coulomb gauge, as can be seen from  Table~\ref{tab:we_sw}, is the absence of large cancellations between these momentum-space contributions. Nevertheless, this observation is extremely useful for obtaining high-precision results for low-$Z$ systems.

Clearly, acceleration tricks reduce extrapolation uncertainties and allow for a more accurate evaluation of the many-potential contributions. Using either the two-potential scheme or the SC scheme reduces the numerical errors and allows one to obtain up to two additional significant digits. As a rule, the most accurate many-potential contribution can be obtained by using a combination of the Coulomb gauge and the SC acceleration scheme.

The GF approach provides a more accurate treatment of the many-potential contribution compared to the DKB approach. This results mainly not from the lower value of $|\kappa_\text{max}|$ employed in this case, but from the fact that  the DKB method  needs the  approximation with respect to the size of the basis set.  The low accuracy of the DKB approach does not detract from the already mentioned advantages, namely, the ability to exclude a specific state and construct Green's functions for a wider class of systems that do not possess spherical symmetry. In some sense, the FBS method (not necessarily in the DKB realization) is more flexible and easier to implement. Therefore, if high accuracy is not required, it is an excellent method for calculating the SE correction. 

Returning to the comparison with the benchmark calculations~\cite{hen}, we note that our results obtained by means of different methods are in good agreement with each other and those from Ref.~\cite{hen}. Small deviations have been revealed. However, the reason for them is unclear for us.

\begin{figure}[H]
    \begin{minipage}[h]{0.49\textwidth}
        \center{\includegraphics[width=0.99\textwidth]{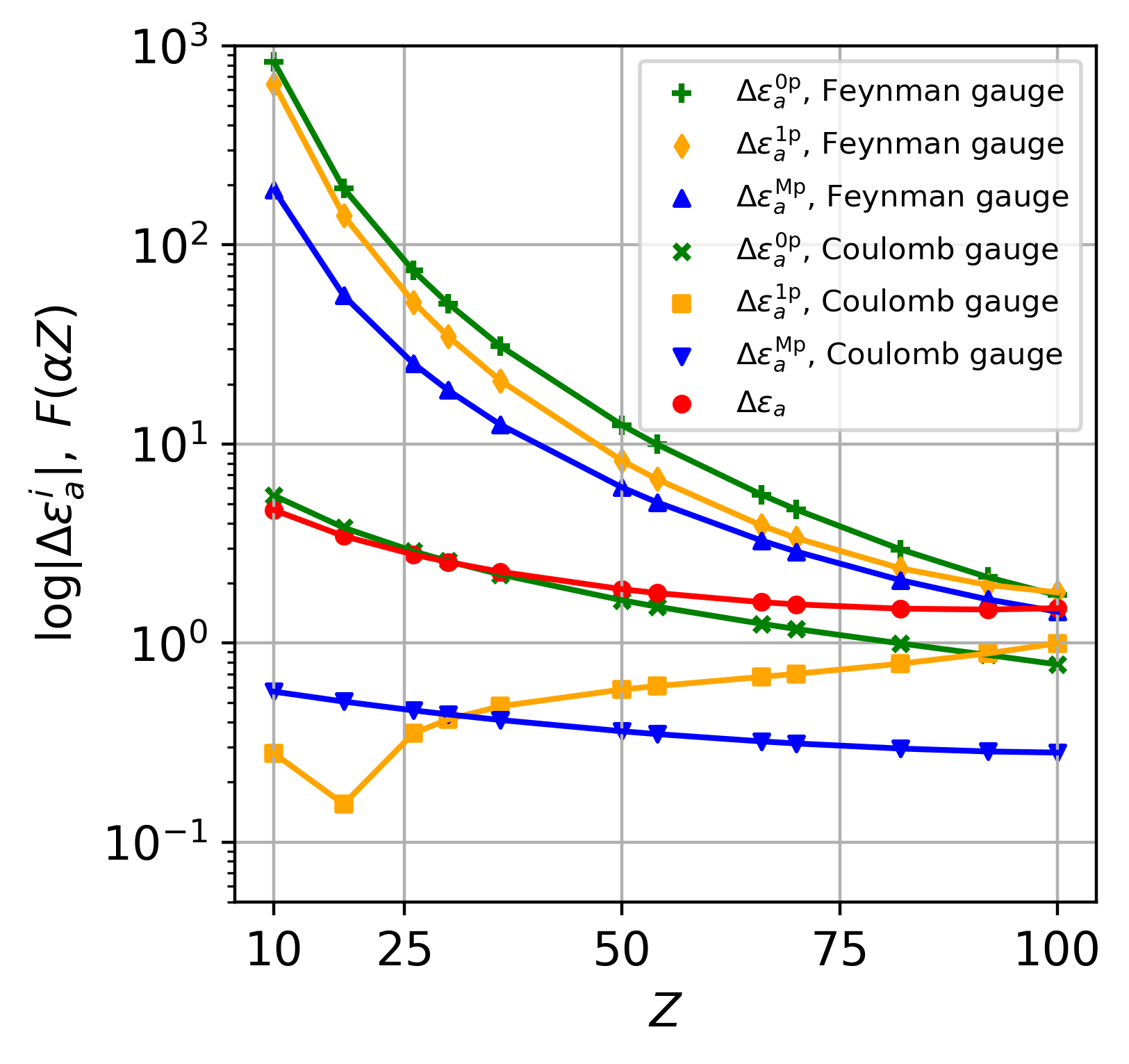} \\ (a)}
    \end{minipage}
    \hfill
    \begin{minipage}[h]{0.49\textwidth}
        \center{\includegraphics[width=0.99\textwidth]{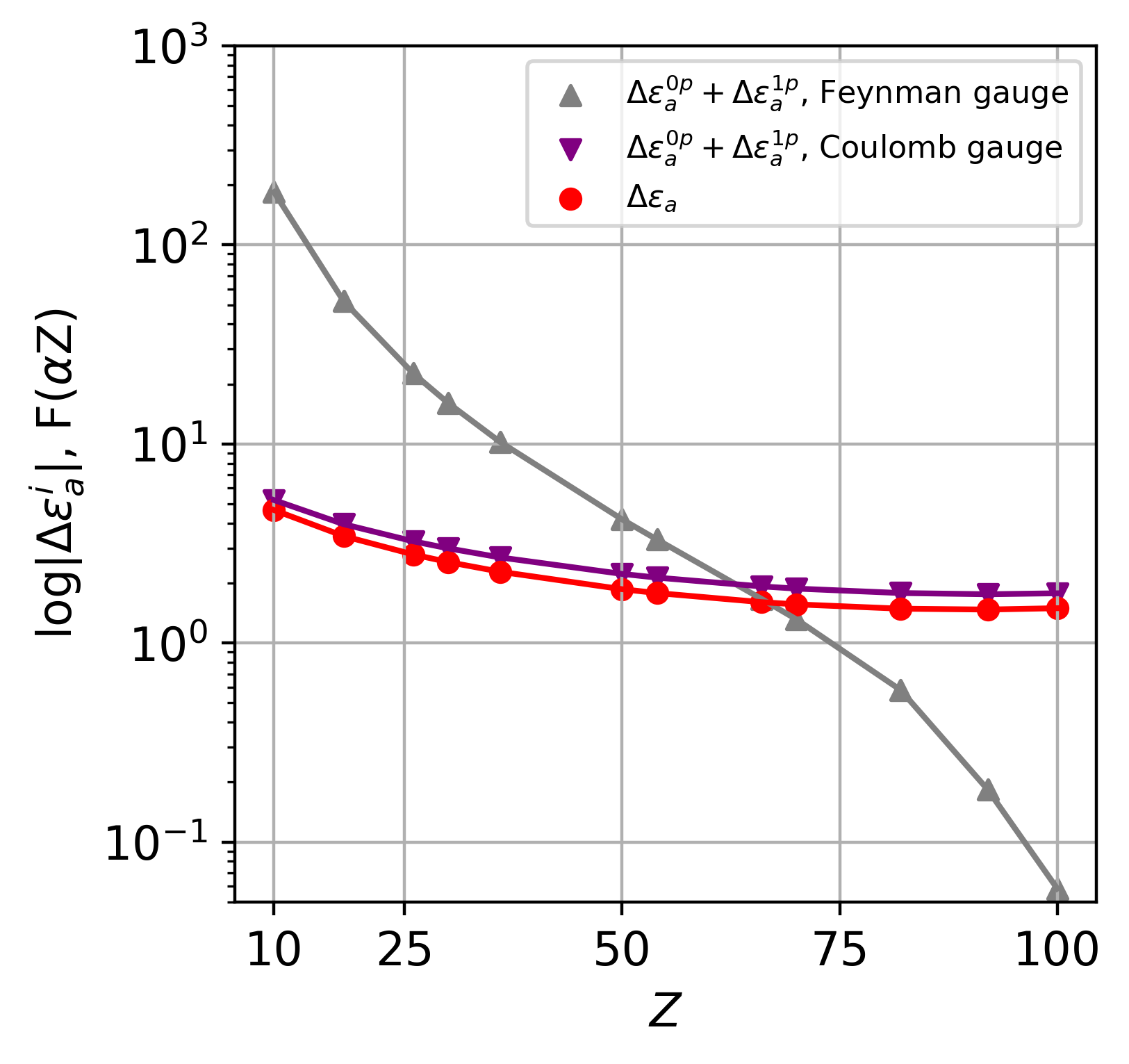} \\ (b)}
    \end{minipage}
    \caption{Individual contributions in two gauges to the self-energy correction for the $1S_{1/2}$ state (in terms of dimensionless function $F(\alpha Z)$ defined in Eq.~(\ref{eq:FaZ}))  as a function of the nuclear charge $Z$. The zero-, one-, and many-potential contributions are shown in Fig.~\ref{fig:01m_terms}(a). Sums of zero- and one-potential contributions are shown in Fig.~\ref{fig:01m_terms}(b). The total self-energy corrections, which are gauge-independent, are shown in both graphs. Note that for negative values  their absolute values are  taken. For convenience, the markers are connected with straight lines.
    }
    \label{fig:01m_terms}
\end{figure}

For a broader comparison, another set of calculations has been performed. Namely, we consider the SE correction for the ground $1S_{1/2}$ state in hydrogen-like ions with $Z \in \{10, 18, 26, 36, 54, 82, 92, 100\}$. In this case, we 
compare our results with those from Ref.~\cite{se_yerokh}. Therefore, the nuclear parameters are taken from Ref.~\cite{se_yerokh}. Average values of rms nuclear radii from there were used. An estimation of the uncertainty of the SE correction with respect to the nuclear radius and nuclear model is beyond the scope of the current paper, the interested reader can find it in the original work~\cite{se_yerokh}. The values of fundamental constants, however, have been adopted from Ref.~\cite{codata_2018}. The corresponding results are summarized in the Table~\ref{tab:we_yerokh}, which is organized similarly to the Table~\ref{tab:we_sw}.
Overall, the results shown in Table~\ref{tab:we_yerokh} confirm the findings obtained earlier. The relative smallness of the Coulomb-gauge many-potential contribution is illustrated by  Fig.~\ref{fig:01m_terms}(a), where different contributions to the SE correction for the $1S_{1/2}$ state in both gauges are plotted as  functions of $Z$. This smallness as well as the absence of cancellation between the momentum-space contributions in the Coulomb gauge is demonstrated also in Fig.~\ref{fig:01m_terms}(b), where the sum of the zero- and one-potential contributions is shown. For neon ($Z=10$), a more ``regular'' behavior of the zero- and one-potential contributions in the Coulomb gauge is even more pronounced than for higher-$Z$ systems. From Table~\ref{tab:we_yerokh}, it is seen that our results are in reasonable agreement with the results from Ref.~\cite{se_yerokh}. We conclude that the most accurate calculations of the many-potential contributions can indeed be performed using the combination of the Coulomb gauge and the SC convergence-acceleration scheme.

Using the most effective methods among those examined, namely, the GF approach and the SC convergence-acceleration scheme, we consider the SE corrections for the $n=2$ states: $2S_{1/2}$, $2P_{1/2}$, and $2P_{3/2}$. The SE corrections are calculated for hydrogen-like ions with $Z \in \{10, 18, 26, 36, 54, 82, 92, 100\}$. The corresponding results for both Feynman and Coulomb gauges are given in Table~\ref{tab:we_yerokh_excited_states}. Analyzing the table one can see that the accuracy of the total SE corrections for the excited states in hydrogen-like ions is still limited by the uncertainty of the many-potential contribution calculations. As in the case of the ground state, this contribution in the Coulomb gauge is significantly smaller than in the Feynman gauge. Moreover, the Coulomb gauge as before lacks strong cancellations between the momentum-space contributions.  Finally, we note that our results are found to be in good agreement with the ones from Ref.~\cite{se_yerokh}.

\section{Conclusion}

The self-energy correction for various states of hydrogen-like ions is considered. Two gauges, Feynman and Coulomb ones, are studied, the acceleration tricks are discussed, and a self-consistent set of useful  for the corresponding calculations formulas are collected.

The acceleration schemes do improve the accuracy of  calculations. The combination of the Coulomb gauge and the Sapirstein-Cheng partial-wave-expansion convergence-acceleration  scheme \cite{sap_orig} is most suitable (among those considered) for calculating the self-energy correction in hydrogen-like ions, especially when studying light nuclei. 

We believe that the analysis performed in the present work will facilitate the calculation of a wide range of more complex radiative corrections with the self-energy loop.

\section{Acknowledgments}
This work was supported by the Foundation for the Advancement of Theoretical Physics and Mathematics ``BASIS''.

\bibliographystyle{apsrev4-1}
\bibliography{ref}

\newpage 

\input{contribs_mp}

\newpage

\input{we_sw}



\newpage

\input{we_yerokh}



\newpage 

\include{we_yerokh_excited_states}

\newpage

\appendix
\section{Free-electron self-energy operator}
\label{ap:zero_pot}

We define the free-electron self-energy operator as:

\begin{equation}
\Sigma^{(0)} (\mathrm{p}) = 4 \pi \alpha i \int \frac{d^4 \mathrm{k}}{(2 \pi)^4}  \gamma^\mu \frac{\slashed{\mathrm{p}} - \slashed{\mathrm{k}} + m_e}{(\mathrm{p} - \mathrm{k})^2 - m_e^2 + i 0} \gamma^\nu D_{\mu \nu} (\mathrm{k}),
\label{sigma0}
\end{equation}
where $\slashed{\mathrm{p}} = \mathrm{p}_\mu \gamma^\mu$ and the photon propagator $D_{\mu\nu}$ is
\begin{equation}
D_{\mu \nu}^\text{F} (\mathrm{k}) = - \frac{ g_{\mu \nu}}{\mathrm{k}^2}
\end{equation}
in the Feynman gauge and
\begin{equation}
D_{\mu \nu}^\text{C} (\mathrm{k}) = \frac{1}{\mathrm{k}^2} \left(-g_{\mu \nu} - \frac{k_\mu k_\nu}{\mathbf{k}^2} + \frac{k^0 (k_\mu \delta_{\nu 0} + k_\nu \delta_{\mu 0})}{\mathbf{k}^2} \right)
\label{phprop_coul}
\end{equation}
in the Coulomb gauge. In Eq.~(\ref{phprop_coul}), $\delta_{\nu \mu}$ is the four-dimensional Kronecker delta, $\delta_{\mu \nu} \equiv\delta^\mu_\nu$. The expression (\ref{sigma0}) is UV divergent in both gauges. To eliminate the UV divergences, we apply the renormalization procedure. To make sense of divergent expressions, we use the dimensional regularization
$d=4-2 \varepsilon$. In both gauges, we separate out the UV divergent part by writing the self-energy in the form:
\begin{equation}
\Sigma^{(0)} (\mathrm{p}) = \delta m - \frac{\alpha}{4 \pi} \Delta_\varepsilon (\slashed{\mathrm{p}} - m_e) + \Sigma^{(0)}_R (\mathrm{p}),
\label{sigma_general}
\end{equation}
where $\Delta_\varepsilon = 1/\varepsilon - \gamma_E + \ln 4 \pi + \ln m_e$, $\gamma_E$ is the Euler constant, and the mass counterterm is
\begin{equation}
\delta m = \frac{3 \alpha}{4 \pi} m_e \left(\Delta_\varepsilon + \frac{4}{3}\right).
\end{equation}
Note that following Ref. \cite{base_1999} we use this definition of $\Sigma^{(0)}_R$ instead of writing a more familiar expression in terms of the renormalization constant $Z_2$. In the Feynman gauge, the second approach leads to the IR singularities of individual contributions. In the Coulomb gauge, both approaches are equally convenient, but, for consistency, we write the expression for the free-electron self-energy operator in this form.

The renormalized part of the free-electron self-energy operator on the right-hand side of Eq.~(\ref{sigma_general}) can be written in the form:
\begin{equation}
\Sigma^{(0)}_R (\mathrm{p}) = \frac{\alpha}{4 \pi} \big(a(p_0, p) + \slashed{\mathrm{p}} b(p_0, p) + \gamma^0 c(p_0, p)\big).
\label{sigma0ren}
\end{equation}
The coefficients $a$, $b$ and $c$ are gauge-dependent. In the Feynman gauge, they can be found, e.g., in Ref.~\cite{base_1999}:
\begin{equation}
\begin{gathered}
a(p_0, p)=2m_e \left(1 + \frac{2 \rho}{1-\rho} \ln \rho \right) , \\ 
b(p_0, p) = - \frac{2-\rho}{1-\rho}\left(1+\frac{\rho}{1-\rho} \ln \rho \right) , \\ 
c(p_0, p) = 0,
\end{gathered}
\label{abc_f}
\end{equation}
where $\rho = 1 - {\mathrm{p^2}} / {m_e^2}$. For the Coulomb gauge, they could be derived from the expression obtained in Ref.~\cite{base_adkins}:
\begin{equation}
\begin{gathered}
\Sigma^{(0)}(\mathrm{p}) = \frac{\alpha m_e}{4\pi} (3 \Delta_\varepsilon + 4) - \delta m - \frac{\alpha}{4 \pi} \Delta_\varepsilon (\slashed{\mathrm{p}} - m_e)
+ \frac{\alpha}{4 \pi} \biggl( \frac{19}{6} (\pmb{\gamma} \cdot \mathbf{p}) - \frac{1}{2} \gamma^0 p_0 \\ 
- \int_0^1 \frac{dx}{\sqrt{x}} \ln X [ (1-x) (\mathbf{p} \cdot \pmb{\gamma}) + m_e]+2 \int_0^1 dx \ln Y [(1-x) \slashed{\mathrm{p}} -m_e] + 2 (\pmb{\gamma} \cdot \mathbf{p}) \int_0^1 dx du \sqrt{x} \ln Z \biggr),
\end{gathered}
\label{sigma_free_adk}
\end{equation}
where 
\begin{equation}
\begin{gathered}
X = 1 + \frac{p^2}{m_e^2} (1-x), \\
Y = 1 - \frac{\mathrm{p}^2}{m_e^2} (1-x) - i 0, \\
Z = 1 - \frac{p_0^2}{m_e^2} (1-u) + \frac{p^2}{m_e^2} (1-x u) - i 0.
\end{gathered}
\end{equation}
Some integrals in these expressions can be evaluated analytically, which was done, e.g., in Ref.~\cite{zp_final}. The latter results allows one to write for the coefficients $a$, $b$, and $c$:
\begin{equation}
\begin{gathered}
a(p_0, p) = 2 m_e \left(1 - F_0 + \frac{\rho \log \rho }{1 - \rho}\right), \\ 
b(p_0, p) = \frac{(\rho - 2)(1-\rho+\rho \ln \rho)}{(1 - \rho)^2} - 2 F_2 \rho 
+ \frac{2 m_e^2 (F_1 \rho \ln \rho - F_0)}{p^2},\\
c(p_0, p) = 2 \frac{p_0}{p^2} \big(F_0 m_e^2 - F_1 m_e^2 \rho \ln \rho + F_2 \rho p^2\big),
\end{gathered}
\label{abc_c}
\end{equation}
where $F_0$, $F_1$, and $F_2$ are defined according to
\begin{equation}
\begin{gathered}
F_0 = \frac{\sqrt{p^2 + m_e^2}}{p} \ln \left| \frac{\sqrt{p^2 + m_e^2} +p}{\sqrt{p^2 + m_e^2} - p} \right|- 2,\\
F_1 = \frac{p_0}{p} \ln \left| \frac{p_0 +p}{p_0 -p} \right|- 2,\\
F_2 = \int_0^1 dx \frac{\sqrt{x} \ln X}{X - \rho + i 0}. 
\end{gathered}
\label{fi_zp_cg}
\end{equation}

\section{Free-electron vertex operator}
\label{ap:one_pot}
In order to calculate the one-potential term, one needs the free-electron vertex function given by:
\begin{equation}
\Gamma^{\mu} (\mathrm{p}',\mathrm{p}) = 4 \pi \alpha i \int \frac{d^{4} \mathrm{k}}{(2 \pi )^{4}} \gamma^\sigma \frac{\slashed{\mathrm{p}}'-\slashed{\mathrm{k}}+m_e}{(\mathrm{p}'-\mathrm{k})^ {2}-m_e^{2}} \gamma^{\mu} \frac{\slashed{\mathrm{p}}-\slashed{\mathrm{k}}+m_e}{(\mathrm{p}-\mathrm{k})^ {2}-m_e^{2}} \gamma^\rho D_{\sigma \rho}(\mathrm{k}).
\end{equation}
This expression is UV divergent in both gauges. Using dimensional regularization, it can be written as:
\begin{equation}
\Gamma^{\mu} (\mathrm{p}',\mathrm{p}) = \frac{\alpha}{4 \pi} \Delta_\varepsilon \gamma^{\mu} + \Gamma_R^{\mu}(\mathrm{p}',\mathrm{p}),
\end{equation}
where, as in the case of the free-electron self-energy operator, the UV finite part of $\Gamma_R$ also turns out to be IR finite. For brevity, we discuss below only the time component of the operator, $\Gamma^{0}_R$, for the case when $p_0 = p'_0 = \varepsilon_a$:

\begin{equation}
\begin{gathered}
\Gamma_{R}^{0}(\mathrm{p}',\mathrm{p}) = \frac{\alpha}{4 \pi}\{A \gamma^{0}+\slashed{\mathrm{p}}' (B_{1} + B_{2}) \varepsilon_a+\slashed{\mathrm{p}} (C_{1} + C_{2}) \varepsilon_a + D (\slashed{\mathrm{p}}' \gamma^0 \slashed{\mathrm{p}}) \\ +(H_{1} + H_2) \varepsilon_a + G_{1} \slashed{\mathrm{p}}' \gamma^0 + G_{2} \gamma^0 \slashed{\mathrm{p}} \}.
\end{gathered}
\label{vertex_function}
\end{equation}
In the Feynman gauge:
\begin{equation}
\begin{gathered}
 A= C_{5}-2+\mathrm{p}'^{2} C_{11} + \mathrm{p}^{2} C_{12} +4 (\mathrm{p}' \cdot \mathrm{p} )( C_{00} + C_{11} + C_{12})
+ m_e^{2} ( -2 C_{00} + C_{11} + C_{12}), \\ 
 B_{1}   =-4(   C_{11}   +   C_{23}   ), \;
   B_{2}   =-4(   C_{00}   +   C_{11}   +   C_{12}   +   C_{25}   ), \\
   C_{1}   = B_2, \;
   C_{2}   =-4(   C_{12}   +   C_{24}   ),\\
D=2(   C_{00}   +   C_{11}   +   C_{12}   ), \\
   H_{1}   =4m_e(   C_{00}   +   2C_{11}   ), \;
   H_{2}   =4m_e(   C_{00}   +   2C_{12}   ), \; G_1 = G_2 = 0,
\end{gathered}
\end{equation}
where
\begin{equation}
\begin{gathered}
C_{ij} = \int_{0}^{1}\frac{d y}{(y \mathrm{p}' + (1-y) \mathrm{p})^{2}} S_i K_j, \\
S_{0,1,2} = \{- \ln X', 1-Y' \ln X', -\frac{1}{2}+Y'-{Y'}^{2} \ln X'\}, \\
K_{0,1,2,3,4,5} = \{1, y, 1-y, y^2, (1-y)^2, y(1-y)\}, \\
C_{5} = -\int_{0}^{1} dy \ln (y^{2}\mathrm{k}^{2} \slash m_e^{2} - y \mathrm{k}^{2} / m_e^{2} + 1 ), \\ 
X'=1+\frac{1}{Y'}, \; Y'=\frac{m_e^2 - y \mathrm{p}'^2 - (1-y) \mathrm{p}^2}{(y \mathrm{p}' + (1-y) \mathrm{p})^2},
\end{gathered}
\end{equation}
and $\mathrm{k}=\mathrm{p}-\mathrm{p}'$. For the Coulomb gauge, these coefficients can be readily determined from Ref.~\cite{OP_CG}:
\begin{equation}
\begin{gathered}
B_1 = F_{19}- F_{20} + 4 (F_5-F_3) +2 p'^2 (F_{14}-2 F_{17})+ \\ +2 p^2 (F_{13}-F_{14}-2 F_{16}+2 F_{17})+2 (\mathbf{p'} \cdot \mathbf{p}) (F_{13}-F_{16}),\\
C_1 = F_{19} - F_{20} + 4(F_3 - F_2 + F_4 - F_5) + \\
+ 2 p'^2 (F_{13} - F_{14} - F_{16} + F_{17})  + \\ + 2 p^2 (F_{12} - 2 F_{13} + F_{14} - F_{15} + 2 F_{16} - F_{17}) + \\ + 2 (\mathbf {p'} \cdot \mathbf{p}) (F_{12} - F_{13} - F_{15} + F_{16}),  \\ 
D = F_7-F_{10} + 2(F_{19} - F_1) + \\ + 2 p^2 (F_{12}-2 F_{13})+4 F_{13} (\mathbf {p'} \cdot \mathbf{p}) +2 k^2 F_{14}, \\ 
B_2 = C_2 = -D, \\
G_1 = m_e \big[F_{10}-F_{19}+2 p^2 F_{13} - 2 (\mathbf {p'} \cdot \mathbf{p}) F_{13} -2 k^2 F_{14}\big] ,\\
G_2 = m_e \big[F_{10}-F_{19}+2 p^2 (F_{13}-F_{12}) +2 (\mathbf {p'} \cdot \mathbf{p})(F_{12}-F_{13}) +2 k^2 (F_{13}-F_{14})\big], \\
H_1 = 4 m_e (F_2-F_1),  \\ 
H_2 =  - G_1 - G_2, \\
\end{gathered}
\end{equation}
and finally $A$:
\begin{equation}
\begin{gathered}    
A = \varepsilon_a^2 (2 F_1 - F_2) - F_{22} + m_e^2 (2 F_1 - 3 F_2) + p'^2 (F_{11} - F_8 + 4 F_5 - 5 F_3) -\\ 
- 4 (\mathbf{p'} \cdot \mathbf{p})^2 F_{13} + p^2 (F_{10} - F_{11} - 5 F_2 + 5 F_3 + 4 F_4 - 4 F_5 - F_7 + \\ 
+F_8 + (\mathbf{p'} \cdot \mathbf{p}) (-2 F_{12} + 4 F_{13})) + (-4 F_5 + 4 F_6 + 2 F_8 - 2 F_9) k^2 + \\ 
+(\mathbf{p'} \cdot \mathbf{p}) (4 F_1 - 2 F_{19} - 2 F_2 - 2 F_{14} k^2) - \varepsilon_a^2 (B_1+C_1-D).
\end{gathered}
\end{equation}
The coefficients $F_{1-22}$ introduced above can be determined as follows. 
Let us define $B_{i,j}^k$ as
\begin{equation}
B_{i,j}^k = 
\begin{cases}
1\,, & \text{if } i \leq k \leq j \,,\\
0\,, & \text{otherwise} \,.
\end{cases}             
\end{equation}
then for $F_1-F_6$ one obtains
\begin{equation}
\begin{gathered}
F_i = \int_0^1 \frac{du}{\mathrm{t}^2} \Big[ C_1^i \delta_1 + C_2^i - C_3^i \frac{A}{\mathrm{t}^2} \Big] C_4^i,\\
\delta_1 = \ln \left( \frac{\mathrm{t}^2+A}{A} \right), \,\,
C_1^i = [1-B_{2,6}^i \left(1+\frac{A}{\mathrm{t}^2}\right)] [1-B_{4,6}^i \left(1+\frac{A}{\mathrm{t}^2}\right)], \\ 
C_2^i = B_{2,3}^i+\frac{1}{2}B_{4,6}^i, \,\,
C_3^i = B_{4,6}^i, \,\, C_4^i = \{1,1,u,1,u,u^2\}.
\end{gathered}
\end{equation}
Here, $A = m_e^2 - u \mathrm{p}^2 - (1 - u) \mathrm{p}^2$, $\mathrm{t} = u \mathrm{p}' + (1 - u) \mathrm{p}$. For $F_{7-11}$ (below the coefficient $F_7$ is obtained by substituting $i=1$, and so on, i.e., $F_8 \leftrightarrow i = 2$, ...):
\begin{equation}
\begin{gathered}
F_i = \int_0^1 du \Big[  C_5^i \delta_2 - C_6^i \frac{1}{t^2} \Big] C_7^i,\\
\delta_2 = \frac{2}{\sqrt{t^2 C}} \tanh^{-1} \left[ \left( \frac{t^2}{C} \right)^{1/2} \right], \,\,
C_5^i = \frac{C}{t^2} B_{1,3}^i+B_{4,5}^i, \\ 
C_6^i = 2 B_{1,3}^i, \,\,
C_7^i = \{1,u,u^2,1,u\},
\end{gathered}
\end{equation}
where $C = m_e^2 - u \mathrm{p'}^2 - (1-u) \mathrm{p}^2 + t_0^2$. For $F_{12}-F_{21}$ (below $ F_{12} \leftrightarrow i = 1$):
\begin{equation}
\begin{gathered}
F_i = \int_0^1 ds du \Big[  C_8^i \delta_3 + C_9^i \delta_4 \Big] C_{10}^i,\\
\delta_3 = \frac{1}{s (t^2)^2} \left\{ \left( \frac{B}{s t^2} \right)^{1/2} \tanh^{-1} \Big[\left( \frac{s t^2}{B} \right)^{1/2}\Big] - 1 \right\}, \,\, \delta_4 = \frac{1}{t^2 (B-s t^2)}, \\
C_8^i = -3 B_{1,7}^i + 2 t^2 B_{8,10}^i, \,\, C_9^i = B_{1,7}^i, \\ 
C_{10}^i = \{1,u,u^2, s,su,su^2,su^3,1,s,su\},
\end{gathered}
\end{equation}
where $B = m_e^2 - u \mathrm{p'}^2 - (1 - u) \mathrm{p}^2 + s t_0^2$. And, finally, $F_{22}$ is given by
\begin{equation}
F_{22} = \int_0^1 du \ln (u^2\mathrm{k}^2 \slash m_e^2 - u \mathrm{k}^2 \slash m_e^2 + 1).
\end{equation}

\section{Many-potential term}
\label{ap:m_pot}

Reduced matrix elements of the operators $I_\text{F}(\omega, \mathbf{r}_1, \mathbf{r}_2)$ and $I_\text{C}(\omega, \mathbf{r}_1, \mathbf{r}_2)$ can be derived using the PW expansion of the photon propagator. 
Namely, for this aim one needs the standard expansion formulas \cite{Varshalovich:1988:book:eng} for the functions which depend on $r_{12}$, such as
\begin{equation}
\frac{e^{i \omega r_{12}}}{r_{12}} = 4 \pi i \omega \sum_{L=0}^{\infty} \sum_{M=-L}^{L} j_L(\omega r_<) h_L^{(1)}(\omega r_>) Y_{LM}^*(\mathbf{\hat{r}_1}) Y_{LM}(\mathbf{\hat{r}_2}).
\label{pwe_example}
\end{equation}
Here, $j_L$ and $h_L^{(1)}$ are the spherical Bessel function and spherical Hankel functions of the first kind, respectively and $r_> = \max (r_1, r_2 )$ and $r_< = \min (r_1, r_2)$. Applying these formulas and performing the integrations over the angular variables, after some tedious manipulations one can obtain
\begin{equation}
\begin{gathered}
\langle ab || I_\text{F} (\omega ) || cd \rangle_J = \alpha \int_0^\infty dr_{1} r_1^2 \int_0^\infty dr_{2} r_2^2 \, \Big\{ (-1)^{J} G_{J}(\kappa_{a},\kappa_{c}) G_J ( \kappa_{b}, \kappa_{d} ) g_J(\hat{\omega}, r_1, r_2) A_{ac} (r_1) A_{bd} (r_2) \\ 
+ \sum_L (-1)^{L+1} [J] g_{L} (\hat{\omega},r_1,r_2) D^{JL}_{ac} (r_1) D^{JL}_{bd} (r_2)\Big\}
\label{M_el_f}
\end{gathered}
\end{equation}
for the Feynman gauge, and
\begin{equation}
\begin{gathered}
\langle ab || I_\text{C} (\omega ) || cd \rangle_J = \alpha \int_0^\infty dr_{1} r_1^2 \int_0^\infty dr_{2} r_2^2 \, \Big\{ (-1)^{J} G_{J}(\kappa_{a},\kappa_{c}) G_J ( \kappa_{b}, \kappa_{d} ) g_J(0, r_1, r_2) A_{ac} (r_1) A_{bd} (r_2) \\ 
+ \sum_L (-1)^{L+1} a_{JL} g_{L} (\hat{\omega},r_1,r_2) D^{JL}_{ac} (r_1) D^{JL}_{bd} (r_2) \\ 
+ (-1)^{J+1} b_{J} [g_{J}^{\text{ret}} (\hat{\omega},r_1,r_2) D^{JJ+1}_{ac} (r_1) D^{JJ-1}_{bd} (r_2) + g_{J}^{\text{ret}} (\hat{\omega},r_2,r_1) D^{JJ-1}_{ac} (r_1) D^{JJ+1}_{bd} (r_2)]
\Big\}
\label{M_el_c}
\end{gathered}
\end{equation}
for the Coulomb gauge. 
In the above expressions, 
$[J]=2J+1$ and the following functions are defined:
\begin{align}
g_{J} (\omega, r_1, r_2) &= i [J] \omega j_J ({\omega} r_<) h_J^{(1)} ({\omega} r_>), \\
g_{J} (0, r_1, r_2) &= \frac{r_<^J}{r_<^{J+1}} , \\ 
g_{J}^{\text{ret}} (\omega, r_1, r_2) &= 
\begin{cases}
i [J] {\omega} j_{J+1} ({\omega} r_1) h_{J-1}^{(1)} ({\omega} r_2), & \text{for}\ r_1<r_2, \\
i [J] {\omega} j_{J-1} ({\omega} r_2) h_{J+1}^{(1)} ({\omega} r_1) - \dfrac{[J]^2 r_2^{J-1}}{\omega^2 r_1^{J+2}}, & \text{otherwise,}   \\
\end{cases}
\end{align}
The expressions depending on the radial wave functions read as
\begin{align}
A_{ab} (r) &= g_{a} (r) g_{b}(r) + f_{a}(r) f_{b} (r)  \, , \\
D^{JL}_{ab} (r) &= g_{a} (r) f_{b} (r) H_{L}^J (\kappa_{a}, -\kappa_{b}) - f_{a}(r) g_{b}(r) H_{L}^J (-\kappa_{a}, \kappa_{b}) \,.
\end{align}
Finally, the angular coefficients are
\begin{align}
G_{J}(\kappa_{a},\kappa_{b}) &= (-1)^{j_b + 1 \slash 2} \sqrt{[j_a][j_b][l_a][l_b]} {\begin{pmatrix}l_a&J&l_b\\0&0&0\end{pmatrix}} {\begin{Bmatrix}j_a&J&j_b\\l_b&\frac{1}{2}&l_a\end{Bmatrix}}, \\ 
H_L^J (\kappa_a, \kappa_b) &= (-1)^{l_a} \sqrt{6 [j_a][j_b][l_a][l_b]} {\begin{pmatrix}l_a&L&l_b\\0&0&0\end{pmatrix}} {\begin{Bmatrix}j_a&\frac{1}{2}&l_a\\J&1&L\\j_b&\frac{1}{2}&l_b\end{Bmatrix}}, \\ 
a_{JL} &= 
\begin{cases}
J+1, & \text{for}\ L=J-1 , \\
2 J + 1, & \text{for}\ L=J , \\ 
J, & \text{for}\ L=J+1 ,
\end{cases} \\
b_{J} &= \sqrt{J (J+1)} \frac{\sqrt{[J+1][J-1]}}{[J]} .
\end{align}

Note that (\ref{M_el_c}) is consistent with Ref.~\cite{mel_yerokh}, except for the typo in Eq.~(B6) there.

\section{Derivative of the free-electron self-energy operator}
\label{ap:zero_pot_deriv}

Let us represent the second derivative of the function $\Sigma^{(0)}_R$ with respect to the time component of the four-vector $\mathrm{p}$ in the form:
\begin{equation}
\frac{\partial^2 \Sigma_R^{(0)} (\mathrm{p})}{\partial p_0^2}\Bigr|_{p_0 = \varepsilon_a} = \frac{\alpha}{4 \pi} \left\{ (N_1 + N_2 \gamma^0 - N_3 (\pmb{\gamma} \cdot \mathbf{p})\right\},
\end{equation}
where the coefficients $N_1$, $N_2$, and $N_3$ can be obtained by differentiating the formula (\ref{sigma0ren}):

\begin{equation}
\begin{gathered}
N_1 = \frac{\partial^2 a(p_0, p)}{\partial p_0^2} \Bigr|_{p_0 = \varepsilon_a}, \\ 
N_2 = \varepsilon_a \frac{\partial^2 b(p_0, p)}{\partial p_0^2} \Bigr|_{p_0 = \varepsilon_a} + \frac{\partial b(p_0, p)}{\partial p_0} \Bigr|_{p_0 = \varepsilon_a} + \frac{\partial^2 c(p_0, p)}{\partial p_0^2} \Bigr|_{p_0 = \varepsilon_a}, \\
N_3 = \frac{\partial^2 b(p_0, p)}{\partial p_0^2} \Bigr|_{p_0 = \varepsilon_a}.
\end{gathered}
\label{ni}
\end{equation}
In the Feynman gauge, the derivatives of the coefficients $a$ and $b$ are (see \cite{sap_orig, sap_alexey}): 

\begin{equation}
\frac{\partial^2 a(p_0, p)}{\partial p_0^2} = - \frac{8}{m_e (1 - \rho)} \left\{ 1 + \frac{1}{1-\rho} \left[ \ln \rho - \frac{2 p_0^2}{m_e^2} \left( \frac{1 + \rho}{\rho} + \frac{2}{1-\rho} \ln \rho \right) \right] \right\},
\end{equation}

\begin{equation}
\frac{\partial b(p_0, p)}{\partial p_0} = \frac{2 p_0}{m_e^2 (1-\rho)^2} \left\{ 3 - \rho + \frac{2}{1-\rho} \ln \rho \right\},
\end{equation}

\begin{equation}
\frac{\partial^2 b(p_0, p)}{\partial p_0^2} = \frac{2}{m_e^2 (1-\rho)^2} \left\{ 3 - \rho + \frac{2}{1-\rho} \left[ \ln \rho - \frac{p_0^2}{m_e^2} \left( \frac{2 + 5 \rho - \rho^2}{\rho} + \frac{6}{1 - \rho} \ln \rho \right) \right] \right\}.
\end{equation}
In the Coulomb gauge, it is more convenient to differentiate the free-electron self-energy operator in the form (\ref{sigma_free_adk}):
\begin{equation}
\begin{gathered}
\frac{\partial^2 \Sigma^{(0)}_{R}(\mathrm{p})}{\partial p_0^2} = \frac{\alpha}{4 \pi} \Big\{2 \int_0^1 dx \left( \frac{\partial^2 \ln Y}{\partial p_0^2} [(1-x) \slashed{\mathrm{p}} -m_e] + 2 \frac{\partial \ln Y}{\partial p_0} (1-x) \gamma^0 \right) + \\ + 2 (\pmb{\gamma} \cdot \mathbf{p}) \int_0^1 dx du \sqrt{x} \frac{\partial^2 \ln Z}{\partial p_0^2} \Big\}.
\end{gathered}
\label{sigma_free}
\end{equation}
Then, for the coefficients $N_1$ - $N_3$ we have:
\begin{equation}
\begin{gathered}
N_1 = - 2 m_e \int_0^1 dx \frac{\partial^2 \ln Y}{\partial p_0^2} \Bigr|_{p_0 = \varepsilon_a}, \\
N_2 = 2 \int_0^1 dx (1-x) \left( \varepsilon_a \frac{\partial^2 \ln Y}{\partial p_0^2} \Bigr|_{p_0 = \varepsilon_a} + 2 \frac{\partial \ln Y}{\partial p_0} \Bigr|_{p_0 = \varepsilon_a}\right), \\
N_3 = 2 \int_0^1 dx \left( (1-x) \frac{\partial^2 \ln Y}{\partial p_0^2} \Bigr|_{p_0 = \varepsilon_a} - \sqrt{x} \int_0^1 du \frac{\partial^2 \ln Z}{\partial p_0^2} \Bigr|_{p_0 = \varepsilon_a}\right),
\end{gathered}
\label{ni_c}
\end{equation}
where
\begin{equation}
\begin{gathered}
\frac{\partial \ln Y}{\partial p_0} = \frac{2 p_0 (x-1)}{m_e^2 - (p_0^2 - p^2) (1-x) - i 0}, \\
\frac{\partial^2 \ln Y}{\partial p_0^2} = \frac{2 (x-1) (m_e^2 + (p_0^2 + p^2) (1-x))}{(m_e^2 - (p_0^2 - p^2) (1-x) - i 0)^2}, \\
\frac{\partial^2 \ln Z}{\partial p_0^2} = \frac{2 (u-1) (m_e^2 + p_0^2(1-u) + p^2 (1-x u))}{(m_e^2 - p_0^2 (1-u) + p^2 (1-x u) - i 0)^2}.
\end{gathered}
\end{equation}  
For the coefficient $Z$, the second derivative can be integrated analytically with respect to $u$:
\begin{equation}
\begin{gathered}
\int_0^1 du \left( \frac{\partial^2 \ln Z}{\partial p_0^2} \right) = 2 \Big[ \frac{p_0^4 - 3 p_0^2 (m_e^2 + p^2) - x p^2 (m_e^2 + p^2 - 3 p_0^2)}{m_e^2 \rho (p_0^2 - x p^2)^2}\\
- \frac{(m_e^2 + p^2 (1-x))(3 p_0^2 + x p^2)}{(p_0^2 - x p^2)^3} \ln \left( \frac{m_e^2 + p^2 - p_0^2}{m_e^2 + p^2 (1-x)} \right)
\Big].
\end{gathered}
\end{equation}
Some integrals over $x$ from the expressions (\ref{ni_c}) can be evaluated analytically using the following master integrals:
\begin{equation}
\begin{gathered}
\int_0^1 dy \frac{y}{(m_e^2 - (p_0^2 - p^2) y)^2} = \frac{1}{p_0^2 - p^2} \left( \frac{ \ln \rho }{p_0^2 - p^2} + \frac{1}{m_e^2 \rho} \right), \\
\int_0^1 dy \frac{y^2}{(m_e^2 - (p_0^2 - p^2) y)^2} = \frac{m_e^2}{(p_0^2 - p^2)^3} \left( \frac{1}{\rho} + 2 \ln \rho - \rho \right), \\
\int_0^1 dy \frac{y^3}{(m_e^2 - (p_0^2 - p^2) y)^2} = \frac{1}{(p_0^2 - p^2)^4} ( p_0^4 - 2 (p_0^2 + 2 m_e^2) p^2 + \frac{2 m_e^4}{\rho} + 6 m_e^4 \ln \rho \\ + 4 m_e^2 p_0^2 + (p^2)^2 - 2 m_e^4), \\ 
\int_0^1 dy \frac{y^2}{m_e^2 - (p_0^2 - p^2) y} = - \frac{1}{2} \left( \frac{1}{p_0^2 - p^2} + \frac{2 m_e^2}{(p_0^2 - p^2)^2} + \frac{2 m_e^4 \ln \rho}{(p_0^2 - p^2)^3} \right).
\end{gathered}
\end{equation}






\end{document}

%% file: contour.tex
\begin{tikzpicture}[scale=0.5]
  \draw[thick, ->] (-8,0) -- (12,0) node[below] {$\text{Re}(\omega)$} coordinate (x axis); 
  \draw[thick, ->] (0,-10) -- (0,10) node[right] {$\text{Im}(\omega)$} coordinate (y axis); 
  \draw[very thick][blue] (12,0.2) arc [start angle=0, end angle=90, x radius=9, y radius=9.8]; 
  \draw[very thick][->][blue] (3, 0) -- (3, 5) node[right]{}; 
  \draw[very thick][blue] (3, 5) -- (3, 10); 

  \draw[very thick][blue] (3, 0) arc [start angle=360, end angle=188, x radius=1.25, y radius = 0.75];
  \draw[very thick][blue] (0.52, -0.13) arc [start angle=30, end angle=200, x radius = 0.3, y radius = 0.3] node[left] {};
  \draw[very thick][blue] (3, -0.74) arc [start angle=330, end angle=187.5, x radius = 1.624, y radius = 0.98]; 
  
  \draw[very thick][blue] (3, -0.74) -- (3, -5); 
    \draw[very thick, ->][blue] (3, -10) -- (3, -5) node[right] {}; 
  \draw[very thick][blue] (3, -10) arc [start angle=270, end angle =180, x radius = 11, y radius = 9.8]; 
  
  \fill[darkgray] (-0.275, 0.275) circle (3pt); 
  \draw[thick, dashed][darkgray] (-0.275, 0.275) .. controls (-3, 0.15) and (-6, 0.15) .. (-8, 0.15);

  \fill[darkgray] (0.275, -0.275) circle (3pt); 
  \draw[thick, dashed][darkgray] (0.275, -0.275) .. controls (1, -1.2) and (2, -1.25) .. (3, -0.5);
  \draw[thick, dashed][darkgray] (3,-0.5) .. controls (3.5, -0.15) and (5, -0.15) .. (12, -0.15);
  
  \fill (7, -2) circle (3pt); 
  \draw[thick] (7, -2) -- (12, -2);

  \fill (2, 2) circle (3pt); 
  \fill (1.96, 2) circle (0pt) node[above] {\tiny{$\varepsilon_a- \varepsilon_{1s}$}};
  \fill (0.8, 2) circle (3pt);
  \fill (0, 2) circle (3pt);
  \fill (-0.64, 2) circle (3pt);
  \fill (-1.1520, 2) circle (3pt);
  \fill (-1.5, 2) circle (3pt);
  \fill (-1.7, 2) circle (3pt);
  \fill (-1.8, 2) circle (3pt);
  \fill (-1.85, 2) circle (3pt);
  \fill (-2.0, 2) circle (3pt);
  \draw[thick] (-2, 2) -- (-8, 2);
  
\end{tikzpicture}

%% file: contribs_mp.tex
{\setlength\LTcapwidth{\textwidth}
\renewcommand{\arraystretch}{1.8}
\begin{table}[h]
    \centering
    \caption{\label{tab:contribs_mp}
         Contributions to the self-energy correction for the ground $1S_{1/2}$ state of hydrogen-like argon and uranium obtained in the Feynman (F) and Coulomb (C) gauges (in eV). The individual terms of the partial-wave expansion for the many-potential contribution as well as
        the results of applying the extrapolation procedure are shown. The nuclear-charge distribution is described by the homogeneously-charged-sphere model with the root-mean-square radii taken from Ref.~\cite{hen}.
         }
{
{\fontsize{9pt}{9pt}\selectfont{
    \begin{tabular}{
                l@{\quad}
                S[table-format=-3.6(2),group-separator={},group-minimum-digits=3]
                S[table-format=-3.6(2),group-separator={},group-minimum-digits=3]
                S[table-format=-3.6(2),group-separator={},group-minimum-digits=3]
                S[table-format=-3.6(2),group-separator={},group-minimum-digits=3]
                }
\hline
\hline
& \multicolumn{2}{c}{$Z = 18$} & \multicolumn{2}{c}{$Z = 92$} \\
\hline 
& \multicolumn{1}{c}{F} & \multicolumn{1}{c}{C} & \multicolumn{1}{c}{F} & \multicolumn{1}{c}{C} \\ 
\hline
$\Delta \varepsilon_a^{0\text{p}}$ & -67.924 836  & 1.341 667  & -516.318 629 & 210.068 167 \\
$\Delta \varepsilon_a^{1\text{p}}$ & 49.511 442  & 0.054 770  & 472.000 467 & 213.738 955\\
\hline
$\Delta \varepsilon_a^{\text{Mp}}(k),$ $k = 1$ 
     &18.954 139   &               -0.098 735    &    392.056 752   &    -54.030 900  \\ 
$2$  & 0.447 224            &       -0.044 326    &   2.922 927       &  -11.168 699  \\ 
$3$  & 0.115 257             &      -0.011 632    &   2.001 276     &    -1.902 586    \\
$4$  & 0.046 658            &       -0.006 453   &    0.915 478      &   -0.685 643    \\
$5$  & 0.023 375            &       -0.004 153   &    0.478 148      &   -0.330 469    \\
$6$  & 0.013 267            &       -0.002 872   &    0.278 392      &   -0.186 004    \\
$7$  & 0.008 186          &         -0.002 080   &    0.175 634      &   -0.115 450    \\
$8$  & 0.005 368          &         -0.001 558   &    0.117 699      &   -0.076 706    \\
$9$  & 0.003 688          &         -0.001 199   &    0.082 646      &   -0.053 600    \\
$10$ & 0.002 630          &         -0.000 943  &     0.060 225       &  -0.038 946    \\
$11$ & 0.001 933          &         -0.000 754  &     0.045 228       &  -0.029 195    \\
$12$ & 0.001 458          &         -0.000 613  &     0.034 822       &  -0.022 452    \\
$13$ & 0.001 124          &         -0.000 505  &     0.027 378       &  -0.017 638    \\
$14$ & 0.000 882         &          -0.000 420  &     0.021 913       &  -0.014 110    \\
$15$ & 0.000 704         &          -0.000 353  &     0.017 811       &  -0.011 464    \\
$16$ & 0.000 570         &          -0.000 300  &     0.014 672       &  -0.009 441    \\
$17$ & 0.000 468         &          -0.000 257  &     0.012 229       &  -0.007 868    \\
$18$ & 0.000 388         &          -0.000 221  &     0.010 300       &  -0.006 626    \\
$19$ & 0.000 325         &          -0.000 192  &     0.008 756       &  -0.005 632    \\
$20$ & 0.000 275         &          -0.000 168  &     0.007 506     &    -0.004 827  \\
$21$ & 0.000 234         &          -0.000 147  &     0.006 483     &    -0.004 169  \\
$22$ & 0.000 201         &          -0.000 130  &     0.005 638     &    -0.003 625  \\
$23$ & 0.000 174         &          -0.000 115  &     0.004 933     &    -0.003 172  \\
$24$ & 0.000 152         &          -0.000 103  &     0.004 341     &    -0.002 792  \\
$\sum_{k=1}^{24}$ & 19.628 680  & -0.178 229 & 399.311 187 & -68.732 014 \\
$\sum_{k=25}^{\infty}$ [extr.] & 0.001595(16) & -0.001300(15) & 0.049936(33)  & -0.032109(30) \\
\hline 
$\Delta \varepsilon_a$ & 1.216893(16) & 1.216897(15) & 355.042972(33) &
355.042991(30) \\
\hline
\hline
\end{tabular}
}}}
\end{table}
}

%% file: we_sw.tex
{\setlength\LTcapwidth{\textwidth}
\renewcommand{\arraystretch}{2.0}
\begin{table}[h]
    \centering
    \caption{\label{tab:we_sw}
         Individual contributions to the self-energy correction for the $1S_{1/2}$  state of hydrogen-like ions in the Feynman (F) and Coulomb (C) gauges (in eV). The many-potential contribution is calculated in three ways: directly (DIR), using the two-potential (TP) scheme, and using the Sapirstein-Cheng (SC) approach. 
         The nuclear-charge distribution is described by the homogeneously-charged-sphere model with the root-mean-square radii taken from Ref.~\cite{hen}. The values of fundamental constants are adopted from the same Ref.~\cite{hen}.
         }
{
\resizebox{\textwidth}{!}{
{\fontsize{12pt}{12pt}\selectfont{
    \begin{tabular}{
                l@{\quad}
                l@{\quad}
                S[table-format=3.9(2),group-separator={\;},group-minimum-digits=3]
                S[table-format=3.9(2),group-separator={\;},group-minimum-digits=3]
                l@{\quad}
                l@{\quad}
                S[table-format=3.9(2),group-separator={\;},group-minimum-digits=3]
                S[table-format=3.9(2),group-separator={\;},group-minimum-digits=3]
                S[table-format=3.9(2),group-separator={\;},group-minimum-digits=3]
                S[table-format=3.9(2),group-separator={\;},group-minimum-digits=3]
                 }
\hline
\hline 

\multicolumn{1}{l}{$Z$} & 
\multicolumn{1}{l}{Gauge $\quad$} & 
\multicolumn{1}{c}{$\Delta \varepsilon_a^{0\text{p}}$} &
\multicolumn{1}{c}{$\Delta \varepsilon_a^{1\text{p}}$} &
\multicolumn{1}{l}{Scheme $\quad$} & 
\multicolumn{1}{l}{Method $\quad$} & 
\multicolumn{1}{c}{$\Delta \varepsilon_a^{2\text{p}}$ / $\Delta \tilde{\varepsilon}_a^{2\text{p}}$} &
\multicolumn{1}{c}{$\Delta \varepsilon_a^{(3+)\text{p}}$ / $\Delta \tilde{\varepsilon}_a^{(3+)\text{p}}$} &
\multicolumn{1}{c}{$\Delta \varepsilon_a^{\text{Mp}}$} &
\multicolumn{1}{c}{$\Delta \varepsilon_a$} \\
\hline
{\multirow{12}{*}{\tablenum{18}}} 				& {\multirow{5}{*}{F}} 				& {\multirow{5}{*}{\tablenum{-67.924836710}}} 				& {\multirow{5}{*}{\tablenum{49.511442330}}} 				& DIR & DKB 				& {--} 				& {--} 				& 19.6299(55) 				& 1.2165(55) \\
 &  				&  				& & DIR & GF 				& {--} 				& {--} 				& 19.630288(16) 				& 1.216893(16) \\
\cline{5-10}
 & 				& & 				& TP & GF 				& 8.9605440(13) 				& 10.66975705(68) 				& 19.6303010(15) 				& 1.2169066(15) \\
\cline{5-10}
 & 				& & 				& SC & DKB 				& {\multirow{2}{*}{\tablenum{8.722252397}}} 				& 10.90804(23) 				& 19.63029(23) 				& 1.21690(23) \\
 & & & 				& SC & GF 				&  				& 10.90804907(50) 				& 19.63030147(50) 				& 1.21690709(50) \\
\cline{3-10}
 & Ref.~\cite{hen} 				&  -67.924 837 74(5)  				&  49.511 443 05(6)  				& DIR & 				& {--} 				& {--} 				&  19.630 296(10)  				&  1.216 90(1)   \\
\cline{2-10}
 				& {\multirow{5}{*}{C}} 				& {\multirow{5}{*}{\tablenum{1.341667275}}} 				& {\multirow{5}{*}{\tablenum{0.054770246}}} 				& DIR & DKB 				& {--} 				& {--} 				& 
                -0.17954(29) 				& 1.21689(29) \\
 &  				&  				& & DIR & GF 				& {--} 				& {--} 				& -0.179540(15) 				& 1.216898(15) \\
\cline{5-10}
 & 				& & 				& TP & GF 				& -0.15555076(88) 				& -0.02397987(14) 				& -0.17953063(89) 				& 1.21690689(89) \\
\cline{5-10}
 & 				& & 				& SC & DKB 				& {\multirow{2}{*}{\tablenum{-0.218341867}}} 				& 0.03878(12) 				& -0.17956(12) 				& 1.21688(12) \\
 & & & 				& SC & GF 				&  				& 0.03881153(20) 				& -0.17953033(20) 				& 1.21690719(20) \\
\cline{3-10}
 & Ref.~\cite{hen} 				&  1.341 668 068(1)  				&  0.054 770 997(7)  				& DIR & 				& {--} 				& {--} 				&  -0.179 538(3)  				&  1.216 901(3)   \\
\hline
{\multirow{12}{*}{\tablenum{54}}} 				& {\multirow{5}{*}{F}} 				& {\multirow{5}{*}{\tablenum{-285.092638464}}} 				& {\multirow{5}{*}{\tablenum{190.356023847}}} 				& DIR & DKB 				& {--} 				& {--} 				& 145.7359(42) 				& 50.9993(42) \\
 &  				&  				& & DIR & GF 				& {--} 				& {--} 				& 145.733863(48) 				& 50.997249(48) \\
\cline{5-10}
 & 				& & 				& TP & GF 				& 55.33424742(67) 				& 90.3996264(62) 				& 145.7338738(62) 				& 50.9972592(62) \\
\cline{5-10}
 & 				& & 				& SC & DKB 				& {\multirow{2}{*}{\tablenum{54.420120868}}} 				& 91.3125(14) 				& 145.7327(14) 				& 50.9960(14) \\
 & & & 				& SC & GF 				&  				& 91.3137527(60) 				& 145.7338736(60) 				& 50.9972590(60) \\
\cline{3-10}
 & Ref.~\cite{hen} 				&  -285.092 638 6(1)  				&  190.356 052 0(3)  				& DIR & 				& {--} 				& {--} 				&  145.733 90(8)  				&  50.997 31(8)   \\
\cline{2-10}
 				& {\multirow{5}{*}{C}} 				& {\multirow{5}{*}{\tablenum{43.590610210}}} 				& {\multirow{5}{*}{\tablenum{17.387958323}}} 				& DIR & DKB 				& {--} 				& {--} 				& -9.9794(45) 				& 50.9992(45) \\
 &  				&  				& & DIR & GF 				& {--} 				& {--} 				& -9.981316(33) 				& 50.997253(33) \\
\cline{5-10}
 & 				& & 				& TP & GF 				& -6.72705681(91) 				& -3.2542520(41) 				& -9.9813088(42) 				& 50.9972597(42) \\
\cline{5-10}
 & 				& & 				& SC & DKB 				& {\multirow{2}{*}{\tablenum{-7.403692141}}} 				& -2.57771(19) 				& -9.98141(19) 				& 50.99716(19) \\
 & & & 				& SC & GF 				&  				& -2.5776170(22) 				& -9.9813091(22) 				& 50.9972594(22) \\
\cline{3-10}
 & Ref.~\cite{hen} 				&  43.590 621 48(6)  				&  17.387 986 7(3)  				& DIR & 				& {--} 				& {--} 				&  -9.981 343(16)  				&  50.997 27(2)   \\
\hline
{\multirow{12}{*}{\tablenum{92}}} 				& {\multirow{5}{*}{F}} 				& {\multirow{5}{*}{\tablenum{-516.318629247}}} 				& {\multirow{5}{*}{\tablenum{472.000467757}}} 				& DIR & DKB 				& {--} 				& {--} 				& 399.41(10) 				& 355.09(10) \\
 &  				&  				& & DIR & GF 				& {--} 				& {--} 				& 399.361134(33) 				& 355.042972(33) \\
\cline{5-10}
 & 				& & 				& TP & GF 				& 144.424321(12) 				& 254.936813(21) 				& 399.361134(24) 				& 355.042972(24) \\
\cline{5-10}
 & 				& & 				& SC & DKB 				& {\multirow{2}{*}{\tablenum{149.189230301}}} 				& 250.22(11) 				& 399.41(11) 				& 355.09(11) \\
 & & & 				& SC & GF 				&  				& 250.171906(27) 				& 399.361136(27) 				& 355.042975(27) \\
\cline{3-10}
 & Ref.~\cite{hen} 				&  -516.318 598(4)  				&  472.000 597(6)  				& DIR & 				& {--} 				& {--} 				&  399.361 2(2)  				&  355.043 2(2)   \\
\cline{2-10}
 				& {\multirow{5}{*}{C}} 				& {\multirow{5}{*}{\tablenum{210.068167293}}} 				& {\multirow{5}{*}{\tablenum{213.738955307}}} 				& DIR & DKB 				& {--} 				& {--} 				& -68.807(81) 				& 355.000(81) \\
 &  				&  				& & DIR & GF 				& {--} 				& {--} 				& -68.764132(30) 				& 355.042991(30) \\
\cline{5-10}
 & 				& & 				& TP & GF 				& -31.829252(29) 				& -36.9348802(21) 				& -68.764132(29) 				& 355.042991(29) \\
\cline{5-10}
 & 				& & 				& SC & DKB 				& {\multirow{2}{*}{\tablenum{-23.349279138}}} 				& -45.463(84) 				& -68.812(84) 				& 354.995(84) \\
 & & & 				& SC & GF 				&  				& -45.4148642(36) 				& -68.7641433(36) 				& 355.0429793(36) \\
\cline{3-10}
 & Ref.~\cite{hen} 				&  210.068 220 5(7)  				&  213.739 094(4)  				& DIR & 				& {--} 				& {--} 				&  -68.764 3(1)  				&  355.043 0(1)   \\
\hline 
\hline 
\end{tabular}
}}}
}
\end{table}
}

%% file: we_yerokh.tex
{\setlength\LTcapwidth{\textwidth}
\renewcommand{\arraystretch}{1.8}
\begin{table}[h]
    \centering
    \caption{\label{tab:we_yerokh}
         Individual contributions to the self-energy correction for the $1S_{1/2}$  state of hydrogen-like ions in the Feynman (F) and Coulomb (C) gauges in terms of dimensionless function $F(\alpha Z)$ defined in Eq.~(\ref{eq:FaZ}). The many-potential contribution is calculated in three ways: directly (DIR), using the two-potential (TP) scheme, and using the Sapirstein-Cheng (SC) approach. 
         For $Z=10$, the homogeneously-charged-sphere  model is used, while for other values of $Z$ the Fermi model is used.Nuclear root-mean-square radii are taken from Ref.~\cite{se_yerokh}. The values of fundamental constants are adopted from the Ref.~\cite{codata_2018}.
         }
{
\resizebox{\textwidth}{!}{
{\fontsize{12pt}{12pt}\selectfont{
    \begin{tabular}{
                l@{\quad}
                l@{\quad}
                S[table-format=3.9(2),group-separator={\;},group-minimum-digits=3]
                S[table-format=3.9(2),group-separator={\;},group-minimum-digits=3]
                l@{\quad}
                l@{\quad}
                S[table-format=3.9(2),group-separator={\;},group-minimum-digits=3]
                S[table-format=3.9(2),group-separator={\;},group-minimum-digits=3]
                S[table-format=3.9(2),group-separator={\;},group-minimum-digits=3]
                S[table-format=3.9(2),group-separator={\;},group-minimum-digits=3]
                 }
\hline
\hline 

\multicolumn{1}{l}{$Z$} & 
\multicolumn{1}{l}{Gauge $\quad$} & 
\multicolumn{1}{c}{$\Delta \varepsilon_a^{0\text{p}}$} &
\multicolumn{1}{c}{$\Delta \varepsilon_a^{1\text{p}}$} &
\multicolumn{1}{l}{Scheme $\quad$} &
\multicolumn{1}{l}{Method $\quad$} &
\multicolumn{1}{c}{$\Delta \varepsilon_a^{2\text{p}}$ / $\Delta \tilde{\varepsilon}_a^{2\text{p}}$} &
\multicolumn{1}{c}{$\Delta \varepsilon_a^{(3+)\text{p}}$ / $\Delta \tilde{\varepsilon}_a^{(3+)\text{p}}$} &
\multicolumn{1}{c}{$\Delta \varepsilon_a^{\text{Mp}}$} &
\multicolumn{1}{c}{$\Delta \varepsilon_a$} \\
\hline
{\multirow{12}{*}{\tablenum{10}}} 				& {\multirow{5}{*}{F}} 				& {\multirow{5}{*}{\tablenum{-828.249501962}}} 				& {\multirow{5}{*}{\tablenum{644.228141485}}} 				& DIR & DKB 				& {--} 				& {--} 				& 188.676(35) 				& 4.654(35) \\
 &  				&  				& & DIR & GF 				& {--} 				& {--} 				& 188.6758(19) 				& 4.6544(19) \\
\cline{5-10}
 & 				& & 				& TP & GF 				& 89.836001(29) 				& 98.8394687(35) 				& 188.675470(30) 				& 4.654110(30) \\
\cline{5-10}
 & 				& & 				& SC & DKB 				& {\multirow{2}{*}{\tablenum{87.239747265}}} 				& 101.43557(26) 				& 188.67532(26) 				& 4.65396(26) \\
 & & & 				& SC & GF 				&  				& 101.4357390(76) 				& 188.6754863(76) 				& 4.6541258(76) \\
\cline{3-10}
 & Ref.~\cite{se_yerokh} 				&  				&  				&  & 				&  				&  				&  				& 4.654 129 \\
\cline{2-10}
 				& {\multirow{5}{*}{C}} 				& {\multirow{5}{*}{\tablenum{5.502181541}}} 				& {\multirow{5}{*}{\tablenum{-0.278283767}}} 				& DIR & DKB 				& {--} 				& {--} 				& -0.57006(39) 				& 4.65384(39) \\
 &  				&  				& & DIR & GF 				& {--} 				& {--} 				& -0.56973(90) 				& 4.65416(90) \\
\cline{5-10}
 & 				& & 				& TP & GF 				& -0.519429(28) 				& -0.05035060(57) 				& -0.569780(28) 				& 4.654118(28) \\
\cline{5-10}
 & 				& & 				& SC & DKB 				& {\multirow{2}{*}{\tablenum{-0.791761730}}} 				& 0.22198(10) 				& -0.56978(10) 				& 4.65412(10) \\
 & & & 				& SC & GF 				&  				& 0.2219924(10) 				& -0.5697693(10) 				& 4.6541285(10) \\
\cline{3-10}
 & Ref.~\cite{se_yerokh} 				&  				&  				&  & 				&  				&  				&  				& 4.654 129 \\
\hline
{\multirow{12}{*}{\tablenum{18}}} 				& {\multirow{5}{*}{F}} 				& {\multirow{5}{*}{\tablenum{-192.239149447}}} 				& {\multirow{5}{*}{\tablenum{140.126027893}}} 				& DIR & DKB 				& {--} 				& {--} 				& 55.5622(83) 				& 3.4491(83) \\
 &  				&  				& & DIR & GF 				& {--} 				& {--} 				& 55.55713(13) 				& 3.44400(13) \\
\cline{5-10}
 & 				& & 				& TP & GF 				& 25.3599041(38) 				& 30.1972745(44) 				& 55.5571786(58) 				& 3.4440571(58) \\
\cline{5-10}
 & 				& & 				& SC & DKB 				& {\multirow{2}{*}{\tablenum{24.685497419}}} 				& 30.871677(93) 				& 55.557175(93) 				& 3.444053(93) \\
 & & & 				& SC & GF 				&  				& 30.8716828(40) 				& 55.5571803(40) 				& 3.4440587(40) \\
\cline{3-10}
 & Ref.~\cite{se_yerokh} 				&  				&  				&  & 				&  				&  				&  				& 3.444 059(1) \\
\cline{2-10}
 				& {\multirow{5}{*}{C}} 				& {\multirow{5}{*}{\tablenum{3.797152555}}} 				& {\multirow{5}{*}{\tablenum{0.155009412}}} 				& DIR & DKB 				& {--} 				& {--} 				& -0.508141(60) 				& 3.444021(60) \\
 &  				&  				& & DIR & GF 				& {--} 				& {--} 				& -0.508151(54) 				& 3.444011(54) \\
\cline{5-10}
 & 				& & 				& TP & GF 				& -0.4402358(25) 				& -0.06786747(86) 				& -0.5081033(26) 				& 3.4440587(26) \\
\cline{5-10}
 & 				& & 				& SC & DKB 				& {\multirow{2}{*}{\tablenum{-0.617945599}}} 				& 0.109780(67) 				& -0.508166(67) 				& 3.443996(67) \\
 & & & 				& SC & GF 				&  				& 0.10984308(95) 				& -0.50810252(95) 				& 3.44405945(95) \\
\cline{3-10}
 & Ref.~\cite{se_yerokh} 				&  				&  				&  & 				&  				&  				&  				& 3.444 059(1) \\
\hline
{\multirow{12}{*}{\tablenum{26}}} 				& {\multirow{5}{*}{F}} 				& {\multirow{5}{*}{\tablenum{-74.244868274}}} 				& {\multirow{5}{*}{\tablenum{51.657591344}}} 				& DIR & DKB 				& {--} 				& {--} 				& 25.371051(13) 				& 2.783774(13) \\
 &  				&  				& & DIR & GF 				& {--} 				& {--} 				& 25.371017(39) 				& 2.783740(39) \\
\cline{5-10}
 & 				& & 				& TP & GF 				& 11.08013293(71) 				& 14.2909085(31) 				& 25.3710414(32) 				& 2.7837645(32) \\
\cline{5-10}
 & 				& & 				& SC & DKB 				& {\multirow{2}{*}{\tablenum{10.809479388}}} 		& 14.561555(10) 				& 25.371034(10) 				& 2.783757(10) \\
 & & & 				& SC & GF 				&  				& 14.5615623(24) 				& 25.3710417(24) 				& 2.7837648(24) \\
\cline{3-10}
 & Ref.~\cite{se_yerokh} 				&  				&  				&  & 				&  				&  				&  				& 2.783 765 \\
\cline{2-10}
 				& {\multirow{5}{*}{C}} 				& {\multirow{5}{*}{\tablenum{2.891040178}}} 				& {\multirow{5}{*}{\tablenum{0.351512542}}} 				& DIR & DKB 				& {--} 				& {--} 				& -0.45869(11) 				& 2.78386(11) \\
 &  				&  				& & DIR & GF 				& {--} 				& {--} 				& -0.458806(35) 				& 2.783747(35) \\
\cline{5-10}
 & 				& & 				& TP & GF 				& -0.37722823(51) 				& -0.08155936(71) 				& -0.45878759(87) 				& 2.78376513(87) \\
\cline{5-10}
 & 				& & 				& SC & DKB 				& {\multirow{2}{*}{\tablenum{-0.497070732}}} 				& 0.038237(52) 				& -0.458833(52) 				& 2.783719(52) \\
 & & & 				& SC & GF 				&  				& 0.03828315(98) 				& -0.45878758(98) 				& 2.78376514(98) \\
\cline{3-10}
 & Ref.~\cite{se_yerokh} 				&  				&  				&  & 				&  				&  				&  				& 2.783 765 \\
\hline
{\multirow{12}{*}{\tablenum{36}}} 				& {\multirow{5}{*}{F}} 				& {\multirow{5}{*}{\tablenum{-31.021248683}}} 				& {\multirow{5}{*}{\tablenum{20.805926964}}} 				& DIR & DKB 				& {--} 				& {--} 				& 12.49483(88) 				& 2.27951(88) \\
 &  				&  				& & DIR & GF 				& {--} 				& {--} 				& 12.494636(22) 				& 2.279314(22) \\
\cline{5-10}
 & 				& & 				& TP & GF 				& 5.16845914(15) 				& 7.3261842(19) 				& 12.4946433(19) 				& 2.2793216(19) \\
\cline{5-10}
 & 				& & 				& SC & DKB 				& {\multirow{2}{*}{\tablenum{5.054839989}}} 				& 7.439761(59) 				& 12.494601(59) 				& 2.279279(59) \\
 & & & 				& SC & GF 				&  				& 7.4398034(15) 				& 12.4946434(15) 				& 2.2793217(15) \\
\cline{3-10}
 & Ref.~\cite{se_yerokh} 				&  				&  				&  & 				&  				&  				&  				& 2.279 322 \\
\cline{2-10}
 				& {\multirow{5}{*}{C}} 				& {\multirow{5}{*}{\tablenum{2.208571631}}} 				& {\multirow{5}{*}{\tablenum{0.481092901}}} 				& DIR & DKB 				& {--} 				& {--} 				& -0.41025(10) 				& 2.27941(10) \\
 &  				&  				& & DIR & GF 				& {--} 				& {--} 				& -0.410348(16) 				& 2.279317(16) \\
\cline{5-10}
 & 				& & 				& TP & GF 				& -0.31535000(13) 				& -0.09499264(77) 				& -0.41034264(78) 				& 2.27932189(78) \\
\cline{5-10}
 & 				& & 				& SC & DKB 				& {\multirow{2}{*}{\tablenum{-0.388558512}}} 				& -0.021798(21) 				& -0.410357(21) 				& 2.279308(21) \\
 & & & 				& SC & GF 				&  				& -0.02178418(49) 				& -0.41034270(49) 				& 2.27932184(49) \\
\cline{3-10}
 & Ref.~\cite{se_yerokh} 				&  				&  				&  & 				&  				&  				&  				& 2.279 322 \\
\hline
\hline
\end{tabular}
}}}
}
\end{table}
}

\setcounter{table}{2}

{\setlength\LTcapwidth{\textwidth}
\renewcommand{\arraystretch}{1.8}
\begin{table}[h]
\centering
\caption{
Self-energy, $1S_{1/2}$ state \textit{(continued)}.
     }
{
\resizebox{\textwidth}{!}{
{\fontsize{12pt}{12pt}\selectfont{
    \begin{tabular}{
                l@{\quad}
                l@{\quad}
                S[table-format=3.9(2),group-separator={\;},group-minimum-digits=3]
                S[table-format=3.9(2),group-separator={\;},group-minimum-digits=3]
                l@{\quad}
                l@{\quad}
                S[table-format=3.9(2),group-separator={\;},group-minimum-digits=3]
                S[table-format=3.9(2),group-separator={\;},group-minimum-digits=3]
                S[table-format=3.9(2),group-separator={\;},group-minimum-digits=3]
                S[table-format=3.9(2),group-separator={\;},group-minimum-digits=3]
                 }
\hline
\hline 

\multicolumn{1}{l}{$Z$} & 
\multicolumn{1}{l}{Gauge $\quad$} & 
\multicolumn{1}{c}{$\Delta \varepsilon_a^{0\text{p}}$} &
\multicolumn{1}{c}{$\Delta \varepsilon_a^{1\text{p}}$} &
\multicolumn{1}{l}{Scheme $\quad$} &
\multicolumn{1}{l}{Method $\quad$} &
\multicolumn{1}{c}{$\Delta \varepsilon_a^{2\text{p}}$ / $\Delta \tilde{\varepsilon}_a^{2\text{p}}$} &
\multicolumn{1}{c}{$\Delta \varepsilon_a^{(3+)\text{p}}$ / $\Delta \tilde{\varepsilon}_a^{(3+)\text{p}}$} &
\multicolumn{1}{c}{$\Delta \varepsilon_a^{\text{Mp}}$} &
\multicolumn{1}{c}{$\Delta \varepsilon_a$} \\
\hline
{\multirow{12}{*}{\tablenum{54}}} 				& {\multirow{5}{*}{F}} 				& {\multirow{5}{*}{\tablenum{-9.961275619}}} 				& {\multirow{5}{*}{\tablenum{6.651147516}}} 				& DIR & DKB 				& {--} 				& {--} 				& 5.09205(49) 				& 1.78192(49) \\
 &  				&  				& & DIR & GF 				& {--} 				& {--} 				& 5.0920137(53) 				& 1.7818856(53) \\
\cline{5-10}
 & 				& & 				& TP & GF 				& 1.933405159(24) 				& 3.15860971(91) 				& 5.09201487(91) 				& 1.78188677(91) \\
\cline{5-10}
 & 				& & 				& SC & DKB 				& {\multirow{2}{*}{\tablenum{1.901467097}}} 				& 3.190533(21) 				& 5.092000(21) 				& 1.781872(21) \\
 & & & 				& SC & GF 				&  				& 3.19054778(77) 				& 5.09201487(77) 				& 1.78188677(77) \\
\cline{3-10}
 & Ref.~\cite{se_yerokh} 				&  				&  				&  & 				&  				&  				&  				& {1.781 887(2)(2)} \\
\cline{2-10}
 				& {\multirow{5}{*}{C}} 				& {\multirow{5}{*}{\tablenum{1.523080462}}} 				& {\multirow{5}{*}{\tablenum{0.607564155}}} 				& DIR & DKB 				& {--} 				& {--} 				& -0.348739(67) 				& 1.781906(67) \\
 &  				&  				& & DIR & GF 				& {--} 				& {--} 				& -0.3487586(38) 				& 1.7818860(38) \\
\cline{5-10}
 & 				& & 				& TP & GF 				& -0.235050090(32) 				& -0.11370765(47) 				& -0.34875774(47) 				& 1.78188688(47) \\
\cline{5-10}
 & 				& & 				& SC & DKB 				& {\multirow{2}{*}{\tablenum{-0.258689306}}} 				& -0.090087(20) 				& -0.348777(20) 				& 1.781868(20) \\
 & & & 				& SC & GF 				&  				& -0.09006848(27) 				& -0.34875778(27) 				& 1.78188683(27) \\
\cline{3-10}
 & Ref.~\cite{se_yerokh} 				&  				&  				&  & 				&  				&  				&  				& {1.781 887(2)(2)} \\
\hline
{\multirow{12}{*}{\tablenum{82}}} 				& {\multirow{5}{*}{F}} 				& {\multirow{5}{*}{\tablenum{-2.952357937}}} 				& {\multirow{5}{*}{\tablenum{2.371873437}}} 				& DIR & DKB 				& {--} 				& {--} 				& 2.067807(47) 				& 1.487323(47) \\
 &  				&  				& & DIR & GF 				& {--} 				& {--} 				& 2.06777005(56) 				& 1.48728555(56) \\
\cline{5-10}
 & 				& & 				& TP & GF 				& 0.7380290908(61) 				& 1.32974101(32) 				& 2.06777010(32) 				& 1.48728560(32) \\
\cline{5-10}
 & 				& & 				& SC & DKB 				& {\multirow{2}{*}{\tablenum{0.745897727}}} 				& 1.321869(19) 				& 2.067766(19) 				& 1.487282(19) \\
 & & & 				& SC & GF 				&  				& 1.32187237(33) 				& 2.06777010(33) 				& 1.48728560(33) \\
\cline{3-10}
 & Ref.~\cite{se_yerokh} 				&  				&  				&  & 				&  				&  				&  				& {\; 1.487 286(16)(3)} \\
\cline{2-10}
 				& {\multirow{5}{*}{C}} 				& {\multirow{5}{*}{\tablenum{0.994837254}}} 				& {\multirow{5}{*}{\tablenum{0.787437130}}} 				& DIR & DKB 				& {--} 				& {--} 				& -0.29496(11) 				& 1.48731(11) \\
 &  				&  				& & DIR & GF 				& {--} 				& {--} 				& -0.29498882(36) 				& 1.48728557(36) \\
\cline{5-10}
 & 				& & 				& TP & GF 				& -0.153987847(36) 				& -0.14100092(20) 				& -0.29498877(20) 				& 1.48728561(20) \\
\cline{5-10}
 & 				& & 				& SC & DKB 				& {\multirow{2}{*}{\tablenum{-0.132956055}}} 				& -0.162040(11) 				& -0.294996(11) 				& 1.487278(11) \\
 & & & 				& SC & GF 				&  				& -0.16203273(14) 				& -0.29498878(14) 				& 1.48728560(14) \\
\cline{3-10}
 & Ref.~\cite{se_yerokh} 				&  				&  				&  & 				&  				&  				&  				& {\; 1.487 286(16)(3)} \\
\hline
{\multirow{12}{*}{\tablenum{92}}} 				& {\multirow{5}{*}{F}} 				& {\multirow{5}{*}{\tablenum{-2.141311301}}} 				& {\multirow{5}{*}{\tablenum{1.957559518}}} 				& DIR & DKB 				& {--} 				& {--} 				& 1.656244(75) 				& 1.472492(75) \\
 &  				&  				& & DIR & GF 				& {--} 				& {--} 				& 1.65624014(29) 				& 1.47248836(29) \\
\cline{5-10}
 & 				& & 				& TP & GF 				& 0.5989613403(33) 				& 1.05727883(17) 				& 1.65624017(17) 				& 1.47248839(17) \\
\cline{5-10}
 & 				& & 				& SC & DKB 				& {\multirow{2}{*}{\tablenum{0.618726402}}} 				& 1.037484(40) 				& 1.656210(40) 				& 1.472458(40) \\
 & & & 				& SC & GF 				&  				& 1.03751375(23) 				& 1.65624016(23) 				& 1.47248837(23) \\
\cline{3-10}
 & Ref.~\cite{se_yerokh} 				&  				&  				&  & 				&  				&  				&  				& {1.472 50(1)(2)} \\
\cline{2-10}
 				& {\multirow{5}{*}{C}} 				& {\multirow{5}{*}{\tablenum{0.871179702}}} 				& {\multirow{5}{*}{\tablenum{0.886500900}}} 				& DIR & DKB 				& {--} 				& {--} 				& -0.2851905(40) 				& 1.4724901(40) \\
 &  				&  				& & DIR & GF 				& {--} 				& {--} 				& -0.28519206(32) 				& 1.47248854(32) \\
\cline{5-10}
 & 				& & 				& TP & GF 				& -0.132008642(12) 				& -0.153183501(35) 				& -0.285192142(37) 				& 1.472488460(37) \\
\cline{5-10}
 & 				& & 				& SC & DKB 				& {\multirow{2}{*}{\tablenum{-0.096833870}}} 				& -0.188360(17) 				& -0.285194(17) 				& 1.472487(17) \\
 & & & 				& SC & GF 				&  				& -0.188358295(21) 				& -0.285192165(21) 				& 1.472488437(21) \\
\cline{3-10}
 & Ref.~\cite{se_yerokh} 				&  				&  				&  & 				&  				&  				&  				& {1.472 50(1)(2)} \\
 \hline
{\multirow{8}{*}{\tablenum{100}}} 				& {\multirow{3}{*}{F}} 				& {\multirow{3}{*}{\tablenum{-1.737050769}}} 				& {\multirow{3}{*}{\tablenum{1.795888856}}} 				& DIR & GF 				& {--} 				& {--} 				& 1.43730086(13) 				& 1.49613895(13) \\
\cline{5-10}
 & 				& & 				& TP & GF 				& 0.534332311 				& 0.90296855(10) 				& 1.43730086(10) 				& 1.49613894(10) \\
\cline{5-10}
 & 				& & 				& SC & GF 				& {\multirow{1}{*}{\tablenum{0.565473422}}}
 & 0.87182740(16) 				& 1.43730082(16) 				& 1.49613891(16) \\
\cline{3-10}
 & Ref.~\cite{se_yerokh} 				&  				&  				& DIR & 				& 				& 				& 				& {1.496 14(7)(47)} \\
\cline{2-10}
 				& {\multirow{3}{*}{C}} 				& {\multirow{3}{*}{\tablenum{0.780926903}}} 				& {\multirow{3}{*}{\tablenum{0.996469025}}} 				& DIR & GF 				& {--} 				& {--} 				& -0.28125683(39) 				& 1.49613910(39) \\
\cline{5-10}
 & 				& & 				& TP & GF 				& -0.116019006(13) 				& -0.165237923(60) 				& -0.281256929(62) 				& 1.496138999(62) \\
\cline{5-10}
 & 				& & 				& SC & GF 				& {\multirow{1}{*}{\tablenum{-0.068173067}}} 
 & -0.213083879(35) 				& -0.281256946(35) 				& 1.496138982(35) \\
\cline{3-10}
 & Ref.~\cite{se_yerokh} 				&  				&  				& DIR & 				& 				& 				& 				& {1.496 14(7)(47)} \\
\hline
\hline
\end{tabular}
}}}
}
\end{table}
}

%% file: we_yerokh_excited_states.tex

{\setlength\LTcapwidth{\textwidth}
\renewcommand{\arraystretch}{1.53}
\begin{longtable}{
                l@{\quad}
                l@{\quad}
                l@{\quad}
                S[table-format=3.9(2),group-separator={\;},group-minimum-digits=3]
                S[table-format=3.9(2),group-separator={\;},group-minimum-digits=3]
                S[table-format=3.9(2),group-separator={\;},group-minimum-digits=3]
                 }

\caption{\label{tab:we_yerokh_excited_states}
         Contributions to the self-energy corrections for the $2S_{1/2}$, $2P_{1/2}$, and $2P_{3/2}$ states of  hydrogen-like ions. Results are given in terms of the dimensionless function $F(\alpha Z)$ defined in Eq.~(\ref{eq:FaZ}. For $Z=10$, the homogeneously-charged-sphere  model is used, while for other values of $Z$ the Fermi model is used.Nuclear root-mean-square radii are taken from Ref.~\cite{se_yerokh}. The values of fundamental constants are adopted from Ref.~\cite{codata_2018}. The many-potential contribution is obtained within the Sapirstein-Cheng convergence-acceleration scheme. 
         }\\

\toprule
\toprule

\multicolumn{1}{l}{$Z$} & 
\multicolumn{1}{l}{Gauge} & &
\multicolumn{1}{c}{$2S_{1/2}$} &
\multicolumn{1}{c}{$2P_{1/2}$} & 
\multicolumn{1}{c}{$2P_{3/2}$} \\

\midrule

\endfirsthead

\caption[]{Self-energy, excited states \textit{(continued)}.}\\

\toprule
\toprule

\multicolumn{1}{l}{$Z$} & 
\multicolumn{1}{l}{Gauge} & &
\multicolumn{1}{c}{$2S_{1/2}$} &
\multicolumn{1}{c}{$2P_{1/2}$} & 
\multicolumn{1}{c}{$2P_{3/2}$} \\

\midrule  

\endhead
\endfoot
\bottomrule
\bottomrule
\endlastfoot

	10 & 
	F & 
	$\Delta \varepsilon_a^{0\text{p}}$ & 
	-2075.579237771 & 
	-2196.693661912 & 
	-2192.070770920 \\

	 & 
	 & 
	$\Delta \varepsilon_a^{1\text{p}}$ & 
	1719.050474880 & 
	1818.840183393 & 
	1815.473220245 \\

	 & 
	 & 
	$\Delta \varepsilon_a^{\text{Mp}}$ & 
	361.423177(95) & 
	377.73868(14) & 
	376.727908(77) \\

	 & 
	 & 
	$\Delta \varepsilon_a$ & 
	4.894415(95) & 
	-0.11480(14) & 
	0.130358(77) \\

	 & 
	C & 
	$\Delta \varepsilon_a^{0\text{p}}$ & 
	7.190429621 & 
	2.624890307 & 
	2.775883599 \\

	 & 
	 & 
	$\Delta \varepsilon_a^{1\text{p}}$ & 
	-1.534430545 & 
	-2.119896459 & 
	-2.038942518 \\

	 & 
	 & 
	$\Delta \varepsilon_a^{\text{Mp}}$ & 
	-0.761608(15) & 
	-0.619823(41) & 
	-0.6065870(90) \\

	 & 
	 & 
	$\Delta \varepsilon_a$ & 
	4.894391(15) & 
	-0.114829(41) & 
	0.1303540(90) \\

	 & 
	 & 
	$\Delta \varepsilon_a$, \cite{se_yerokh} & 
		4.894 384(7)        & 
	-0.114 84(2)   & 
	    0.13036(2) \\
	
\hline

	18 & 
	F & 
	$\Delta \varepsilon_a^{0\text{p}}$ & 
	-510.933712531 & 
	-547.071566494 & 
	-543.528241119 \\

	 & 
	 & 
	$\Delta \varepsilon_a^{1\text{p}}$ & 
	405.527774766 & 
	431.854905602 & 
	429.503304379 \\

	 & 
	 & 
	$\Delta \varepsilon_a^{\text{Mp}}$ & 
	109.105831(51) & 
	115.119152(77) & 
	114.165747(15) \\

	 & 
	 & 
	$\Delta \varepsilon_a$ & 
	3.699893(51) & 
	-0.097509(77) & 
	0.140810(15) \\

	 & 
	C & 
	$\Delta \varepsilon_a^{0\text{p}}$ & 
	5.135244411 & 
	2.027253175 & 
	2.157952816 \\

	 & 
	 & 
	$\Delta \varepsilon_a^{1\text{p}}$ & 
	-0.736198502 & 
	-1.519146727 & 
	-1.443991645 \\

	 & 
	 & 
	$\Delta \varepsilon_a^{\text{Mp}}$ & 
	-0.699151(12) & 
	-0.605613(34) & 
	-0.5731412(22) \\

	 & 
	 & 
	$\Delta \varepsilon_a$ & 
	3.699895(12) & 
	-0.097507(34) & 
	0.1408199(22) \\

	 & 
	 & 
	$\Delta \varepsilon_a$, \cite{se_yerokh} & 
		 3.699 892          & 
	-0.097 511(4)  & 
	    0.140819 \\
	
\hline

	26 & 
	F & 
	$\Delta \varepsilon_a^{0\text{p}}$ & 
	-207.664305943 & 
	-224.482210414 & 
	-221.561948805 \\

	 & 
	 & 
	$\Delta \varepsilon_a^{1\text{p}}$ & 
	159.521301465 & 
	169.968233475 & 
	168.173153568 \\

	 & 
	 & 
	$\Delta \varepsilon_a^{\text{Mp}}$ & 
	51.202289(18) & 
	54.437750(31) & 
	53.542405(15) \\

	 & 
	 & 
	$\Delta \varepsilon_a$ & 
	3.059284(18) & 
	-0.076227(31) & 
	0.153610(15) \\

	 & 
	C & 
	$\Delta \varepsilon_a^{0\text{p}}$ & 
	4.030422889 & 
	1.681608238 & 
	1.790274523 \\

	 & 
	 & 
	$\Delta \varepsilon_a^{1\text{p}}$ & 
	-0.321172956 & 
	-1.166082251 & 
	-1.100585293 \\

	 & 
	 & 
	$\Delta \varepsilon_a^{\text{Mp}}$ & 
	-0.6499581(65) & 
	-0.591744(19) & 
	-0.5360707(33) \\

	 & 
	 & 
	$\Delta \varepsilon_a$ & 
	3.0592919(65) & 
	-0.076218(19) & 
	0.1536185(33) \\

	 & 
	 & 
	$\Delta \varepsilon_a$, \cite{se_yerokh} & 
		 3.059 292(1)       & 
	-0.076 218(1)  & 
	    0.153620 \\
	
\hline

	36 & 
	F & 
	$\Delta \varepsilon_a^{0\text{p}}$ & 
	-92.012084971 & 
	-100.545376092 & 
	-98.122650921 \\

	 & 
	 & 
	$\Delta \varepsilon_a^{1\text{p}}$ & 
	68.478913119 & 
	72.554183239 & 
	71.176248946 \\

	 & 
	 & 
	$\Delta \varepsilon_a^{\text{Mp}}$ & 
	26.118135(10) & 
	27.946190(13) & 
	27.118204(15) \\

	 & 
	 & 
	$\Delta \varepsilon_a$ & 
	2.584964(10) & 
	-0.045003(13) & 
	0.171802(15) \\

	 & 
	C & 
	$\Delta \varepsilon_a^{0\text{p}}$ & 
	3.191617572 & 
	1.402984217 & 
	1.483773405 \\

	 & 
	 & 
	$\Delta \varepsilon_a^{1\text{p}}$ & 
	-0.003718984 & 
	-0.871022395 & 
	-0.823673062 \\

	 & 
	 & 
	$\Delta \varepsilon_a^{\text{Mp}}$ & 
	-0.6029277(25) & 
	-0.5769556(70) & 
	-0.4882934(27) \\

	 & 
	 & 
	$\Delta \varepsilon_a$ & 
	2.5849709(25) & 
	-0.0449938(70) & 
	0.1718069(27) \\

	 & 
	 & 
	$\Delta \varepsilon_a$, \cite{se_yerokh} & 
		 2.584 972          & 
	-0.044 991     & 
	    0.171808 \\
	
\hline

	54 & 
	F & 
	$\Delta \varepsilon_a^{0\text{p}}$ & 
	-32.616716171 & 
	-36.324811900 & 
	-34.414650841 \\

	 & 
	 & 
	$\Delta \varepsilon_a^{1\text{p}}$ & 
	23.445352616 & 
	24.194742894 & 
	23.197273689 \\

	 & 
	 & 
	$\Delta \varepsilon_a^{\text{Mp}}$ & 
	11.331970(13) & 
	12.155676(19) & 
	11.4259381(82) \\

	 & 
	 & 
	$\Delta \varepsilon_a$ & 
	2.160606(13) & 
	0.025607(19) & 
	0.2085609(82) \\

	 & 
	C & 
	$\Delta \varepsilon_a^{0\text{p}}$ & 
	2.346175319 & 
	1.103617134 & 
	1.136073664 \\

	 & 
	 & 
	$\Delta \varepsilon_a^{1\text{p}}$ & 
	0.361787949 & 
	-0.516470137 & 
	-0.523758607 \\

	 & 
	 & 
	$\Delta \varepsilon_a^{\text{Mp}}$ & 
	-0.5473526(17) & 
	-0.5615339(54) & 
	-0.4037522(13) \\

	 & 
	 & 
	$\Delta \varepsilon_a$ & 
	2.1606107(17) & 
	0.0256131(54) & 
	0.2085629(13) \\

	 & 
	 & 
	$\Delta \varepsilon_a$, \cite{se_yerokh} & 
		 {2.160 612(2)(3)}  & 
	0.025 617      & 
	    0.208563 \\
	
\hline

	82 & 
	F & 
	$\Delta \varepsilon_a^{0\text{p}}$ & 
	-11.119815486 & 
	-12.768564627 & 
	-11.152418631 \\

	 & 
	 & 
	$\Delta \varepsilon_a^{1\text{p}}$ & 
	8.151898860 & 
	7.689867435 & 
	6.797162744 \\

	 & 
	 & 
	$\Delta \varepsilon_a^{\text{Mp}}$ & 
	5.0332808(93) & 
	5.284306(16) & 
	4.6269758(34) \\

	 & 
	 & 
	$\Delta \varepsilon_a$ & 
	2.0653642(93) & 
	0.205609(16) & 
	0.2717199(34) \\

	 & 
	C & 
	$\Delta \varepsilon_a^{0\text{p}}$ & 
	1.703650832 & 
	0.860750394 & 
	0.829247308 \\

	 & 
	 & 
	$\Delta \varepsilon_a^{1\text{p}}$ & 
	0.875870907 & 
	-0.073015197 & 
	-0.274656464 \\

	 & 
	 & 
	$\Delta \varepsilon_a^{\text{Mp}}$ & 
	-0.5141558(23) & 
	-0.5821238(57) & 
	-0.28287037(79) \\

	 & 
	 & 
	$\Delta \varepsilon_a$ & 
	2.0653659(23) & 
	0.2056114(57) & 
	0.27172047(79) \\

	 & 
	 & 
	$\Delta \varepsilon_a$, \cite{se_yerokh} & 
		 {\;\;2.065 367(23)(5)}   		 & 
	{0.205 613(1)(0)}     	 & 
	0.271721						 \\
	
\hline

	92 & 
	F & 
	$\Delta \varepsilon_a^{0\text{p}}$ & 
	-8.389628927 & 
	-9.729683232 & 
	-8.095704176 \\

	 & 
	 & 
	$\Delta \varepsilon_a^{1\text{p}}$ & 
	6.410146198 & 
	5.762419098 & 
	4.775608621 \\

	 & 
	 & 
	$\Delta \varepsilon_a^{\text{Mp}}$ & 
	4.1499859(48) & 
	4.2841266(80) & 
	3.6151366(45) \\

	 & 
	 & 
	$\Delta \varepsilon_a$ & 
	2.1705032(48) & 
	0.3168625(80) & 
	0.2950411(45) \\

	 & 
	C & 
	$\Delta \varepsilon_a^{0\text{p}}$ & 
	1.553629935 & 
	0.800352700 & 
	0.755392208 \\

	 & 
	 & 
	$\Delta \varepsilon_a^{1\text{p}}$ & 
	1.134051703 & 
	0.128027677 & 
	-0.216677107 \\

	 & 
	 & 
	$\Delta \varepsilon_a^{\text{Mp}}$ & 
	-0.51717660(74) & 
	-0.6115153(27) & 
	-0.24367249(13) \\

	 & 
	 & 
	$\Delta \varepsilon_a$ & 
	2.17050503(74) & 
	0.3168651(27) & 
	0.29504261(13) \\

	 & 
	 & 
	$\Delta \varepsilon_a$, \cite{se_yerokh} & 
		 {\!\!\!2.170 52(2)(2)}     	 & 
	{0.316 869(2)(2)}     	 & 
	{0.295 043(0)(1)}				 \\
	
\hline

	100 & 
	F & 
	$\Delta \varepsilon_a^{0\text{p}}$ & 
	-6.953875290 & 
	-8.114515401 & 
	-6.400104490 \\

	 & 
	 & 
	$\Delta \varepsilon_a^{1\text{p}}$ & 
	5.597189494 & 
	4.824500696 & 
	3.682814800 \\

	 & 
	 & 
	$\Delta \varepsilon_a^{\text{Mp}}$ & 
	3.6813603(44) & 
	3.7350259(74) & 
	3.0308200(39) \\

	 & 
	 & 
	$\Delta \varepsilon_a$ & 
	2.3246745(44) & 
	0.4450112(74) & 
	0.3135304(39) \\

	 & 
	C & 
	$\Delta \varepsilon_a^{0\text{p}}$ & 
	1.438749738 & 
	0.749905361 & 
	0.704955752 \\

	 & 
	 & 
	$\Delta \varepsilon_a^{1\text{p}}$ & 
	1.413259460 & 
	0.344771111 & 
	-0.177247595 \\

	 & 
	 & 
	$\Delta \varepsilon_a^{\text{Mp}}$ & 
	-0.52733313(74) & 
	-0.6496630(26) & 
	-0.21417647(12) \\

	 & 
	 & 
	$\Delta \varepsilon_a$ & 
	2.32467606(74) & 
	0.4450134(26) & 
	0.31353168(12) \\

	 & 
	 & 
	$\Delta \varepsilon_a$, \cite{se_yerokh} & 
		 {\!\!\!\!\!\!2.324 7(1)(8)}     & 
	  {\!\!\!0.445 01(1)(9)}  & 
	     {\!\!\!0.313 53(0)(2)}	 \\
	
\end{longtable}
}